\documentclass[traditabstract]{aa} 
\usepackage{graphicx}
\usepackage{txfonts}
\begin{document}
   \title{Modelling the spectral energy distribution of galaxies.\thanks
{We dedicate this paper to the memory of Angelos Misiriotis, sorely
missed as a friend, collaborator and exceptional scientist.}} 

   \subtitle{V. The dust and PAH emission SEDs of disk galaxies}

   \author{Cristina C. Popescu\inst{1}
          \and
          Richard J. Tuffs\inst{2}
          \and
          Michael A. Dopita\inst{3}
          \and
          J\"org Fischera\inst{3}
          \and
          Nikolaos D. Kylafis\inst{4,5}
          \and
          \newline Barry F. Madore\inst{6}
          }

   \institute{Jeremiah Horrocks Institute for Astrophysics and Supercomputing, 
              University of Central Lancashire,
              PR1 2HE, Preston, UK\\
              \email{cpopescu@uclan.ac.uk}
         \and
             Max Planck Institut f\"ur Kernphysik, Saupfercheckweg 1, D-69117
             Heidelberg, Germany\\
             \email{Richard.Tuffs@mpi-hd.mpg.de}
          \and
              Research School of Astronomy \& Astrophysics, 
              The Australian National University, Cotter Road,
              Weston Creek ACT 2611, Australia
           \and
              University of Crete, Physics Department and Institute of 
              Theoretical and Computational Physics,  
              71003 Heraklion, Crete, Greece
            \and
             Foundation for Research and Technology - Hellas, 71110 Heraklion,
             Crete, Greece  
            \and
              Observatories of the Carnegie Institution of Washington, 
              813 Santa Barbara Street, Pasadena, CA 91101
              }

   \date{Received; accepted}

 
  \abstract
  {We present a self-consistent model of the spectral energy distributions (SEDs)
of spiral galaxies from the ultraviolet (UV) to the mid-infrared
(MIR)/far-infrared (FIR)/submillimeter (submm) based on a full radiative
transfer calculation of the propagation of starlight in galaxy disks. This
model predicts not only the total integrated energy absorbed in the UV/optical
and re-emitted in the infrared/submm, but also the colours of the dust emission
based on an explicit calculation of the strength and colour of the UV/optical
radiation fields heating the dust, and incorporating a full calculation of the
stochastic heating of small dust grains and PAH molecules.

The geometry of the translucent components of the model is empirically
constrained using the results from the radiation transfer analysis of Xilouris
et al. on spirals in the middle range of the Hubble sequence, while the
geometry of the optically thick components is constrained from physical
considerations with a posteriori checks of the model predictions with
observational data. Following the observational constraints, the model has both
a distribution of diffuse dust associated with the old and young disk stellar
populations as well as a clumpy component arising from dust in the parent
molecular clouds in star forming regions. In accordance with the fragmented
nature of dense molecular gas in typical star-forming regions, UV light from
massive stars is allowed to either freely stream away into the diffuse medium
in some fraction of directions or be geometrically blocked and locally absorbed
in clumps.

These geometrical constraints enable the dust emission to be predicted in terms
of a minimum set of free parameters: the central face-on dust opacity in the
B-band $\tau^f_B$, a clumpiness factor $F$ for the star-forming regions, the
star-formation rate $SFR$, the normalised luminosity of the old stellar
population $old$ and the bulge-to-disk ratio $B/D$. We show that these
parameters are almost orthogonal in their predicted effect on the colours of
the dust/PAH emission. In most practical applications $B/D$ will actually not
be a free parameter but (together with the angular size ${\theta}_{gal}$ and
inclination $i$ of the disk) act as a constraint derived from morphological
decomposition of higher resolution optical images. This also extends the range
of applicability of the model along the Hubble sequence. We further show that
the dependence of the dust emission SED on the colour of the stellar photon
field depends primarily on the ratio between the luminosities of the young and
old stellar populations (as specified by the parameters $SFR$ and $old$) rather
than on the detailed colour of the emissions from either of these populations.
The model is thereby independent of a priori assumptions of the detailed
mathematical form of the dependence of SFR on time, allowing UV/optical SEDs to
be dereddened without recourse to population synthesis models.

Utilising these findings, we show how the predictive power of radiative
transfer calculations can be combined with measurements of ${\theta}_{gal}$,
$i$ and $B/D$ from optical images to self-consistently fit
UV/optical-MIR/FIR/submm SEDs observed in large statistical surveys in a fast
and flexible way, deriving physical parameters on an object-by-object basis. We
also identify a non-parametric test of the fidelity of the model in practical
applications through comparison of the model predictions for FIR colour and
surface brightness with the corresponding observed quantities. This should be
effective in identifying objects such as AGNs or star-forming galaxies with
markedly different geometries to those of the calibrators of Xilouris et al.
The results of the calculations are made available in the form of a large
library of simulated dust emission SEDs spanning the whole parameter space of
our model, together with the corresponding library of dust attenuation
calculated using the same model. 
}


   \maketitle
%

\section{Introduction}
\label{sec:introduction}
Copious quantities of interstellar dust are present in the disks of all
metal rich star-forming galaxies. This dust pervades all
components of the interstellar medium (ISM), ranging from the
diffuse ionised and neutral medium filling most of the volume
of the gaseous disk, through embedded neutral and molecular
clouds of intermediate sizes and densities, down to the dense
cloud cores on sub-parsec scales which are the sites of formation
of stars. The ubiquity and high abundance (relative to the
available pool of interstellar metals) of grains
is a fundamental result of the solid state being
the favoured repository for refractory elements
in all but the coronal component of the ISM, and affects the very
nature of galaxies. In particular, dust plays a major role
in determining the thermodynamic balance
of the ISM, through photoelectric heating and inelastic interactions
with gas species, influencing the propensity of galaxies
to accrete, cool and condense gas into stars (see e.g. Popescu \& Tuffs
2010). In view of the physical role played by dust particles in the
process of star formation, it can be regarded as a minor perversity of
nature that, by virtue of their strong interaction cross section with
stellar photons, the very same particles strongly inhibit and distort
our view of the resulting stellar populations, thus preventing a straightforward
confrontation of theories for star formation in galaxies with
observations.

Statistical studies of the luminosity and colour distributions of
large samples of galaxies seen at different orientations
(most notably the attenuation-inclination relation) have shown the severity 
of this effect even in the optical bands, and not just in exceptionally opaque
systems, but in the major part of the population of star-forming disk
galaxies in the local Universe
(Driver et al. 2007, Shao et al. 2007, Choi 2007, Driver et al. 2008,
Unterborn \& Ryden 2008, Padilla \& Strauss 2008, Cho \& Park 2009,
Maller et al. 2009, Ganda et al. 2009, Masters et al. 2010). This means
that we not only need to solve the problem of deriving intrinsic stellar
emissivities from newly formed stars emitting
predominantly in the ultraviolet, but also need to decode the
effect of dust on the amplitude and colours of optical light to
derive the SF history of galaxies. This is fundamentally an
ill-constrained problem using broad-band optical data alone, due to the
age/metallicity/opacity degeneracy (Silva \& Elston 1994, Worthey 1994, 
Li \& Han 2008). 
Even for galaxies bright
enough for constraints on dust attenuation to be made through optical
spectroscopy, for example from measurements of the emission from
hydrogen recombination lines, large uncertainties still prevail
due to the complex geometry of dust in galaxies and the different
effect of this dust on the light from stellar populations of different ages.

In recent years the advent of spaceborne infrared astronomy has meant that
we have now the capability of viewing the absorbed energy of the starlight
in galaxies, which would seem to offer a solution to unravelling
the effects of dust, at least from an observational perspective.
Star-forming disk galaxies are observed to have most of
this absorbed energy re-radiated in the far-infrared (FIR), with a
significant amount also in the mid-infrared (MIR) and in the Polycyclic
Aromatic Carbon (PAH) bands (see review of Sauvage et al. 2005). Simple
energy balance measurements indicate that indeed a large fraction of the stellar
light is absorbed even in relatively quiescent spiral galaxies in the local
universe (Popescu \& Tuffs 2002, Xu et al. 2006, Driver et al.
2008). Commonly, the total re-radiated energy is estimated by
fitting the (typically sparsely sampled) measurements of PAH/dust emission 
either with phenomenological SED models (Devriendt et
al. 1999, Dale \& Helou 2002, Draine \& Li 2007) or with SED
models that use templates (Xu et al. 1998, Sajina et al. 2006,
Marshall et al. 2007, da Cunha et al. 2008). However, this
approach is insufficient to derive the intrinsic UV and optical
emission of galaxies, even in cases where the total infrared emission
can be accurately estimated. 

There are several reasons why obtaining information about the
relative absorption probabilities of UV/optical photons as a function of
wavelength is a very complex problem, which
cannot be solved by means of the application of single wavelength
dependent correction factors as is the case for extinction corrections
for stars
\footnote{We note that attenuation is an integral property of an
extended distribution (e.g. a galaxy) of light and should not be confused
with the extinction measured for point sources (e.g. single stars).
While the
extinction is simply proportional to the column density of dust and its
wavelength dependence is determined by the optical properties of the
grains, the attenuation depends on the distribution  of dust and stars, the
variation of dust properties and the orientation of the galaxies}.

Firstly, as already noted, the ISM is highly structured,
with dust present in both large scale and small scale structures.
This strongly affects the dependence of attenuation on UV/optical wavelength,
even for uniform geometrical distributions of stars, as shown by the
study of attenuation of starlight in a turbulent ISM with a log-normal density
distribution by Fischera \& Dopita (2005). Similarly, various
studies have shown that
different mass fractions of dust in clumps and diffuse structures can
affect the amplitude and wavelength dependence both of the UV/optical
attenuation and of the dust emission
(Kuchinski et al. 1998, Bianchi et al. 2000a, Witt \& Gordon 2000,
Misselt et al. 2001, Misiriotis \& Bianchi 2002, Pierini et al. 2004,
Tuffs et al. 2004, Bianchi 2008). All these studies indicate that
models for the attenuation of starlight and its re-emission
in the infrared will only have predictive power if independent
constraints on the geometry of the dust-bearing structures in the ISM are 
available.

Secondly, stellar populations of different ages
have systematically different geometrical relations to these dust structures
(Silva et al. 1998, Charlot \& Fall 2000, Popescu et al. 2000)
due to the disruption of birth clouds through the action of stellar
winds and supernovae, the migration of stars away from these clouds
with time and the subsequent dynamical evolution of stellar populations
on Gyr timescales from a thin disk to a thick disk configuration (Wielen 1977).

Thirdly, the response of grains to light is not just a question
of the wavelength and flux of the photons hitting them
but also of their optical properties
(dependent on shape, size and composition; see review by Draine 2010).
A particularly complex aspect of this is the
account that should in principle be made of any systematic differences
in composition on location in the ISM, such as for example the
transition between pure refractory species in the diffuse ISM irradiated
by UV to ices and ice-coated grains in opaque molecular clouds.
Furthermore, even supposing the composition and size distribution is
known, the stochastic emission of small dust grains responding to
impulsive heating has to be calculated throughout the volume, which is
computationally challenging.

Arguably, the most fundamental and challenging of these problems is
the determination of the geometry of the dust-bearing structures
in the ISM, since knowledge of this is a pre-requisite to obtaining
a self-consistent solution embracing the amplitude and geometry of
the stellar populations and the optical properties of the grains.
Where FIR data is available, a powerful way of constraining the geometry of
dust on all scales is to use the fact that grains act as test 
particles with FIR colours characteristic of the intensity and colour
of the interstellar radiation fields (ISRF), 
which in turn are a strong function of morphological 
structures within galaxies. 
For example grains locally heated by the strong radiation
fields within embedded star formation regions emit an infrared radiation 
that peaks around 60\,${\mu}$m, while grains heated by the diffuse radiation 
fields in the disk of galaxies will emit an infrared radiation that typically 
peaks well longwards of 100\,${\mu}$m, but also exhibit
substantial emission in the MIR
region due to the stochastic heating of small grains (see Sauvage et
al. 2005). In general using dust grains as tracers of the ISRF is a key way to 
constrain both opaque and translucent components of the ISM on all spatial
scales,  and the technical implementation of this approach is to use radiative
transfer (RT) calculations. Much effort has been
put into this problem of developing efficient RT codes, and there are now
several available in the literature (see Kylafis \& Xilouris 2005 for a
comprehensive review on radiative transfer techniques used in galaxy 
modelling). Specific RT calculations of the integrated
MIR/FIR/submm emission from grains heated by starlight in disk galaxies
have been made for various specifications of the geometry of stars and dust
by Siebenmorgen \& Kr\"ugel 1992, Silva et al. 1998, Popescu et al. 2000,
Efstathiou \& Rowan-Robinson 2003, Piovan et al. 2006, Bianchi 2008.

Further empirical constraints on the geometry of dust in galaxies
are possible for systems where major dust-bearing components of the ISM
are translucent, thus allowing information about the distributions of dust
and stars on scales of hundreds of parsecs to be inferred from
images of the optical emission.  This approach, which also relies on
modelling of the data with RT techniques, was adopted by 
Xilouris et al. (1999) in their investigation of the large scale
diffuse components in edge-on spiral galaxies, revealing
reproducible trends in the geometry of stars with respect to the dust.
The analysis of Xilouris et al. was also notable in that it
was also able to measure the
extinction law for the visible component of dust for these systems,
showing it to be consistent with that of dust in the diffuse ISM of the
Milky Way in the observed optical/NIR range. In principle the geometric
constraints on the distributions of stellar emissivity and dust derived
from optical observations of translucent regions
can be extended to more obscured components of the ISM, such as
for example the molecular layer in the disk out of which stars are forming,
by making use of the physical connection between gas, stars and dust to
further constrain the geometry of the problem. 
Overall, our knowledge of galaxies, at least in the local Universe,
suggests that the specification of geometry does not need to be
completely ad hoc and can be empirically constrained by the colour of the
integrated dust emission, by the optical measurements of structure, and by
considerations of physical plausibility. This approach can be directly
applied to the panchromatic observations to decode their information into
basic physical parameters of galaxies, and we
therefore call this approach {\it decoding observed panchromatic SEDs} 
(see Popescu \& Tuffs 2010). 

The only alternative to invoking empirical constraints on the
distributions of dust and stars in galaxies is to calculate these
geometry from first principles, using numerical simulations
of how galaxies form and evolve.
This approach was followed by Chakrabarti et al. (2008),
Chakrabarti \& Whitney (2009), Jonsson et al. (2009) and
we call it {\it encoding predicted physical quantities}. A particular
advantage of the encoding approach is in its application to
high redshift galaxies with more complex geometries due
to frequent mergers and interactions, where hydrodynamical simulations are 
ideal for describing structures on scales greater than ca. 100\,pc
(Jonsson et al. 2009), at least in a statistical sense.
The coupling of cosmological 
simulations with SED modelling tools is moreover an obvious way to
interpret the cosmological evolution of luminosity functions of both the
direct and re-radiated components of stellar light, something which
has already been done using semi-analytic models of structure
formation (Almeida et al. 2010). On the other hand, the encoding
process relies on the assumption that the theory used to predict the relation
between gas, dust and stars in galaxies is complete.
Furthermore, deriving information on galaxies on an
object-by-object base is better suited to the decoding process, at least for
evolved disk galaxies in the local Universe with more uniform geometries.

In this paper we follow the decoding approach, using
an empirical determination of geometry to self-consistently
predict the UV/optical attenuation of starlight and the corresponding
dust/PAH emission SEDs of star-forming galaxies in the local Universe.
To this end we use an updated and enhanced version of the
original model of Popescu et al. (2000, hereafter Paper~I)  which
incorporates the geometrical constraints on dust and stars in spiral galaxies
found by Xilouris et al. (1999). Specifically, we
calculate a comprehensive library of  spatially integrated
dust/PAH re-emission SEDs\footnote{All
available in electronic format at CDS data base}
of disk galaxies as a function of a minimal set of physical parameters
needed to predict the dust/PAH emission. We further describe how this
set of dust/PAH re-emission SEDs can be self-consistently
combined with a new library of UV/optical dust attenuations 
calculated using the same model (an update of the existing library from 
Tuffs et al. 2004; hereafter Paper~III) to invert an observed set of 
broad-band photometry of a
galaxy spanning the UV/optical - FIR/submm range to derive the
intrinsic (i.e. as would be observed in the absence of dust)
UV/optical emission of the galaxy. In particular we demonstrate
that this analysis can, without significant loss in accuracy, be
done independent of a priori assumptions of
the detailed mathematical form of the dependence of SFR on time. This
approach allows the UV/optical SED to be dereddened in a fast and flexible
way without recourse to population synthesis models. The latter models
can then be compared to the dereddened UV/optical SED to derive the
SF history in a further, independent step, avoiding bias due to the
age/metallicity-opacity degeneracy.

The primary motivation to develop this approach was to extend the
applicability of a fully self-consistent RT solution to the large 
statistical samples of optically selected local Universe star-forming
galaxies spanning a full range of mass, morphology and environment 
for which, thanks to facilities such as GALEX, WISE, Spitzer, AKARI
and Herschel (e.g. Driver et al. 2009, Eales et al. 2010, Martin 2010) 
integrated photometry is now for the first time
becoming routinely available across the full UV-submm range.\footnote{Here we 
emphasise once more that our model has been
calibrated for local Universe galaxies, and therefore its applicability
should mainly lie within low redshift galaxies. The models are also
targeted to spiral galaxies. We have not
attempted to model elliptical galaxies, starburst galaxies or AGN
nuclei. Models for starburst galaxies
can be found in Rowan-Robinson \& Efstathiou 1993, Kr\"ugel \& Siebenmorgen
1994, Silva et al. 1998, Efstathiou et al. 2000, Takagi et al. 2003,
Dopita et al. 2005, Siebenmorgen \& Kr\"ugel 2007, Groves et al. 2008. 
Models
for starburst dwarfs can be found in Galliano et al. (2003).
Models for AGN torus were presented by Pier \& Krolik 1992,
Efstathiou \& Rowan-Robinson 1990, Granato \& Danese 1994,
Efstathiou \& Rowan-Robinson 1995, Nenkova et al. 2002,
Dullemond \& van Bemmel 2005, Fritz et al. 2006, H\"onig et al. 2006,
Schartmann et al. 2008, Nenkova et al. 2008.}
In addition to showing how the UV/optical attenuation is constrained
through the observed colour and amplitude of the observed PAH/dust
emission spectrum, we also describe how to utilise morphological
information from high resolution optical observations of galaxies 
(such as linear sizes of disks and the bulge-disk decompositions)
which are expected to come from the next generation of wide field imaging 
spectroscopic surveys of local Universe galaxies (e.g. Driver et al. 2009).
In keeping with the decoding approach that we have adopted, we envisage that
the model will be well suited to the establishment of empirical
relations between SF activity and SF history of optically selected
galaxies with a large range of properties of interest, such as 
dynamical mass (both of the parent DM halo and in disk), baryonic gas content,
dust content, relativistic particle content, and environment. Amongst 
other applications the model should also be useful for evaluating the
relation between broad-band and spectroscopic measures of both SFR and 
opacity in spiral galaxies so that these quantities may be measured
in cases where a full panchromatic coverage of the UV-submm SED is not
yet available.

Although we have calibrated the model using the optical observations of the
galaxies from Xilouris et al. (1999) which are all spirals in the middle range
of the Hubble sequence, we have incorporated a bulge-to-disk luminosity ratio
parameter which will extend the applicability of the model to galaxies of
earlier and later morphological types. We have also identified a non-parametric
test to evaluate the fidelity of this model by virtue of its prediction of the 
colour-surface-brightness relation for spiral galaxies. This test will allow 
galaxies with systematic differences from the geometry adopted in this model 
to be identified.  

This paper is organised as follows.
In Sect.~\ref{sec:calculation} we describe our model for the dust emission,
together with the physical motivation associated with its underlying
assumptions, and formulate the mathematical framework underpinning 
the quantitative comparison of dust emission with the attenuation
calculations presented in this paper. We draw particular attention to the
changes in the model compared to the original
formulation in Paper~I. Specifically, we describe the inclusion of
PAH molecules, needed to extend the applicability of the model
to shorter (mid-IR) wavelengths, and which also led to a revision
to the overall grain size distribution to preserve consistency
with the extinction and emission properties of interstellar dust.
A further major update to the model 
is the inclusion of the contribution to dust heating 
of the young stellar population at optical wavelengths, 
avoiding the artificial cut-off of the emission of this
population beyond the UV range in the original version of 
the model from Paper~I. In this section we further describe
a revised treatment of the emission from 
star-formation regions, incorporating a wavelength-dependence of
the escaping UV light to achieve consistency with their attenuation
properties as formulated in Paper~III and including an updated template 
for the corresponding infrared emission using the recent model of Groves et al. 
(2008). 
In Sect.~\ref{sec:illustration} we illustrate our model for the case of the 
well studied galaxy NGC~891. 
A more general quantitative analysis of the effects of the
main assumptions of the model on the predicted spatially
integrated SEDs as well as tests on the fidelity of the model are given in 
Sect.~\ref{sec:assumptions}.
We then present our library of simulated dust emission SEDs in 
Sect.~\ref{sec:library} and discuss the predicted variation of these SEDs 
with the main parameters of the model in Sect.~\ref{sec:predictions}.
We summarize the main results of the paper in Sect.~\ref{sec:summary}.

The updated calculations of attenuation of stellar light for the diffuse 
component are given in Appendix~\ref{sec:attenuation}. To facilitate the use of
this library of attenuation in combination with the dust emission calculations 
we give in Appendix~\ref{sec:composite} formulae for the composite attenuation of 
stellar light from the different geometrical components of stellar emissivity 
in terms of the parameters used for the prediction of the dust emission.
For an easy access to potential users of the 
model we give in Appendix~\ref{sec:howto} a step-by-step guide on how to use the
model for fitting real data, together with the full mathematical 
formalism. In this appendix we also describe the non-parametric
test to objectively judge the efficacy of the model by virtue of its
prediction of the colour-luminosity relation for spiral galaxies.

\section{Calculation of the infrared SEDs}
\label{sec:calculation}

We calculate the infrared SEDs using an updated version of the model
from Paper~I which was applied to
the interpretation of dust emission from edge-on galaxies by 
Misiriotis et al. (2001; Paper~II) and also used to
make predictions for the effects of dust in the optical range
in Paper~III and by M\"ollenhoff et al.  
(2006; Paper~IV). 
In the following we describe the changes in detail, giving their
physical motivation, and showing how they interlink with the mathematical 
formulation of the model.

\subsection{The distribution of stars and dust}
\label{subsec:distribution}

Star-forming galaxies are fundamentally inhomogeneous, containing highly 
obscured massive star-formation regions, as well as more extended large scale 
distributions of stars and dust. It is important to emphasize that the
large-scale distribution of diffuse dust plays a major role in mediating the
propagation of photons in galaxy disks and dominates the total bolometric
output of dust emission. The discovery of this diffuse component
was one of the highlights of the Infrared Space Observatory
(ISO)(see Tuffs \& Popescu  2006 and Sauvage et al. 2006 for reviews on the ISO
science legacy on normal nearby galaxies). This result was obtained both from
analysing the integrated properties of infrared emission from galaxies (Popescu
et al. 2002) but also from resolved studies of nearby galaxies 
(Haas et al. 1998, Hippelein et al. 2003, Tuffs \& Gabriel 2003, 
Popescu et al. 2005) and is now being confirmed by the
infrared data from Spitzer (P\'erez-Gonz\'alez et al. 2006, Hinz et al. 2006,
Dale et al. 2007, Bendo et al. 2008, Kennicutt et al. 2009)
and AKARI (Suzuki et al. 2007). Our model being empirically motivated we tried
to incorporate all available observational constrains provided by the data. 
Accordingly our model includes both a diffuse component and a localised 
component representing the star-formation complexes.

\begin{figure*}
\centering
\includegraphics[width=12cm]{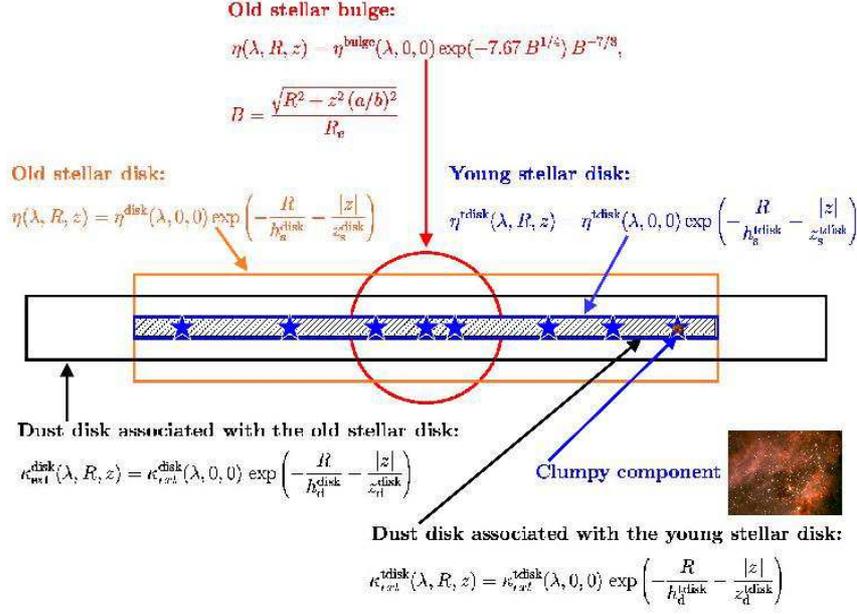}
\caption{Schematic representation of the geometrical distributions of stellar 
and dust emissivity together with a mathematical prescription of the stellar
emissivities and dust opacities used in the model. Here, and in the main body
of the text we use the superscript 
``disk'', ``bulge'' and ``tdisk''  for all the quantities respectively 
describing the disk (the old stellar disk plus the associated dust disk,
 whether the latter is sometimes referred to as  the ``first dust disk''), 
the bulge
and the thin  disk (the young stellar disk plus the 
associated dust disk, whether the latter is sometimes referred to as  the 
``second dust disk'').}
\label{fig:schematic}
\end{figure*}

A schematic picture of the geometrical components of the model is given
in Fig.~\ref{fig:schematic}, together with a mathematical prescription of the 
stellar emissivities and dust opacities used in the model. 

The large scale distribution of stars and dust are approximated as
continuous spatial functions of stellar emissivity and dust opacity, which we
refer to as ``diffuse'' distributions. We have separate distributions for the
old and young stellar populations, and we also consider separate distributions
for diffuse dust associated with these populations.

The old stellar population resides in a disk and a bulge, with its 
emissivity described by a double exponential and a de Vaucouleurs distribution,
respectively:

\begin{eqnarray}\nonumber
\eta(\lambda,R,z) = {\eta^{\rm disk}}(\lambda,0,0) 
\exp \left( - \frac{R}{{h^{\rm disk}_{\rm s}}} - 
\frac{|z|}{{z^{\rm disk}_{\rm s}}} \right)
\end{eqnarray}
\begin{eqnarray}\label{eq:oldemissivity}
~~~~~~~~~~+{\eta^{\rm bulge}(\lambda,0,0)} 
\exp (-7.67\,B^{1/4})\,B^{-7/8}, 
\end{eqnarray}
\begin{eqnarray}\label{eq:bulgeparam}
B = \frac{\sqrt{R^2 + z^2\,({a/b})^2}}{{R_e}},
\end{eqnarray}
where $R$ and $z$ are the cylindrical coordinates, 
$\eta^{\rm disk}(\lambda,0,0)$ is the stellar emissivity at the centre of the 
disk,
$h^{\rm disk}_{\rm s}$, $z^{\rm disk}_{\rm s}$ are the scalelength and 
scaleheight of the disk, $\eta^{\rm bulge}(\lambda,0,0)$ is the 
stellar emissivity at the centre of the bulge, $R_e$ is the effective radius 
of the bulge, and $a$ and $b$ are the semi-major and semi-minor axes of the 
bulge. 

The total luminosity of the old stellar population in the disk is then given by:
\begin{eqnarray}\label{eq:oldlum}
L^{\rm disk}_{\nu}  = 4\,\pi\,\eta^{\rm disk}(\lambda,0,0)\,
z^{\rm disk}_{\rm s}\,(h^{\rm disk}_{\rm s})^2
\end{eqnarray}
whereas for the bulge there is no exact analytical formula for the spatially 
integrated luminosity. 

The dust
associated with the old stellar population (``the first dust disk'') is
also described by an exponential disk:
\begin{eqnarray}\label{eq:dustemissivity1}
\kappa^{\rm disk}_{\rm ext}(\lambda,R,z) = {\kappa^{\rm disk}_{ext}
(\lambda,0,0)}\,\exp \left( - \frac{R}{h^{\rm disk}_{\rm d}}- 
\frac{|z|}{z^{\rm disk}_{\rm d}} \right),
\end{eqnarray}
where $\kappa^{\rm disk}_{\rm ext}(\lambda,0,0)$ is the extinction coefficient
 at the centre of the disk and $h^{\rm disk}_{\rm d}$ and 
$z^{\rm disk}_{\rm d}$ are the scalelength and scaleheight of the dust 
associated with the old stellar disk.

In our model the parameters describing the geometry of the old stellar 
population and of the first dust disk are empirically constrained from 
resolved optical and near-IR images via the results of the modelling procedure 
of Xilouris et al. (1999). For edge-on 
systems these calculations completely determine the scale heights 
and lengths of the exponential disk of old stars and associated diffuse dust, 
as well as the effective radius and ellipticity of the dustless stellar 
bulge. This is feasible for edge-on systems since the scale height of the dust 
is less than that of the stars. By analysing
a sample of 5 nearby edge-on galaxies with morphological classification in the
range of Sb-Sc, Xilouris et al. found that the
scalelength of the dust was larger than that of the stellar disk, that the
scaleheight of the stars was larger than that of the dust and that the
scalelength of the stellar disk decreases with increasing wavelength. In this
way Xilouris et al. was also able to find
a general relation between the scaleheights and scalelength of old stars and
dust and the dependence of this relation on wavelength. The derived relation
 enabled us to fix the relative distribution of old stars and dust in our 
model. In Papers~I and II  we verified that the derived geometrical parameters 
from Xilouris et al. correctly predicted the dust emission SEDs of the 5 
nearby edge-on galaxies. Observationally it was also confirmed that the dust
disk not only has a larger scalelength than the stellar disk 
(Alton et al. 1998, Davies et al. 1999), but was detected to physically
extend well beyond the stellar disk (Popescu \& Tuffs 2003; see also Popescu
et al. 2002 and Hinz et al. 2006 for dwarf galaxies). From resolved studies 
of galaxies it was also found that there is a
large scale distribution of diffuse dust having a face-on opacity that 
decreases with radius (Boissier et al. 2004, Popescu et al. 2005,
P\'erez-Gonz\'alez et al. 2006, Boissier et al. 2007, 
Mu\~noz-Mateos et al. 2009).

The geometrical parameters of our model are listed
in Table~\ref{tab:geometry},  where all the length
parameters are normalised to the B-band scalelength of the disk, 
$h^{disk}_s(B)$. In our calculations we take 
$h^{disk}_s(B)=h^{\rm disk}_{s, ref}=5670$\,pc, the fixed reference 
scalelength of our model galaxy, as derived for NGC~891.
 
The ``young'' stellar population and associated dust are also specified by 
exponential disks, which are taken to have small scaleheights (thin disks,
thereby the superscript 'tdisk'),
which we shall refer to as the ``young stellar disk'' and the 
``second dust disk'':
\begin{eqnarray}\label{eq:youngemissivity}
\eta^{\rm tdisk}(\lambda,R,z) = \eta^{\rm tdisk}(\lambda,0,0) \exp 
\left( - \frac{R}{h^{\rm tdisk}_{\rm s}} - \frac{|z|}{z^{\rm tdisk}_{\rm s}} 
\right)
\end{eqnarray}
\begin{eqnarray}\label{eq:dustemissivity2}
\kappa^{\rm tdisk}_{ext}(\lambda,R,z) = 
\kappa^{\rm tdisk}_{ext}(\lambda,0,0)\,\exp 
\left( - \frac{R}{h^{\rm tdisk}_{\rm d}}- \frac{|z|}{z^{\rm tdisk}_{\rm d}} 
\right)
\end{eqnarray}
where $\eta^{\rm tdisk}(\lambda,0,0)$ is the stellar emissivity at the centre 
of the thin disk, $h^{\rm tdisk}_{s}$ and $z^{\rm tdisk}_{s}$ are the 
scalelength and scaleheight of the thin disk, 
$\kappa^{\rm tdisk}_{ext}(\lambda,0,0)$ is the extinction coefficient at
the centre of the thin disk and $h^{\rm tdisk}_{d}$ and $z^{\rm tdisk}_{d}$ are
the scalelength and scaleheight of the dust associated with the young stellar
disk.

The total luminosity of the young stellar disk is given by:
\begin{eqnarray}\label{eq:youngluminosity}
L^{\rm tdisk}_{\nu}  = 4\,\pi\,\eta^{\rm tdisk}(\lambda,0,0)\,
z^{\rm tdisk}_{\rm s}\,(h^{\rm tdisk}_{\rm s})^2
\end{eqnarray}
 
 Unlike the old components, the parameters describing the
geometry of the young components cannot be constrained from images of stellar
light, because the young stellar populations are highly obscured in most cases.
It was thus necessary to constrain these parameters from physical
considerations. The scale height of the young stars was taken to 
be 90 pc (the value for the Milky 
Way) and its scale length is equated to that of the ``old stellar disk'' in 
B-band, $h^{\rm tdisk}=h^{\rm disk}(B)$. The dust associated with the 
young stellar population was fixed to have the same scalelength and 
scaleheight as for the young stellar disk. The reason for this choice 
is that our thin disk of dust was 
introduced to mimic the diffuse component of
dust which pervades the spiral arms, and which occupies approximately the same
volume as that occupied by the young stars. This choice is also physically 
plausible, since the star-formation rate is closely connected to the gas 
surface density in the spiral arms, and this gas bears the grains which caused 
the obscuration.
In principle it would be more realistic to place the young
stellar population and associated dust into a spiral pattern rather than into a
disk. However, in practice, as we show in Sect.~\ref{subsec:spiral}, this makes
little difference to the predicted volume integrated dust emission SED and
starlight attenuation of a galaxy. Furthermore, the inclusion of spiral
arms would necessitate another model parameter to describe this more complex
geometry. Also, in most practical applications for the interpretation of the 
integrated emission of galaxies in large statistical samples, photometric
information about the spiral arm pattern is not available. For all these
reasons we consider the approximation of the second dust disk to be  both
reliable and practical. The geometrical parameters of the young stellar disk 
and second-dust disk are listed in Table~\ref{tab:geometry}.

Apart from the geometrical parameters described above, the distributions of
diffuse stellar emissivity and dust are also described in terms of their
amplitudes. The amplitudes of the two dust disks $\kappa_{ext}^{\rm disk}$, $
\kappa_{ext}^{\rm tdisk}$ can be expressed in terms of
the central face-on opacity in the B band,
$\tau^{f,disk}_B$,$\tau^{f,tdisk}_B$, defined by:
\begin{eqnarray}\label{eq:opacity1}
\tau^{f,disk}_B = 2\,\kappa_{ext}^{\rm disk}(\lambda_{B},0,0)\,z_{\rm d}^{\rm disk}
\end{eqnarray}
\begin{eqnarray}\label{eq:opacity2}
\tau^{f,tdisk}_B = 2\,\kappa_{ext}^{\rm tdisk}(\lambda_{B},0,0)\,z_{\rm d}^{\rm tdisk}
\end{eqnarray}

The opacity of the first dust disk can be derived jointly with the
geometrical parameters from the optimisation technique of Xilouris et
al. (1999). The optimisation also determines the extinction law of the diffuse
dust empirically, since the calculations are done independently for each
optical/NIR image. For the galaxies studied by Xilouris the derived extinction
law was consistent with a Milky Way type dust, which is also the type of dust
adopted for our dust model (see Sect.~\ref{subsec:dustmodel} and 
\ref{sec:extinction}). The opacity of the second dust disk was a free
parameter in the calculations from Papers~I and II, and is strongly 
constrained by the level of the submm emission. To minimise the number of 
free parameters, we fix 
\begin{eqnarray}\label{eq:opacityratio}
\tau_{ratio} = \frac{\tau^{f,disk}_B}{\tau^{f,tdisk}_B} 
\end{eqnarray}
to the value 0.387 found for our proto-type
galaxy NGC~891, which is also close to what was found for a second edge-on
galaxy with submm data, NGC~5907, which we modeled in Paper~II. We note here 
that the attenuation-inclination relation predicted
for this fixed ratio of opacities in the two dust disks was found to 
successfully reproduce the observed attenuation-inclination relation of a 
large and statistically complete sample of galaxies from the Millennium Galaxy 
Catalogue Survey (Driver et al. 2007). 
We thus adopt as a 
free parameter of the model {\it the total central face-on 
opacity in the B-band $\tau^{f}_B$}:
\begin{eqnarray}\label{eq:totalopacity}
\tau^{f}_B =  \tau^{f,disk}_B + \tau^{f,tdisk}_B
\end{eqnarray}

The remaining parameters  - the amplitudes of the two stellar disks 
${\eta^{\rm disk}}$, $\eta^{\rm tdisk}$ and of the 
bulge $\eta^{\rm bulge}$- and their link to the free parameters of the model 
are discussed in 
Sect.~\ref{subsec:intrinsic}.

\subsection{The clumpy component}
\label{subsec:clumpy}

An important component of the spatially integrated dust emission from 
star-forming galaxies arises from dust in the birthclouds of massive stars, as 
previously modelled by Silva et al. (1998), Popescu et al. (2000), 
Charlot \& Fall (2000), Efstathiou \& Rowan-Robinson (2003), Jonsson et al. 
(2009). 
Because these clouds
are spatially correlated with their progeny on parsec scales, they are
illuminated by a strong UV-dominated
radiation field of intensity $10-100$ times that in the diffuse
ISM. This gives rise to a localised component of emission from grains
in thermal equilibrium with these intense radiation fields, which, despite
the tiny filling factor of the SF regions in the galaxy can nevertheless
exceed the entire diffuse infrared emission of a galaxy
at intermediate wavelengths (ca. 20 to ca. 60\,${\mu}$m). It is therefore
particularly important to incorporate a clumpy component  of
dust associated with the opaque parent molecular clouds of massive
stars. Following Popescu \& Tuffs (2005) we refer to these clumps as 
``active clumps''. 
The active clumps are assumed to have the same spatial distribution as the 
young stellar disk and the second dust disk. Furthermore, it is assumed that 
the properties of these clumps do not systematically depend on their radial 
location within the galaxy. In reality we expect star formation complexes in 
more
pressurised regions (such as the inner disk) to be more compact and therefore
have warmer FIR colours than their counterparts in low pressure regions (such
as the outer disk), as modelled by Dopita et al. (2005). And indeed this
phenomenon has been observed in HII regions in M~33 (Hippelein et
al. 2003). However, bearing in mind that we will empirically calibrate our
template (see Sect.~\ref{subsec:infraredclumpy}) on the whole ensemble of star 
forming complexes in the Milky Way, the
assumption we make in this regard should not significantly affect our 
predictions for the integrated emission of galaxies.

Since birthclouds of stars are typically fragmented due to the
combined effects of supernovae, stellar winds, and the general motion
of stars away from the clouds, only a certain fraction of the
total luminosity of massive stars in a galaxy will be
locally absorbed. Following the original formulation in Paper~I
we denote this fraction in our model by the ``clumpiness factor'', $F$.
Since birthclouds are completely opaque at all UV/optical/near-IR wavelengths
(e.g. Sievers et al. 1991), $F$ can be physically identified
with the luminosity-weighted mean fraction of directions from the
massive stars, averaged over the lifetime of the stars, which
intersect the birthcloud. This concept, first introduced by Popescu et
al. (2000), allows UV light to freely stream away from star-forming regions in 
some fraction of directions, allowing the diffuse dust to be illuminated with 
more UV photons than would have been the case if the young stars had been 
assumed to be completely cocooned in their parent molecular clouds.
This is in qualitative
accordance both with the high frequency at which counterparts of star-formation
regions seen at 24\,${\mu}$m are detected in the UV, as well as with the
predominance of PAH emission from diffuse dust in galaxies correlated
with cold dust (Bendo et al. 2008). 

A more detailed 
description of the clumpy component is given in Sect.~\ref{subsubsec:escape} 
and \ref{subsec:infraredclumpy}. 

An approximation of our model is that the heating of the grains in the 
active clumps is dominated by photons from the stellar progeny, and we neglect 
any external contribution from the diffuse ambient radiation fields in the 
galaxy. This should be an excellent approximation for spiral galaxies, where the
filling factor of star formation regions is small. We note that
if this approximation were invalid, we would be forced to perform
radiative transfer  calculations with parsec resolution to properly describe
the heating of grains in the optically thick birth clouds, which would
render the calculation of a comprehensive set of SEDs as given in this
paper intractable with current computing resources. To date, the best
resolution achieved with an adaptive grid code when modeling the observed dust 
emission SED from an individual galaxy is about 20\,pc (Bianchi 2008). 
In any case, we would
only expect collective effects, whereby photons from adjacent SF regions
provide a significant fraction of the dust heating, to be significant
in galaxies with very high volume densities of SF regions,
such as the central region of a starburst galaxy. The low filling factors of
opaque clouds in spiral disks is supported by high resolution surveys of the
Galactic Plane with large ground-based telescopes in the submm (e.g. the APEX
Telescope large Area Survey - ATLAS; Schuller et al. 2009).  
At these wavelengths the clouds are optically thin, thus directly tracing the
total column density of dust, yet only sporadic peaks of dust emission can be
seen with local enhancement in visual optical depth much greater than 1. This 
contrasts with surveys in the CO lines (Matsunaga et
al. 2001), which show a more highly filled distribution of emission due to
optical depth effects in the radio molecular line. 
Recently the new Herschel Infrared Galactic Plane Survey (Hi-Gal) 
(Molinari et al. 2010) has revealed the fainter, more extended emission
components which dominate the morphology of the dust emission in the Milky Way,
showing again that the optically thick molecular cores have only a small
filling factor.
In summary, based on all
available observational evidence we
believe it is a reasonable approximation to ignore the external heating of
active clumps.

Another approximation of our model is that the clumpy distribution of dust is 
exclusively associated with the opaque parent molecular clouds of massive 
stars - the active clumps. In 
reality, some of the dust in the diffuse dust disks may also be in clumps 
without internal photon sources, and thus heated only by the diffuse ambient
radiation fields in the disk. Following Paper~I (see also Popescu
\& Tuffs 2005) we refer to such
clumps as ``passive'' or ``quiescent'' clumps. Provided passive clumps are 
optically thin, the transfer of radiation through the disks will be virtually 
identical to that in a homogeneous disk. However, once the 
passive clumps become  optically thick, the self-shielding of grains will 
yield a solution with a reduction in both the attenuation of the stellar light 
and the infrared emission compared to the case where the same mass of grains 
is diffusely distributed. These effects have been quantified by Bianchi et 
al. (2000a) who showed that the shape of the infrared SED of a disk galaxy can 
be 
strongly affected shortwards of $200\,{\mu}$m (see also Misselt et al. 2001 for
passive clumps distributed in a spherical shell geometry). We do not include 
passive clumps in our model for 
both physical and empirical reasons. Physically, it seems likely that most 
passive clumps will be optically thin, since as soon as they become optically 
thick to the impinging external UV light, they will lose their principal 
source of heating (the photoelectric effect) and will be prone to collapse and 
form stars (Fischera \& Dopita 2008). Empirically, this presumption is 
supported by the recent findings of Holwerda et al. (2007a,b), that the 
structure of the diffuse ISM consists of optically thin dusty
clouds. Furthermore, the change in the shape of the predicted dust emission 
SEDs imposed by the incorporation of passive clumps tends to provide a colder 
solution than needed to fit real data (Bianchi 2008).

\subsection{Constraints on the intrinsic SEDs of the stellar populations}
\label{subsec:intrinsic}

As mentioned in Sect.~\ref{subsec:distribution}, the amplitudes of the stellar 
populations of the two
stellar disks (${\eta^{\rm disk}}(\lambda,0,0$) and 
${\eta^{\rm tdisk}}(\lambda,0,0)$) are parameters that are yet to be
determined. In other words we want to find a solution for the
attenuation of stellar light and dust emission SED that can fit observed SEDs to
provide resulting intrinsic stellar SEDs. At the same time, in our modelling 
procedure we deliberately do not want to use population synthesis models to 
fully fix the intrinsic SED, as we want to be as free as possible of any 
assumptions about the SF history. Nonetheless, some constraints on the input 
SEDs need to be made to avoid having the amplitude of the stellar SED a free
parameter at each of the sampled wavelengths listed in 
Table~\ref{tab:stellarsed}.
Since the main factors shaping the dust emission SEDs are the
total luminosity of the young stellar population and of the old stellar
populations, we choose to have these two quantities as free parameters of our 
model and produce calculations for all combinations of these two
variables. We then assume that each of the two components (young and old
stellar populations) have a fixed wavelength dependence (a fixed
template SED), thus reducing the number of free parameters describing the 
intrinsic SEDs
of the stellar populations to two. We will show in 
Sect.~\ref{subsec:approximation} that these
assumptions have a negligible effect on the predictions for the dust emission 
SEDs and the associated parameters $\tau^{f}_B$ and $F$. In turn, the solution
for the wavelength dependence of the attenuation in the UV-optical range, which
depends on $\tau^{f}_B$, $F$ (and inclination), will be similarly secure, thus
allowing the true dereddened stellar emission SED to be recovered. In effect we
are using the fact that the dependence of the dust emission SED 
on the colour of the stellar photon field depends primarily on the 
ratio between the luminosities of the young and old stellar populations 
rather than on the detailed colour of the emissions from either of these 
populations. This way of constructing the model will allow SF histories to be
extracted from the dereddened stellar emission SED in a separate step from the 
extraction of $\tau^{f}_B$ and $F$.

In constructing the template SEDs of the stellar populations we consider
the term ``optical'', ``UV'', ``ionising UV'' and ``non-ionising UV'' to denote 
the wavelength ranges ${\lambda} \ge 4430\,\AA$, ${\lambda} < 4430\,\AA$, 
${\lambda} \le 912\,\AA$,  $912 < {\lambda} < 4430\,\AA$, respectively,
thereby marking the boundary between the UV and optical regime in the B band.

\subsubsection{The template SED of the old stellar population in the disk}
\label{subsubsec:templateold}

In defining the fixed shape of the SED of the old stellar population in the
disk we only 
consider optical radiation and neglect any contribution in the
UV.
The precise wavelength dependence of this SED was fixed to the empirical 
relation obtained by Xilouris et al. 
(1999) from fitting simulated images produced by radiative transfer
calculations to the observed B, V, I, J, K images\footnote{In the K and H 
bands we rescaled the luminosities derived by
Xilouris et al. (1998) to more recent, higher quality data from the 2MASS 
survey.} of NGC~891.  The SED of the old stellar
population in the disk, as ``calibrated'' on NGC~891 is given as the light blue 
curve in Fig.~\ref{fig:sed_ngc891}.
In applications to other galaxies this curve needs to be 
scaled according to the total output of the old stellar population in the disk 
of the modelled galaxy. For this purpose we use a unitless parameter $old$
defined as
\begin{eqnarray}\label{eq:old}
old=\frac{L^{disk}}{L_{unit}^{old}} 
\end{eqnarray}
where $L_{unit}^{old}=2.241 \times 10^{37}$\,W, which corresponds to 10 
times the luminosity of the non-ionising UV photons produced by a 
$1{\rm M}_{\odot}/$yr young stellar population, as defined in Sect.~
\ref{subsubsec:templateyoung}. Thus
$old$ is the normalised luminosity of the old stellar populations in the disk
and is adopted to be one of the free parameters of the model.
In this scheme the best fit solution for NGC~891 corresponds to $old=0.792$.

The intrinsic spectral luminosity densities of the old stellar 
population in the disk corresponding to $old=1$, $L^{old}_{\nu,\,unit}$, used 
in our calculations are listed in Table~\ref{tab:stellarsed}. 
The spectral luminosity density of the old stellar population in the disk of a 
model galaxy is then given by:
\begin{eqnarray}\label{eq:oldsed}
L^{disk}_{\nu} = old \times L^{old}_{\nu,\,unit}
\end{eqnarray}
where $L^{disk}_{\nu}$ has been previously defined in terms of 
$\eta^{\rm disk}(\lambda,0,0)$ in Eq.~\ref{eq:oldlum},

\begin{eqnarray}\label{eq:lunitolddef}
L_{unit}^{old} = \int_{opt} L^{old}_{\lambda,\,unit}\,d{\lambda}
\end{eqnarray}
and the value of $L_{unit}^{old}$ can be derived by integrating over
wavelength\footnote{Throughout the paper all spectral integrations are done
  over wavelength rather than over frequency. This ensures self-consistency in
  cases where functions are sparsely sampled in wavelength.} the values of 
$L^{old}_{\nu,\,unit}$ from Table~\ref{tab:stellarsed}.

As already mentioned, we will show in Sect.~\ref{subsec:approximation}
that, although the template SED of the old stellar population used
in the model is fixed to a single shape (here defined
empirically on the galaxy NGC~891), it is still valid to use
it to model the dust emission SEDs of other galaxies which may have very
different stellar emission SEDs.

\subsubsection{The template SED of the old stellar population in the bulge}
\label{subsubsec:templatebulge}

The wavelength dependence of the stellar luminosity produced by the bulge
stellar population is simply linked to that of the old stellar population in the
disk via a wavelength independent {\it bulge-to-disk ratio B/D}, one of the
free parameters in our model.

\begin{eqnarray}\label{eq:bulgesed}
L^{bulge}_{\nu} = (B/D) \times L^{disk}_{\nu}
\end{eqnarray}

\subsubsection{The template SED of the young stellar population}
\label{subsubsec:templateyoung}

The wavelength dependence of the stellar luminosity produced by the young
stellar population in the UV cannot be constrained 
empirically, as this population
is heavily obscured by dust. Because of this we used
population synthesis models. We used the models from Kotulla 
et al. (2009), making standard assumptions for the star-formation history, 
namely an exponentially declining star-formation rate with a time constant 
${\tau}=5\,$Gyr, solar metallicity and a Salpeter IMF with an upper mass 
cut-off of $100\,M_{\odot}$. We emphasise that, although arbitrary,
this choice of parameters will not significantly bias the
model dereddening of the observed UV/optical SEDs, since, as we will show in
Sect.~\ref{subsec:approximation}, using different assumed shapes for the
stellar emissivity SED does not significantly
affect the predictions for the dust emission SEDs.

We also consider that a fraction $f_{ion-uv}=0.3${\footnote{The fraction of ionising UV photons that is
  absorbed by dust in HII regions exhibits a broad range of values, varying from
  0.3-0.7 (Inoue et al. 2001; Inoue 2001). However, even if this fraction 
approaches unity, their contribution to the dust emission of the 
star-forming complexes is still only at the percent level, because the
intrinsic luminosity of the ionising UV photons is so much smaller than that 
of the non-ionising UV photons (Bruzual \& Charlot 1993). 
This means their contribution to the total dust emission is even less.
}
of the ionising UV photons 
(emitting shortwards of 
$912\,{\AA}$) from the massive stars within the HII regions are locally absorbed
by dust. This makes only a minor contribution to the dust heating, an order of
magnitude less than the contribution of the non-ionising UV photons.

A new feature of the model is the consideration of the optical emission from
the young stellar population embedded in the second dust-disk. Previously the
emissivity function of this population had been artificially truncated
longwards of the B band.
In order to fix the shape of the SED template for the
young stellar population in the optical we again make use of the best fit
solution to NGC~891. Specifically, the optical radiation from this template
at each wavelength sampling point was fixed to be the residual between the
prediction of the population synthesis model for the best fit of NGC~891 
and the empirical SED of the old stellar population (derived from the best fit 
of NGC~891 as explained above)\footnote{
Physically, this approach corresponds to the dynamical heating of
stellar populations born in the young stellar disk and their
transfer over time into the larger scale height old stellar disk
due to inelastic scatterings with spiral arms
and/or giant molecular clouds.}
The resulting optical part of the SED template
of the young stellar population was then combined with the UV part (which is
simply the population synthesis SED for the best fit SFR for NGC~891)
to fix the shape of the template over the full UV/optical range. For use in
other model galaxies, the overall amplitude of this SED of the young stellar
population, as ``calibrated'' on NGC~891, needs to be 
scaled according to the $SFR$ of that galaxy, adopting $SFR$ as a free
parameter of the model.

For logistical purposes only, we chose as a 
unit for the luminosity of the young
stellar population $L^{young}_{unit}=4.235 \times 10^{36}$\,W, which is the 
luminosity produced by a  $SFR$ of $1\,{\rm M}_{\odot}/$yr 
(for the standard assumptions about the SF history described before) in the
range $(912.-50000.)\,\AA$. $L^{young}_{unit}$ can be given as a sum 
of the luminosity of the non-ionising UV photons 
$L_{unit,\,uv}^{young}=2.241 \times 10^{36}$\,W and the luminosity of the 
optical  photons emitted by the young stellar disk
$L_{unit,\,opt}^{young}=1.994 \times 10^{36}$\,W. As already introduced in 
Sect.~\ref{subsubsec:templateold}, we also 
defined the unit for the luminosity of the old stellar population 
$L_{unit}^{old}$ to correspond to 
10 times the luminosity of the non-ionising UV photons produced by a 
$1\,{\rm M}_{\odot}/$yr young stellar population. In addition we also define 
$L_{unit,\,ion-uv}^{young}=0.267 \times 10^{36}$\,W
for the luminosity of the ionising UV photons. We note that
$L_{unit,\,ion-uv}^{young}$ is not included in the definition of
$L^{young}_{unit}$. 

 We then parameterised the calculations in terms of $SFR$
\begin{eqnarray}\label{eq:sfr}
\frac{SFR}{1\,{\rm M}_{\odot}\,{\rm yr^{-1}}} = \frac{L^{tdisk}}{L^{young}_{unit}}
\end{eqnarray}
where the $SFR$ is linked to the normalised luminosity of the young stellar
population in the thin disk. 

In an equivalent way we can link the normalised luminosity of the young stellar
population in the UV to the SFR:

\begin{eqnarray}\label{eq:sfr_UV}
\frac{SFR}{1\,{\rm M}_{\odot}\,{\rm yr^{-1}}} = 
\frac{L^{tdisk}_{uv}}{L^{young}_{unit,\,uv}}
\end{eqnarray} 
where $L_{unit,\,uv}^{young}=2.241 \times 10^{36}$\,W, as defined previously,
and

\begin{eqnarray}\label{eq:sfr_ion-UV}
\frac{SFR}{1\,{\rm M}_{\odot}\,{\rm yr^{-1}}} = 
\frac{L^{tdisk}_{ion-uv}}{L^{young}_{unit,\,ion-uv}}
\end{eqnarray}

The intrinsic spectral luminosity densities of the young stellar population
 corresponding to $SFR=1\,{\rm M}_{\odot}\,{\rm yr}^{-1}$, 
$L^{young}_{\nu,\,unit}$,  used in our calculations 
are listed in Table~\ref{tab:stellarsed}. 

The spectral luminosity density of the young stellar population in the thin
disk of a model galaxy is then given by:
\begin{eqnarray}\label{eq:youngsed}
L^{tdisk}_{\nu} = \frac{SFR}{1\,{\rm M}_{\odot}\,{\rm yr}^{-1}} \times 
L^{young}_{\nu,\,unit}
\end{eqnarray}
where $L^{tdisk}_{\nu}$ has been previously defined in terms of 
$\eta^{\rm tdisk}(\lambda,0,0)$ in Eq.~\ref{eq:youngluminosity},

\begin{eqnarray}\label{eq:lunityoungdef}
L_{unit}^{young} = \int_{UV+opt} L^{young}_{\lambda,\,unit}\,d{\lambda}
\end{eqnarray}
and the value of $L_{unit}^{young}$ can be derived by integrating over
wavelength the values of $L^{young}_{\nu,\,unit}$ from
Table~\ref{tab:stellarsed}.

\subsection{The dust model}
\label{subsec:dustmodel}

The dust model - i.e. the prescription for the optical properties of the
grains, the chemical composition and the grain size distribution - was
updated to include PAH molecules, which are regarded as the carriers of the
unidentified infrared emission features commonly seen in
the mid-IR emission spectrum of star-forming galaxies. 
As shown by Zubko, Dwek \& Arendt 
(2004), there is no unique model incorporating PAHs that can
simultaneously fit the main observational constraints on
such models, namely the extinction and  emission properties of the diffuse 
cirrus in the Milky Way. Several models have been proposed in the literature, 
including the models from Zubko et al. (2004), the model of 
Fischera \& Dopita (2008) which considers an extra component of iron grains 
with the optical properties from Fischera (2004), and the model of 
Weingartner \& Draine 2001 (see also Li \& Draine 2001; Draine \& Li 2007). 
Although not unique in terms of reproducing the properties of the galactic
cirrus, these models do nevertheless differ in some predictions
relevant to the modelling of the infrared emission of spiral galaxies,
particularly with regard to whether conducting particles are included
as a constituent of the dust model.
For example in the model of Fischera \& Dopita a substantial part of the
diffuse 60\,${\mu}$m emission is powered by optical photons absorbed
by iron grains in equilibrium with the ambient radiation fields (see also
Chlewicki \& Laureijs 1988), whereas
in the model of Weingartner \& Draine (2001) this emission 
mainly arises from carbonaceous grains stochastically heated by UV photons.
Potentially, therefore, the choice of grain model can influence
the inferred contributions of young and old stellar populations
in heating the grains. We are however not aware of any direct evidence for the 
existence of
pure metallic grains in the diffuse interstellar medium of galaxies.
Furthermore, simple considerations of potential redox reactions
indicate that any such grains in the diffuse neutral ISM
will be susceptible to oxidation and will revert to the properties of 
non-conducting grains (Duley 1980), like silicates.
Here we substituted the model of Draine \& Lee (1984) and 
Laor \& Draine (1993) used in Papers~I-IV to the latest model from 
Weingartner \& Draine (2001) and Draine \& Li (2007) incorporating a mixture 
of silicate, graphite and PAH molecules. 

We note that the adopted dust model is only used for the calculation of
transfer of radiation through the diffuse interstellar medium, where the 
properties of the grains are reasonably well constrained. When including the 
infrared emission from star-forming complexes, which contain dense clouds 
where grains may form ices and other complex compounds, we use a template model 
SED function which is observationally constrained, as described in 
Sect.~\ref{subsec:infraredclumpy}. 
This sidesteps the uncertainties in the optical properties of grains in dense 
clouds.

For the parameters of the model we consider Case A and $R_V=3.1$ from
Weingartner \& Draine (2001), with the updates from Draine \& Li (2007)
(revised size distribution and optical constants for PAHs). 
The parameters of the model considered here are summarised in 
Table~\ref{tab:dustparam}.

Following Li \& Draine (2001) we allow the relative abundance of neutral and 
ionised PAHs to vary with molecule size but to be independent of position in 
the galaxy (or intensity of the radiation field). The ionisation fraction
($\Phi_{ion}$) of PAHs was fixed according to the average over the three 
phases of the diffuse ISM given in Fig.~7 of Li \& Draine (2001).
The PAH-size distribution was then multiplied with $\Phi_{ion}$ and 
$1- {\Phi}_{ion}$, respectively, to obtain the size distribution of ionised 
and neutral PAH-molecules, respectively.

Since for the dust and PAH emission we accurately take into account the 
stochastic heating of the grains by the ambient radiation fields 
(see Sect.~\ref{subsec:calculationtemperature}), we also need to derive the 
heat capacities for our dust 
grains. The heat capacities for silicate and graphite grains
needed to derive the temperature distributions are summarised in Popescu et
al. (2000). The heat capacities for the PAH-molecules were calculated
using Eq.~15 of Li \& Draine (2001) and Eq.~5 from Li \& Draine (2002).

Finally, to facilitate the incorporation of  future improvements 
in the dust model/inclusion of new dust models, we have constructed a 
flexible interface between the code for calculating dust emission and the 
input dust model.


Because we have modified the dust model we also had to redo the calculations of
attenuation from Paper~III. From the perspective of the extinction properties
of the two dust models (the old version and the new updated version), they are
both designed to reproduce the observed extinction curve of the
Milky Way. However the discrepancy arises from the change in the relative
contribution of absorption and scattering to the total extinction. 
This will give rise to a small but non-negligible difference in the overall
energy balance (around $10\%$ for NGC~891 - see 
Appendix~\ref{sec:attenuation}). 
In Appendix~\ref{sec:attenuation} we briefly
describe the attenuation calculations obtained with the revised dust model and
compare with the results from Paper~III.

\subsection{Calculation of radiation fields}
\label{subsec:calculationradiation}

In the formulation of our model an explicit calculation of radiation 
fields is only done for the diffuse component. There are two channels by which 
the diffuse interstellar medium can be illuminated by stellar light, since,
as described in Sec.~\ref{subsec:distribution}, our model has a diffuse and a 
clumpy distribution of
stars and dust. The first channel is through the illumination of the diffusely 
distributed dust by the smoothly distributed population of old stars. The 
second channel is through the escape of radiation from the young stars, out of 
the localised star-forming complexes.

Thus, the first step in the calculation of
radiation fields in the diffuse medium is to determine the amplitude and SED of
the stellar light which escapes from star-forming complexes.

\subsubsection{The escape fraction of stellar light from the clumpy component}
\label{subsubsec:escape}

In Sect.~\ref{subsec:clumpy} we introduced the factor $F$, which can be 
physically identified
with the luminosity-weighted mean fraction of directions from the
massive stars, averaged over the lifetime of the stars, which
intersect the birth-cloud. The high opacity assumed by our model for
the birth clouds bestows a grey body absorption characteristic for the locally 
absorbed radiation in any individual star-formation region. There is 
nevertheless a wavelength dependence
of the probability of absorption of stellar photons because stars of
different masses survive for different times, such that lower mass and
less blue stars spend a higher proportion of their lifetime radiating
when they are further away from their birthclouds, and because
of the progressive disruption of the clouds by supernovae and stellar winds.
To calculate this wavelength dependence we follow the analysis of
Appendix A of Paper~III, thus preserving consistency between the
model of the attenuation of light from the young stellar population
given in Paper~III and the model of the infrared emission from
star-forming regions given here. Denoting
the wavelength
dependence of the locally absorbed radiation by the function $F_{\lambda}$,
the spectrum and amplitude of the
starlight from the young stellar population in the thin disk with
spectral luminosity density $L^{tdisk}_{\lambda}$ which is intercepted by
a parent cloud is thus given by:

\begin{eqnarray}\label{eq:localised}
L^{local}_{\lambda} = F_{\lambda} \times L^{tdisk}_{\lambda}
\end{eqnarray}
where
\begin{eqnarray}\label{eq:ffactor}
F_{\lambda} = F \times f_{\lambda}
\end{eqnarray}
and $f_{\lambda}$ is tabulated in Table A.1 of Paper~III. As described
in Paper~III, the tabulated
values of $f_{\lambda}$ correspond to a secular decrease of the
solid angle subtended by the parent clouds on a timescale of
$3 \times10^7$\,yr. For completeness we remind the reader that 
$L^{tdisk}_{\lambda}$ is defined
in terms of the free parameter of the model, $SFR$, in Eq.~\ref{eq:youngsed} (see Sect.~\ref{subsubsec:templateyoung}).
\footnote{We note that $L^{tdisk}_{\lambda}$  and $L^{tdisk}_{\nu}$ denote the
  same physical quantities, expressed respectively as spectral densities in
  wavelength and frequency.} 

The amplitude and spectrum of the light from the
young stellar population which illuminates the diffuse dust,
is given by
\begin{eqnarray}\label{eq:diffuse}
L^{tdisk,\,diff}_{\lambda} = 1 - L^{local}_{\lambda}
\end{eqnarray}
Since $F_{\lambda}$ must satisfy $F_{\lambda} \le 1.0$ at all wavelengths,
our model has an intrinsic constraint on the clumpiness
factor of
\begin{eqnarray}\label{eq:conditionf}
F \le 1.0/max(f_{\lambda}) = 0.61
\end{eqnarray}

Physically, this constraint corresponds to the requirement that
the ``porosity factor'' $p_0$ of the birthcloud,  used in Eq.~A2, A12 and A13
of Paper~III to denote the fragmentation of the cloud,
is always less than unity. This limit denotes a complete
lack of fragmentation of the cloud due to the action of supernovae and stellar
winds, such that the probability of escape of photons into the
diffuse ISM is determined only by the migration of stars away from their
birthclouds.

\subsubsection{Combining radiation fields from the young and old stellar
  populations}
\label{subsubsec:combining_radiation}

Since all the diffuse stellar components are attenuated by the same 
distribution of
dust, the diffuse radiation fields seen by each grain can be considered to be 
a sum of the radiation fields produced by each stellar component. Thus we 
calculated the diffuse radiation fields separately for the three main stellar 
components of our model,
the young stellar disk, the old stellar disk, and the bulge. Then the 
individual diffuse radiation fields can be combined to produce the total 
diffuse radiation fields of our desired model galaxy. This is the same concept 
as that used in Paper~III for the calculation of the attenuation of
stellar light. 

To achieve this we first ran the radiative transfer calculations for each
stellar component, each wavelength and each given optical depth
$\tau^f_B$, using the intrinsic spectral luminosity densities
$L^{old}_{\nu,\,unit}$ and $L^{young}_{\nu,\,unit}$, as tabulated in 
Table~\ref{tab:stellarsed}. 
This resulted in the calculation of the unit energy densities for the disk 
$u_{\lambda,\, unit}^{disk}(R,z,\tau^f_B)$, for the thin disk  
$u_{\lambda,\, unit}^{tdisk}(R,z,\tau^f_B)$ and for the bulge 
$u_{\lambda,\, unit}^{ bulge}(R,z,\tau^f_B)$ for the unit stellar luminosities 
(i.e. $L^{old}_{\nu,\,unit}$ for the disk or the bulge and
$L^{young}_{\nu,\,unit}$ for the thin disk).

Then the unit radiation fields were scaled according to the luminosity
of each diffuse stellar component, according to the parameters of the 
model. 
For the old stellar disk (only diffuse) the resulting radiation fields 
$u_{\lambda}^{disk}(R,z,\tau^f_B,old)$ were obtained from:
\begin{eqnarray}\label{eq:radiationfields_disk}
u_{\lambda}^{disk}(R,z,\tau^f_B,old) = old\times 
u_{\lambda,\, unit}^{disk}(R,z,\tau^f_B)
\end{eqnarray}
where  $old$ is the normalised luminosity of the old stellar population, 
defined by Eq.~\ref{eq:old}. For the young 
stellar disk $u_{\lambda}^{tdisk}(R,z,\tau^f_B,SFR,F)$ is derived from:
\begin{eqnarray}\label{eq:radiationfields_tdisk}
u_{\lambda}^{tdisk}(R,z,\tau^f_B,SFR,F) = 
\frac{SFR}{1\,{\rm M}_{\odot}\,{\rm yr^{-1}}}\times(1-F{f_\lambda})\times 
\nonumber \\
\times\, u_{\lambda,\, unit}^{tdisk}(R,z,\tau^f_B)
\end{eqnarray}
where $SFR$ is related to the normalised luminosity of the young stellar
populations in Eq.~\ref{eq:sfr}.
For the bulge the radiation fields are given by:
\begin{eqnarray}\label{eq:radiationfields_bulge}
u_{\lambda}^{bulge}(R,z,\tau^f_B,old,B/D) = old \times 
B/D  \times 
\, u_{\lambda,\, unit}^{bulge}(R,z,\tau^f_B)\nonumber \\
\end{eqnarray}

The total diffuse radiation fields $u_{\lambda}(R,z,\tau^f_B,SFR,F,old,B/D)$ are given by:
\begin{eqnarray}\label{eq:radiationfields_total}
u_{\lambda}(R,z,\tau^f_B,SFR,F,old,B/D) = u_{\lambda}^{disk}(R,z,\tau^f_B,old) + 
\nonumber \\
+ u_{\lambda}^{tdisk}(R,z,\tau^f_B,SFR,F) + 
u_{\lambda}^{bulge}(R,z,\tau^f_B,old,B/D)  
\end{eqnarray}

This formulation provides an exact treatment of the spatial variation of the
colour of the radiation field in galaxies, which is a
crucial factor in shaping the infrared SEDs.

In principle this formalism implies that we would need to create a library of
radiation fields calculated for each possible value of the $F$ factor, implying
a different wavelength dependence $(1-Ff_{\lambda})$ of the fraction of photons 
escaping from the clumpy component into the diffuse one, making the problem 
quite
intractable. In practice, however, it is possible to simplify the problem if we
make the assumption that the wavelength dependence $(1-Ff_{\lambda})$ is fixed
to the $(1-F_{cal}f_{\lambda})$ corresponding to a fiducial calibration value 
$F_{cal}=0.35$, which we consider to be a typical value for spiral galaxies in 
the local universe. This intermediate value of $F_{cal}$
(theoretical values of $F$ can vary from 0 to 0.61; Paper III) is 
consistent with the observation that 60-70$\%$ of the dust emission from spiral 
galaxies emanates from the diffuse component (e.g. Sauvage et
al. 2006). Furthermore, detailed analysis of face-on spiral galaxies with
resolved star-formation regions in the Spitzer 24\,${\mu}$m and the GALEX FUV
bands (e.g. Calzetti et al. 2005)
have shown an almost one-to-one correspondence between the direct and
re-radiated light coming from these complexes, albeit with a large variance
in the ratio of the two emissions. Thus there were essentially no
Spitzer sources with no FUV counterparts (meaning that $F$ is unlikely to take
the value 0.61), but also no FUV sources with no Spitzer counterparts (meaning
that $F$ is unlikely to take the value 0). We note that the value of $F_{cal}$ 
is only used to determine the exact wavelength dependence of that
part of the light from the young stellar population which escapes the
star-formation complexes and actually illuminates
the diffuse dust. Although the computed colour of this light will be
inaccurate if $F$ differs from $F_{cal}$, corresponding
inaccuracies in the predictions for the dust emission SEDs are unlikely to
be more than a few percent, as demonstrated in
Sect.~\ref{subsec:approximation}, where we quantify the effect on the dust
emission of changing the spectral shape of the stellar emissivity.
We note that, for the same reason, when dereddening the
observed UV/optical SED, the wavelength dependence
of the component of attenuation from the
clumpy component can use the exact solution for $F$ found in
the optimisation.


The effective reddening law of the continuum photons escaping the 
star-formation regions corresponding to $F_{cal}$ is given in 
Table~\ref{tab:flambda}.

The use of the calibration factor $F_{cal}$ means that in practice all the
equations that describe the illumination of the diffuse dust by the young
stellar disk need to be rescaled to accommodate different values of $F$ than 
those used in the calibration. For this we define a correction factor for the
diffuse component $corr^{d}(F)$:
\begin{eqnarray}\label{eq:correction_diffuse_F}
corr^{d}(F) = (1-F) \times \frac{L^{young}_{unit,\, UV}}{\int_{UV} 
L^{young}_{\lambda,\,    unit}\, (1-F_{cal}\,f_{\lambda})\, d{\lambda}}
\end{eqnarray} 
In this case Eq.~\ref{eq:radiationfields_tdisk} becomes:

\begin{eqnarray}\label{eq:radiationfields__rescale}
u_{\lambda}^{tdisk}(R,z,\tau^f_B,SFR,F) = 
\frac{SFR}{1\,{\rm M}_{\odot}\,{\rm yr^{-1}}}\,(1-F_{cal} f_{\lambda})\,
corr^{d}(F)\times \nonumber\\
\times u_{\lambda,\,unit}^{tdisk}(R,z,\tau^f_B) 
\end{eqnarray} 
In an analog way we define a correction factor for the localised component
$corr^{l}(F)$:
\begin{eqnarray}\label{eq:correction_localised_F}
corr^{l}(F) = F \times \frac{L^{young}_{unit,\, UV}}{\int_{UV} 
L^{young}_{\lambda,\, unit}\, F_{cal}\,f_{\lambda}\, d{\lambda}}
\end{eqnarray} 
in which case  Eq.~\ref{eq:localised} becomes:

\begin{eqnarray}\label{eq:localised_rescaled}
L^{local}_{\lambda} = \frac{SFR}{1\,{\rm M}_{\odot}\,{\rm yr^{-1}}}
F_{cal}\,f_{\lambda}\,corr^{l}(F)
  \times L^{young}_{\lambda,\,unit}
\end{eqnarray}
  
\subsubsection{The radiative transfer calculations}

One of the key elements of any self-consistent model for the SEDs of galaxies
is the incorporation of a radiative transfer (RT) code that can be used to
determine the radiation fields in galaxies and their appearance in the
UV/optical bands. Four methods have 
been proposed so far in 
RT models, though only two of them have been broadly
used in the context of galaxy modelling, namely ray-tracing methods 
(Kylafis \& Bahcall 1987, Silva et al. 1998, 
Semionov \& Vansevi\u cius 2005a, Semionov \& Vansevi\u cius 2006) and 
Monte-Carlo (MC) techniques (Witt et al. 1992, Bianchi et al. 1996, 
de Jong 1996, Gordon et al. 2001, Baes \& Dejonghe 2002, Baes et al. 2003, 
Baes et al. 2005, Jonsson 2006, Baes 2008, Bianchi 2008, 
Chakrabarti \& Whitney 2008). 
\footnote{There are also RTs 
initially developed for use in other astrophysical contexts 
(Mattila 1970, Witt 1977, 
Yusef-Zadeh 1984, Wood et al. 1996, Lucy 1999, Wolf et al. 1999, 
Bjorkman \& Wood 2001, Steinacker et al. 2003, Wolf 2003, Juvela 2005, 
Steinacker et al. 2006, Fritz et al. 2006).} 

Since our model makes explicit calculations
of radiation fields in galaxies we use an ray-tracing code, which is better 
suited to this type of problem. As in Paper~I we use the Kylafis \& Bahcall 
(1987) code, updated and revised according to the specific needs of our 
modelling technique.

The code of Kylafis \& Bahcall (1987) was originally developed to model the
surface brightness distribution of edge-on galaxies with cylindrical symmetry.
This utilises a ray-tracing method that exactly calculates
the direct and first order scattered-light and uses the method of
scattered intensities to approximate the higher order scattered light
(see review of Kylafis \& Xilouris 2005). Several modifications
were incorporated for the present application, the most significant
of which we briefly describe here.

Firstly, the code used was generalised
and optimised for application to any cylindrically symmetric distribution
of emissivity and opacity, such as the thin disk representing the
young stellar population and associated dust in our standard calculations,
and the annular representation of this disk that we use in the
tests made in Sect.~\ref{subsec:spiral} to evaluate the second disk 
approximation to
spiral structure. To achieve this, we incorporated an in-ray sampling
scheme which automatically adapts to variations in emissivity and
opacity encountered along the rays. Further, in regions of low emissivity, 
where the method of
scattered intensities becomes less accurate, the maximum order of the
scattered light had to be reduced. The reduction was done progressively
accordingly to the emissivity encountered, such that only
the fully accurate first order scattering term was utilised in the regions
with the smallest emissivity. This is a good approximation for the
geometries of dust and stars considered here, in which low emissivities
are never encountered in regions of high opacities.
Lastly, for application to the calculation of radiation
energy densities the code was further modified
to calculate the vector radiation field sampled in each direction of a
predefined set of 216 directions.

To provide a more frequent sampling of the inner part of the galaxy the 
spatial positions were arranged on an
irregular grid with sampling intervals in radial direction ranging
from 50\,pc in the centre to up to 2\,kpc in the outer disk and with sampling
intervals in vertical direction ranging from 50\,pc in the plane to up to 
500\,pc in the outer halo. 
This scheme was necessary to properly sample the radiation in
the more optically thick central regions, where, as can be
seen from Fig.~\ref{fig:radiation_radial}, the dust associated with the thin 
stellar disk in conjunction with the strong positional dependence
of the emissivities of the young stellar population (in $z$)
and the inner emissivity cusp of the bulge (in both $R$ and $z$) can
lead to quite small scale structures even in the diffuse radiation fields.
We made some further tests with higher sampling, ranging from 25\,pc to up
to 1\,kpc in radial direction and from 25\,pc to up 250\,pc in vertical
direction. This made a less than $1\%$ difference in the results with a large
increase in the computation time. We have therefore concluded that the optimum
sampling of the radiation fields is the ${\Delta}R=50$\,pc - 2\,kpc and
${\Delta}z=50$\,pc - 500\,pc scheme.

The database of integrated SEDs presented in this paper does not incorporate
the effect of the reabsorption of infrared light emitted by grains
in the galaxy, as this process has only a marginal
effect on the integrated dust emission, and would entail a significant
increase in complexity and decrease in speed of the calculations.
This is a good approximation for the present problem. For example,
for the most opaque disks we consider ($\tau^f_B\,=\,$8), only
the $\sim\,3\,{\mu}$m dust emission shows a significant depression in the
integrated emission. However we do calculate
reabsorption of infrared light when we produce maps of galaxies seen at higher
inclination, and this process has been incorporated in the calculation
of the profiles presented in  Sect.~\ref{subsec:second-dustdisk}.

The radiation fields illuminating the dust in the diffuse disk
are calculated taking into account anisotropic scattering,
using the albedo and anisotropy of the scattering phase function
as defined in Sect.~\ref{sec:extinction} and the geometry of the dust
as defined by Eqs.~\ref{eq:dustemissivity1} and \ref{eq:dustemissivity2} in 
Sect.~\ref{subsec:distribution}. At each UV/optical
wavelength\footnote{The radiation transfer calculations were performed
at each of the wavelengths given in Table~\ref{tab:stellarsed},
with the exception of the longest wavelength at
50000\,$\AA$, where the radiation fields could be extrapolated sufficiently
accurately from the radiation fields at 22000\,$\AA$ assuming a
Rayleigh-Jeans law.
The sampling points for the calculations are the same as those used for the
calculation of attenuation in Paper~III, except for the addition of points at
1500 \& 3650\,$\AA$ to improve the definition of the stellar emissivity
SEDs. }
of Table~\ref{tab:stellarsed} and each optical depth $\tau^f_B$
a separate calculation was made for each of
the three stellar components, with smoothly distributed
emissivities given by Eq.~\ref{eq:oldemissivity} for the old stellar population
in the disk and in the bulge,
and Eq.~\ref{eq:youngemissivity} for the young stellar population in the thin 
disk. This adequately samples the wavelength variation in
the optical constants of grains; tests done on this indicate
uncertainties at the level of a percent. Concerning the
corresponding error due to the finite sampling of the stellar SEDs,
the reader is referred to Sect.~\ref{subsec:approximation},
in which we show that the exact colours of
either young or old component of the stellar SEDs only affect the resulting dust
emission SEDs at the level of a percent. Overall, therefore,
we consider that uncertainties
introduced by our finite wavelength sampling on the predicted
dust emission SEDs are minor.

When calculating the heating of the
grains, and therefore of the wavelength-integrated energy absorbed, the 
radiation fields were interpolated on a fine wavelength grid
containing 500 data points logarithmically sampled between 912 and 50000\,$\AA$.
Tests done with 1000 data points showed negligible
differences in the results, therefore we decided to fix the wavelength 
sampling to 500. Energy balance tests done within a
common wavelength sampling indicate 
that our models conserve energy to within better than 1 percent.

\subsection{Calculation of grain temperature distributions}
\label{subsec:calculationtemperature}

As described in detail in Popescu et al. (2000) our calculation incorporates 
an explicit treatment of the temperature fluctuations undergone by small 
grains whose cooling timescales are shorter than the typical time interval 
between impacts of UV photons. This so-called ``stochastic heating''
process determines the amplitude and colour of the bulk of the diffuse emission
from most spiral galaxies in the shortest infrared wavelength bands
($<100\,{\mu}$m). Thus, since for each position in the galaxy the energy
density of the radiation fields heating the grains has been derived 
(see Sect.~\ref{subsec:calculationradiation}),  we can calculate the 
probability distribution of
temperature $P_{i}(a,T)$ for each grain size $a$ and composition $i$ by 
equating the rate of absorption of energy to the rate of 
emission of energy :

\begin{eqnarray}\label{eq:energyconservation}
\pi\, a^2\, c\, \int Q_{abs,\,i}(a,\lambda)\, u_{\lambda}\, d{\lambda} &  = & 4\,\pi\,
(\pi\, a^2)\, \int Q_{abs,\, i}(a,{\lambda})\times \nonumber \\ \times \int B_{\lambda}(a,T)\, P_{i}(a,T)\, dT\,d{\lambda}
\end{eqnarray} where
$c$ is the speed of light and $B_{\lambda}$ is the Planck function.

The
probability distribution of temperature is calculated following Voit (1991),
which combines the numerical integration method of Guhathakurta \& Draine
(1989) and a stepwise analytical approximation. We sample the temperature
distribution in 61 logarithmically spaced points.

\subsection{Calculation of infrared emissivities and spatially
  integrated SEDs in the diffuse component}
\label{subsec:calculationinfrared}

Once the temperature distribution of a grain is known,  
the brightness $I_{{\lambda},\,i}$ of a grain of radius $a$ and composition $i$ 
is then given by:
 
\begin{eqnarray}\label{eq:infraredbrightness}
I_{{\lambda},\, i}(a) & = & Q_{abs,\, i}(a,\lambda)\int B_{\lambda}(a,T)\, P_{i}(a,T)\, dT
\end{eqnarray} where
$I_{{\lambda},\,i}$ is in units  of 
$[{\rm W}\,{\rm m}^{-2}{\rm sr^{-1}}{\rm \AA^{-1}}]$.

Once the infrared brightnesses $I_{{\lambda},\, i}$ are calculated for each
grain size and composition, we can then integrate over the grain size
distribution $n(a)$ to obtain the infrared emissivity per H atom of grains of 
a given composition $i$, $j_{\lambda,\, i}^{\rm H}$:

\begin{eqnarray}\label{eq:infraredemissivity}
j_{\lambda,\, i}^{\rm H} & = & \int_{a_{min}}^{a_{max}} \pi\, a^2\,n(a)\,I_{{\lambda},\, i}(a)\,da
\end{eqnarray} where
$j_{\lambda,\, i}^{\rm H}$ is in units of  
$[{\rm W}\,{\rm sr^{-1}}{\rm \AA^{-1}}{\rm H^{-1}}]$.

Finally, the total infrared emissivity per H atom, $j_{\lambda}^{\rm H}$ is
obtained by summing over the grain composition $i$

\begin{eqnarray}\label{eq:infraredemissivity_total}
j_{\lambda}^{\rm H} & = & \sum_i j_{\lambda,\, i}^{\rm H}
\end{eqnarray}

\begin{figure}
\centering
\includegraphics[width=9cm]{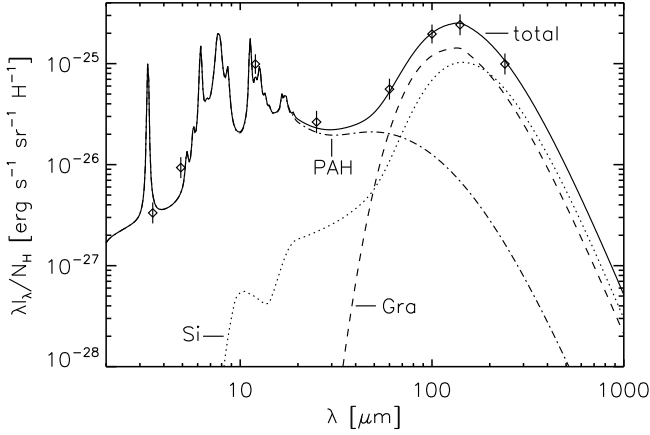}
\caption{The dust and PAH emission SED for the diffuse ISM at high
  Galactic latitude (solid line) calculated using our model. Also plotted 
  here are the COBE data (symbols
  with error
  bars). These are given as an average of the data from the North ecliptic pole
  field and the Lockman Hole field (Arendt et al. 1998) and are further colour
  corrected. The contributions of the different dust compositions to the total
  model SED are as follows: Si (dotted line), Gra (dashed line), PAH
  (dashed-dotted line).}
\label{fig:emission_LIRF}
\end{figure}

In Fig.~\ref{fig:emission_LIRF} we show the calculation for the infrared 
emissivity of grains 
heated by the local interstellar radiation fields (LIRF), with the energy
densities of the radiation fields taken from Mathis, Metzger \& Panagia
(1983). The resulting infrared SED is similar to that presented in Fig.~12 of
Draine \& Li (2007) and indicates that our method for calculating stochastic
heating of grains gives similar results to the method of Draine \& Li. This can
be taken as a benchmark test for the calculation of infrared emission.
In Fig.~\ref{fig:emission_LIRF} we also plotted the colour-corrected 
DIRBE/COBE data for the high 
Galactic latitude. As expected, there is  a good match between the model and 
the observations. In Fig.~\ref{fig:emission_LIRF} we also show the
predicted contribution of the different dust compositions to the total infrared
emissivity.  

Once the total infrared emissivities per H atom have been calculated, 
these can be then scaled to the 
 volume density of dust grains at each position in the galaxy to obtain the 
volume luminosity density  of the first and second dust disks, 
$L_{\lambda, dust}^{disk}(R,z)$ and $L_{\lambda, dust}^{tdisk}(R,z)$, as a function of 
position:
\begin{eqnarray}\label{eq:infraredluminosity_disk1}
L^{disk}_{\lambda, dust}(R,z) =  \frac{4\,\pi}{C_{ext}(B)}\,  
\frac{\tau_B^{f}\,\tau_{ratio}}{2\, z^{\rm disk}_{\rm d}\,(1+\tau_{ratio})}\,
\times\, j_{\lambda}^H(R,z) \times \nonumber \\
\times \exp \left( - \frac{R}{h^{\rm disk}_{\rm d}}- 
\frac{|z|}{z^{\rm disk}_{\rm d}}\right) 
\end{eqnarray}
\begin{eqnarray}\label{eq:infraredluminosity_disk2}
L^{tdisk}_{\lambda, dust}(R,z) =  \frac{4\,\pi}{C_{ext}(B)}\,  
\frac{\tau_B^{f}}{2\, z^{\rm tdisk}_{\rm d}\,(1+\tau_{ratio})}\,
\times\, j_{\lambda}^H(R,z) \times \nonumber \\
\times \exp \left( - \frac{R}{h^{\rm tdisk}_{\rm d}}- 
\frac{|z|}{z^{\rm tdisk}_{\rm d}}\right) 
\end{eqnarray}
where $L^{disk}_{\lambda, dust}$ and $L^{tdisk}_{\lambda, dust}$ are in units
of $[{\rm W}\,{\rm m}^{-3}\AA^{-1}]$ and
$C_{ext}(B)=6.38\times 10^{-22}\,{\rm cm}^2\,{\rm H}^{-1}$ is the extinction
cross-section in the B band (at $4430\,\AA$) for the dust model considered 
here. The spatially integrated SEDs (the spectral 
luminosity density) of the 
diffuse component of the
model galaxy (with the fixed size  $h^{\rm disk}_{s, ref}$), 
 $L_{\lambda,\, dust}^{diff,\, model}$ is then:
\begin{eqnarray}\label{eq:infraredsed_diffuse}
L_{\lambda,\, dust}^{diff, \, model} = 
2\,\pi\,\int_z\int_R (L^{disk}_{\lambda, dust}+
L^{tdisk}_{\lambda, dust})\, R\,dR\,dz
\end{eqnarray}

We checked the fundamental energy balance in the calculations, by comparing the
energy absorbed from the radiation fields with the energy emitted in the
infrared, obtaining consistency to within better than $1\%$.

\subsection{The infrared emission from the star-forming regions}
\label{subsec:infraredclumpy}

In our model we take the total energy absorbed by the birthclouds to
be
\begin{eqnarray}\label{eq:localabsorption}
L^{local}_{abs} = \int_{\lambda} L^{local}_{\lambda}\,d{\lambda} + 
f_{ion-uv}\, L^{tdisk}_{ion-uv}\
\end{eqnarray}
where $L^{local}_{\lambda}$ is defined in Eq.~\ref{eq:localised_rescaled},   
$L^{tdisk}_{ion-uv}$ is defined in Eq.~\ref{eq:sfr_ion-UV} and
$f_{ion-uv}=0.3$.
 
That is, as justified in Sect.~\ref{subsec:clumpy}, we take the heating of the 
grains in the 
birthclouds to be dominated by photons from the stellar progeny, and neglect 
any external contribution from the ambient radiation fields in the galaxy. 

To determine the SED of the reradiated light from dust grains
and PAH molecules in the birthclouds we use a fixed spectral template of a 
photodissociation region (PDR): 
\begin{eqnarray}\label{eq:HIItemplate}
L^{local}_{\lambda,\,dust} = L^{local}_{abs}\, L^{PDR}_{\lambda}
\end{eqnarray}
where $L^{PDR}_{\lambda}$ is the template function for a photodissociation
region, normalised such that

\begin{eqnarray}\label{eq:normalisationtemplate}
\int L^{PDR}_{\lambda}\, d{\lambda} = 1.0
\end{eqnarray}

We use the model of Groves et al. (2008) (see also Dopita et al. 2005),
fitted to broad-band measurements of dust and PAH emission from galactic
star-formation regions, to provide an empirically constrained prediction
for the detailed spectral form of the template $L^{PDR}_{\lambda}$.
Although this model was primarily developed
for use in predicting the SED of starburst galaxies,
its fundamental constituent is a prediction of the PAH/dust and nebular
line emission from the HII region and PDR components of individual
SF regions, and so, with suitable choice for the values of the
model parameters,  is also directly applicable to the prediction
of the PDR emission
from SF regions in spiral galaxies. For the spectral template used here
we only consider the PDR component calculated by Groves et al., ignoring
the emission component from the HII region (as justified below). We also
only include the PAH and dust emission components in the spectral
template, not including the free-free and line emission from the gas phase.

The prime physical motivation for the use of the model
of Groves et al. is that it self-consistently calculates the effect
on the emergent dust and PAH emission SED of the dynamical evolution
of the emitting regions. Specifically, the model considers a
spherically symmetric, fully enclosed mass-loss bubble,
driven by the mechanical energy input through winds and supernovae
of a star cluster as a function of the external density
(parameterised in terms of the pressure of the ambient ISM) and a
``compactness parameter'', $C$, which
scales according to the mean luminosity-weighted flux of photons onto
the PDR. The emission from PAH and dust in the PDR is then self-consistently
determined through a separate radiation transfer calculation (Groves et
al. 2008), as a function of the grain column density in the PDR
(which in the formulation of Groves et al. is actually prescribed through a
combination of the gas column in the PDR and the metallicity).
Being based on a dynamical model, the calculations of Groves et al.
explicitly take into account the expected variation
in colour and amplitude of the incident photon flux on the PDR
due to the increase in radius and evolution of the stellar population,
which acts to broaden the SED of the dust emission from the
large ensemble of SF regions at different evolutionary stage
expected in a spiral galaxy. The model
also considers the photodestruction of PAH, strongly influencing
the relative amplitudes of the PAH and far-IR/submm emission components
from the PDR.

\begin{table}
\caption{The references used for collecting the observed flux densities of the
  57 SF regions plotted in Fig.~\ref{fig:HII_template}.} 
\label{tab:ref}
\begin{tabular}{l}
\hline\hline
Bains et al. (2006)\\
Buckley \& Ward-Thompson (1996)\\
Conti \& Crowther (2004)\\
Gordon (1987)\\
Gordon \& Jewell (1987)\\
Hill et al. (2005)\\
Mooney et al. (1995)\\
Moore et al. (2007)\\
Sievers et al. (1991)\\
Thompson et al. (2006)\\
Ward-Thompson \& Robson (1990)\\
\hline
\end{tabular}
\end{table}

To empirically verify these predictions of the
model, and determine the best model parameter to determine the
PAH and dust emission template from an ensemble of localised SF regions
in spiral galaxies, we compared the model
predictions with observed data (see Table~\ref{tab:ref}) for
the radio-selected sample of star-formation regions in the Milky Way of
Conti \& Crowther (2004). This sample is distributed throughout the
disk of the Milky Way, so is likely to be representative of the
population of SF regions in the galaxy. In Fig.~\ref{fig:HII_template} we plot
the mid-IR/far-IR/submm SEDs of 57 galactic star-forming regions
from Conti \& Crowther, normalised to the 100\,${\mu}$m flux density. Overlaid
on this is the best fit model for the PDR emission from Groves et al.,
found by
searching the parameter space in the compactness factor $C$ and the
dust column density in the PDR (adjusted by varying the product of
metallicity and gas column density). The other major parameter, pressure,
was kept fixed as the PAH and dust emission is completely insensitive
to this parameter which only affects the emission lines
(see Fig.~4 of Groves et al.), which are not required in our template.
In comparing the model with the data we did however take into
account the free-free emission component from the PDRs, since this
emission component will also be represented in the broad-band continuum
data of the Milky Way star formation regions. The free-free emission slightly
raises the level of the emission near 1\,mm above the prediction of the
dust emission template, as shown
by the divergence of the dotted line, which includes the free-free
component from the full line in Fig.~\ref{fig:HII_template}.

\begin{figure}
\includegraphics[width=9.4cm]{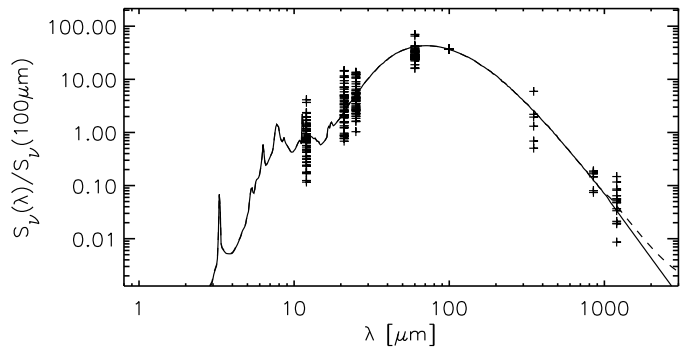}
\caption{MSX and IRAS mid-IR and
far-IR integrated photometry for 57 SF regions from
Conti \& Crowther (2004), supplemented by all
published sub-mm and/or near-mm data for which integrated flux measurements
covering the full angular extent of the source are available (18 sources); 
(see Table~\ref{tab:ref}). Measurements are normalised to the 100\,${\mu}$m 
flux density. The solid line is the adopted model for the PDR dust, PAH
emission from Groves et al. (2008),
having compactness parameter $log(C)=6.5$, metallicity
1.0, and PDR column density $log(N)=22.0$. The dashed-line, which deviates
slightly from the solid line in the submm range, is the same but with the
free-free emission included.}
\label{fig:HII_template}
\end{figure}

The best fitting PDR model is for
compactness parameter $log(C)=6.5$, metallicity
1.0, and hydrogen column density $log(N)=22.0$. Although there is a large
spread in observed colours, indicating the wide range of
evolutionary states and environment of individual SF regions,
overall this model
reproduces the mean colours of the population of
sources over the whole
spectral range covered by the data.
This spectral range extends from
12\,${\mu}$m (including the PAH emission), through 25, 60 and
100\,${\mu}$m (where emission is dominated by
large grains in thermal equilibrium with the
intense local radiation fields), and into the submm/near-mm range
(where the emission is dominated by dust which is cold due to
the self-shielding of grains in the PDR. This latter cold dust emission
component is  difficult to predict theoretically, depending
on the otherwise poorly constrained dust column density, so is
fundamentally an empirically determined quantity. Likewise,
the relatively high value of the compactness factor $C$ is
needed to fit the rather warm observed far-IR/mid-IR colours.
As directly confirmed through high resolution imaging of
the sources this indicates a close proximity (typically
parsec scales) of the exciting star clusters
to the PDRs, most of which are associated with rather
dense fragments of molecular material which will be only slowly
pushed out from the cluster. It is interesting to note that
the fit of the PDR model to the data leaves no room for a
warmer dust emission component from an enclosed HII region, suggesting
that the HII regions typically extend to larger spatial scales. On this
picture, the cloud fragments carrying the PDR emission are embedded
in a diffuse ionised medium which blends
in with the diffuse emission from the second dust disk.

Although, as we argue above, the physical values for the compactness parameter
$C$, metallicity and hydrogen column density used for the fit seem plausible,
we note that there may be room for some systematic uncertainties in these
parameters if the optical properties of the grains differ from the model of 
Weingartner \& Draine used by Groves et al.. The most important 
point however in the context of our use of this template SED 
is that it gives a good fit to a representative sample of star formation regions
in our Milky Way. So even if the physical interpretation of the shape of the
SED may be still somewhat imprecise, the template itself will be the correct
one. In conclusion, we believe that the PDR model of
Groves et al. that best fits the galactic star-forming regions can be
taken as a template SED for the clumpy component of our model.

\subsection{The free parameters for the calculation of the infrared SEDs}
\label{subsec:freeparam}

The spatially integrated SED for the dust emission of a galaxy 
is given by:
\begin{eqnarray}\label{eq:infraredsed_total}
L^{model}_{\lambda,\, dust}(\tau^f_B,SFR,F,old,B/D) =    \nonumber\\
L_{\lambda,\, dust}^{diff, \,model}(\tau^f_B,SFR,F,old,B/D)
+ L_{\lambda,\, dust}^{local}(SFR,F)
\end{eqnarray}
where $L_{\lambda,\, dust}^{diff, \,model}$ is defined by 
Eq.~\ref{eq:infraredsed_diffuse} and 
$L_{\lambda,\, dust}^{local}(SFR,F)$ is defined by Eq.~\ref{eq:HIItemplate}.
From Eq.~\ref{eq:infraredsed_total} one can see that the free parameters of our 
model are, to recap: 
\begin{itemize}
\item
the central face-on opacity in the
B-band $\tau^f_B$ (Eq.~\ref{eq:totalopacity}; Sect.~\ref{subsec:distribution}), 
\item
the clumpiness factor $F$ (Sect.~\ref{subsec:clumpy}), 
\item
the star-formation rate $SFR$ (Sect.~\ref{subsubsec:templateyoung}), 
\item
the normalised luminosity of the old stellar  population $old$ 
(Sect.~\ref{subsubsec:templateold}) and 
\item
the bulge-to-disk ratio $B/D$ (Sect.~\ref{subsubsec:templatebulge}). 
\end{itemize}
In cases where detailed modelling
of the optical/NIR images is also available, e.g. the study of NGC~891, the
$old$ and $B/D$ parameters are independently constrained from the optimisation 
of the optical images, and thus only 3 free parameters are needed for the model:
$\tau^f_B$, $SFR$, $F$.

We note that the free parameters $\tau^f_B$, $F$, and $B/D$ are also free
parameters for the attenuation of the UV/optical stellar light given 
in Paper~III (the fourth and final free parameter that affects the attenuation
is the inclination $i$ of the disk). Therefore the dust emission SEDs 
predicted here can be used in conjunction with the predictions for attenuation 
given in this paper for a self-consistent modelling of the UV/optical-IR/submm 
SEDs.

The parameter $\tau^f_B$ can be related to the total dust mass using the
equation: 
\begin{eqnarray}\label{eq:dustmass}
M_{dust} = \gamma \, h^2_s(B) \, \tau^f_B
\end{eqnarray}
where $\gamma$ is a constant related to the geometry of the distribution of 
dust in galaxies and to the dust model. For our model with  $h^2_s(B)$ in
parsec,  $\gamma = 0.9912 {\rm M}_{\odot}\,{\rm pc}^{-2}$.

\section{Illustration of the model on NGC~891}
\label{sec:illustration}

We illustrate our model, including description of the intermediate steps in the
calculation, using NGC~891. Because of the changes to our model, it is 
important to see if we can now fit the whole SED of NGC~891, 
including the MIR emission dominated by the PAH spectral features.

\begin{figure}
\centering
\includegraphics[width=9.4cm]{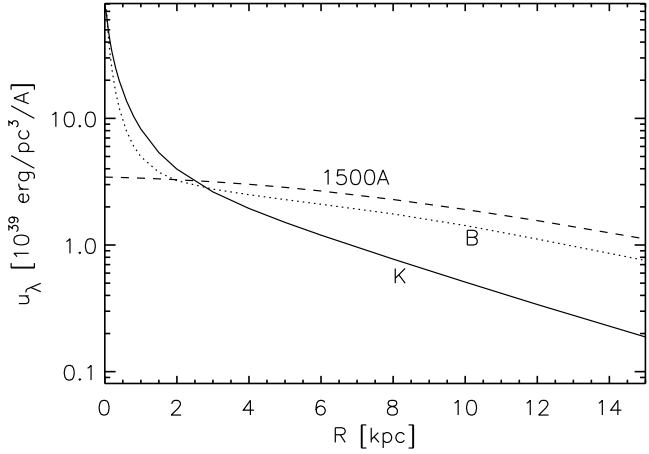}
\caption{Examples of calculated radial profiles of radiation fields in the
  plane of the disk ($z=0$\,pc), for the best fit model of our prototype galaxy
  NGC~891. The different profiles are 
for different wavelengths.}
\label{fig:radiation_radial}
\end{figure}

\begin{figure}
\centering
\includegraphics[width=9.4cm]{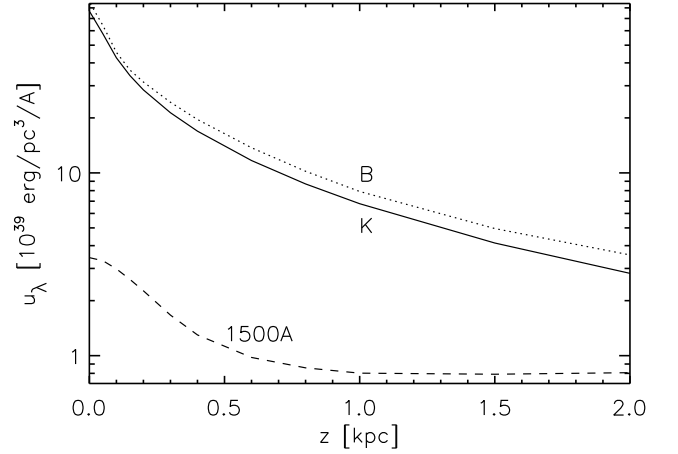}
\caption{Examples of calculated vertical profiles of radiation fields in the
  centre of the disk ($R=0$\,pc), for the best fit model of our prototype
  galaxy NGC~891. The different profiles are 
for different wavelengths.}
\label{fig:radiation_vertical}
\end{figure}

As described in Sect.~\ref{subsec:freeparam}, since the luminosity of the old 
stellar populations
has been derived from the optimisation of the optical/NIR images, we only need
3 free parameters to fit the dust emission SED of NGC~891: 
$\tau^f_B$, $SFR$, $F$. For this we
ran the calculations for trial combinations of these parameters. We note that 
the normalised luminosity of the old stellar population derived for NGC~891 
is $old=0.792$. Throughout this section all the examples shown are for the best
fit model\footnote{We quote the best fit
  $SFR$ to two decimal places because we use $SFR$ as a proxy for the 
  luminosity of the young stellar population (see Eq.~\ref{eq:sfr}).} 
for NGC~891, which is for 
$\tau^{f}_{B}=3.5$, $SFR=2.88$\,M$_{\odot}$/yr and  $F=0.41$.

\subsection{The radiation fields}
\label{subsec:radiation}

The first step in the calculation 
(see Sect.~\ref{subsec:calculationradiation}) is the derivation of the
radiation fields illuminating the diffuse dust, which we illustrate in Figs.~\ref{fig:radiation_radial} and 
\ref{fig:radiation_vertical} through examples of 
radial and vertical profiles of energy densities $u_{\lambda}$ of total 
radiation fields (not to be confused with profiles of stellar emissivities).  
The sharp rise in the inner parts of the
radial profiles in the K and B band is produced by the dominance of the 
radiation
coming from the bulge. We note here the large variation in the colour of the
radiation fields with position, in particular in the radial direction, which, 
as shown in the next figures, will introduce large differences in the shape of 
the FIR SEDs. Thus, models that assume radiation fields with the fixed colour 
of the  local interstellar radiation fields (LIRF) are likely to introduce 
systematic uncertainties in the predictions for the dust emission SEDs.

\subsection{The temperature distributions}
\label{subsec:temperature}

The next step in the calculations 
(see Sect.~\ref{subsec:calculationtemperature}) is the derivation of the 
probability distributions of dust temperature. In Fig.~\ref{fig:temp_distrib} 
we show examples of temperature distributions for 
grains of different sizes and compositions placed at different positions 
within the galaxy. The sizes plotted for each composition reflect the range of 
sizes given by the
dust model considered in this paper. For example sizes smaller than $10\AA$ 
are only considered for PAH molecules. Also, according to the dust model, the
biggest PAHs are those of $100\AA$, while graphites have only sizes
larger than $100\AA$. As expected, small grains exhibit broad probability
distributions. For example PAH molecules are only heated stochastically, as
they have small sizes. Larger grains have narrower probability distributions,
eventually tending towards a delta function, when they start to emit at
equilibrium temperature.

\begin{figure*}
\centering
\includegraphics{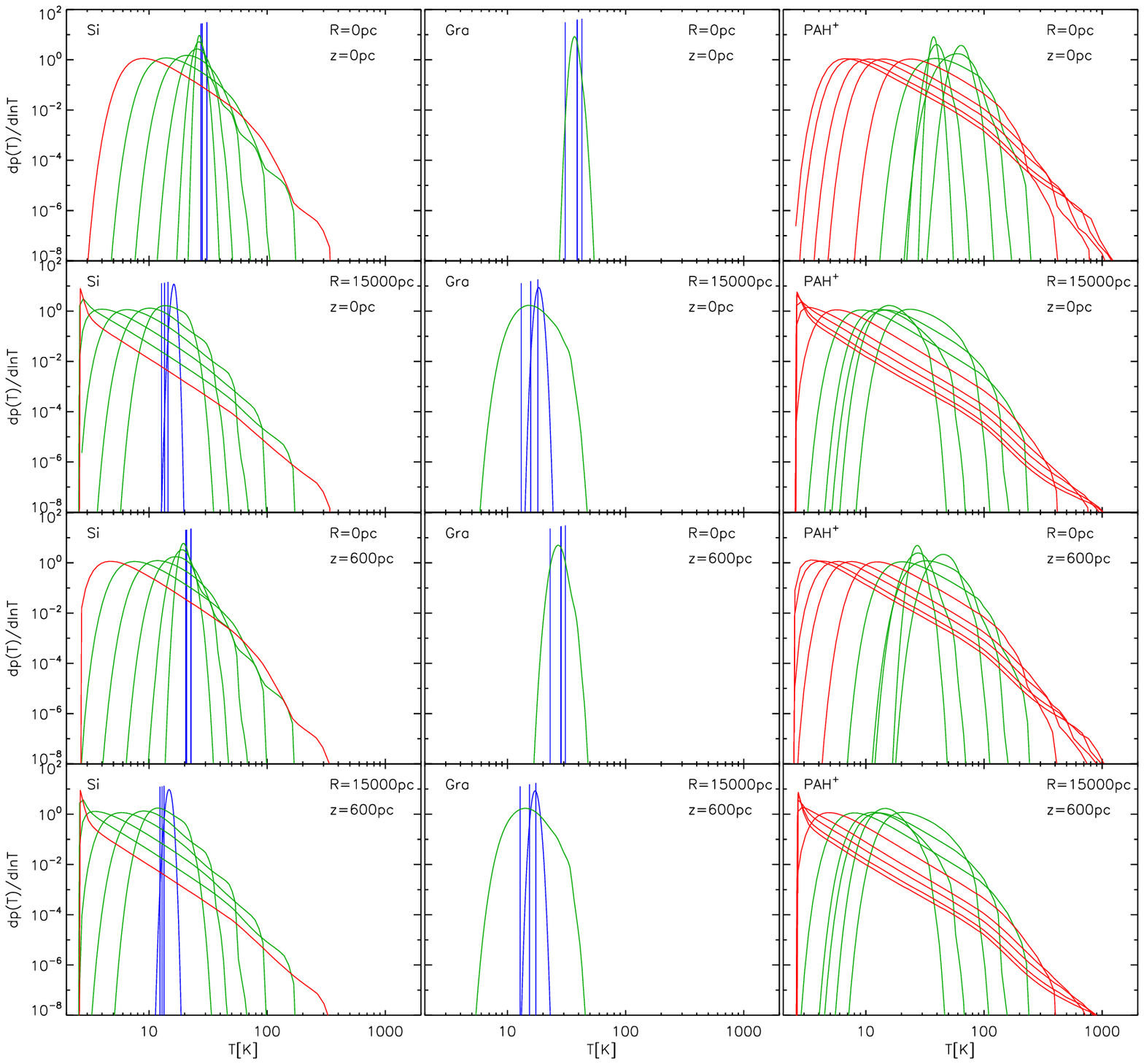}
\caption{Temperature distributions for dust grains of different sizes (plotted
  as different curves in each panel) and 
  various composition: Si (left
  panels), Gra (middle panels) and PAH$^{+}$ (right panels), heated by the
  diffuse radiation fields calculated for the
  best fit model of our prototype galaxy NGC~891. 
 Temperature distributions for PAH$^{0}$ are not plotted in this 
figure. The colour codding
is as follows: red is for grains with radius $a< 0.001\,{\mu}$m 
(0.00035, 0.00040, 0.00050, 0.00063 and 0.00100\,${\mu}$m), green is for
grains with $0.001<a\le0.01$ (0.00158, 0.00251, 0.00398, 0.00631, 0.01000${\mu}$m) and blue
is for grains with $a>0.01$ (0.0316, 0.10000, 0.31623, 
0.7943\,${\mu}$m). The biggest grains have delta function distributions, since
they emit at equilibrium temperature. Going from the top to the
bottom panels the calculations are done for different positions in the model
galaxy: $R=0$\,pc, $z=0$\,pc; $R=15000$\,pc, $z=0$\,pc; $R=0$\,pc, $z=600$\,pc; 
and $R=15000$\,pc, $z=600$\,pc.}
\label{fig:temp_distrib}
\end{figure*}

The figure also illustrates the dependence of the temperature
distributions with position in the galaxy, due to the spatial variation of the
intensity and colour of the radiation fields. For example a graphite grain of
size $0.0316\,{\mu}$m will exhibit a delta function temperature distribution 
if placed in the centre of the galaxy (first and third row from the top, 
middle panels, blue curves in Fig.~\ref{fig:temp_distrib}), 
where the radiation fields
have higher energy densities and redder colours, than if placed in the
outskirts of the galaxy (second and 4th rows from the top, middle panels). Thus, in the 
centre of the galaxy the $0.0316\,{\mu}$m graphite grain will emit at 
equilibrium temperature, while the same grain will start to emit 
stochastically at $R=15000\,$pc.

Finally Fig.~\ref{fig:temp_distrib} shows the dependence of temperature on composition. For example 
a $0.0316\,{\mu}$m grain placed in the centre of NGC~891, will 
exhibit a lower temperature if it had a silicate composition (left upper panel) 
than if it had a graphite composition (middle upper panel), due to the
differences in the optical properties of the grains.

\subsection{The infrared emissivities}
\label{subsec:infrared}

\begin{figure*}
\centering
\includegraphics{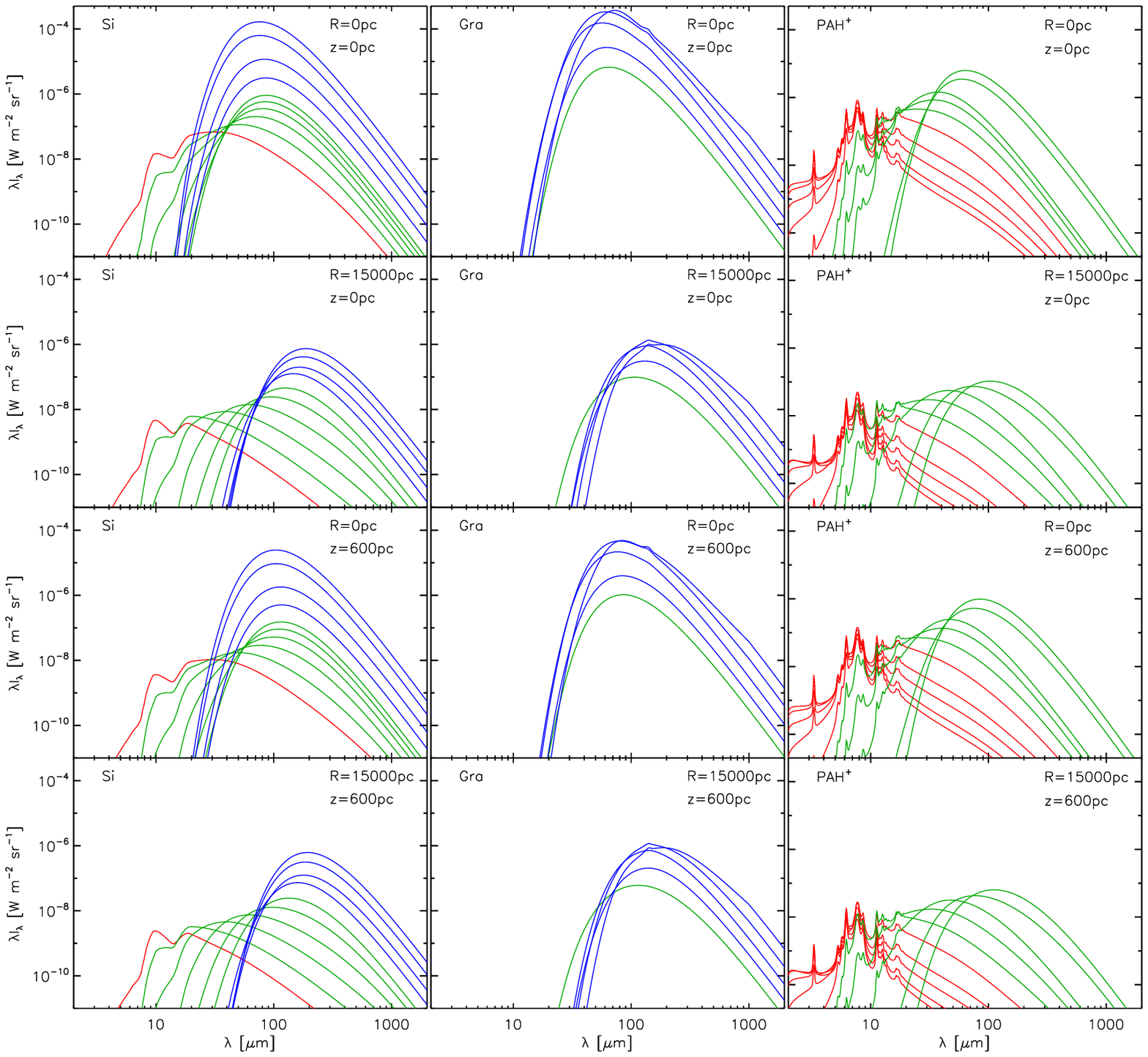}
\caption{Infrared brightnesses for grains of different sizes (plotted
  as different curves in each panel) and 
  various composition: Si (left
  panels), Gra (middle panels) and PAH$^{+}$ (right panels), heated by the
  diffuse radiation fields calculated for the
  best fit model of our prototype galaxy NGC~891. 
Infrared brightnesses for PAH$^{0}$ are not plotted in this 
figure. The colour coding
is as follows: red is for grains with radius $a< 0.001\,{\mu}$m 
(0.00035, 0.00040, 0.00050, 0.00063 and 0.00100\,${\mu}$m), green is for
grains with $0.001<a\le0.01$ (0.00158, 0.00251, 0.00398, 0.00631, 0.01000\,${\mu}$m) and blue
is for grains with $a>0.01$ (0.0316, 0.10000, 0.31623, 
0.7943\,${\mu}$m). The grain sizes considered in this plot are the same as those
plotted in Fig.~\ref{fig:temp_distrib}. Going 
from the top to the
bottom panels the calculations are done for different positions in the model
galaxy: $R=0$\,pc, $z=0$\,pc; $R=15000$\,pc, $z=0$\,pc; $R=0$\,pc, $z=600$\,pc; 
and $R=15000$\,pc, $z=600$\,pc.}
\label{fig:infrared_brightness}
\end{figure*}

In Fig.~\ref{fig:infrared_brightness} we show examples of calculated infrared 
brightnesses (from Eq.~\ref{eq:infraredbrightness}) for grains
of different sizes and compositions placed at different positions within
the diffuse ISM of
our prototype galaxy NGC~891. For the calculations we used the probability 
distributions of temperature shown in Fig.~\ref{fig:temp_distrib}. We also 
show the same positions and  sizes as those from Fig.~\ref{fig:temp_distrib}. 
Following the trends in the temperature distributions, one can
see the dependence of the infrared brightnesses with position in the
galaxy, due to the variation in the intensity and colour of the radiation 
fields. One should note that here we show the absolute brightness of each grain,
meaning that the emission is not weighted to take into account the
abundance of grains according to size or composition;
it simply indicates the response of each dust grain to the radiation fields.

By comparing the pairs of positions, first and third row from the top on one 
hand, and second and 4th row on the other hand, we see
that overall the SEDs show a stronger radial variation than a vertical
variation and that this trend is independent of grain composition and grain 
size. This is primarily a result of the fact that the radiation fields
show a stronger variation in colour in the radial direction than in the 
vertical direction (see again Fig.~\ref{fig:radiation_vertical} and 
\ref{fig:radiation_radial}), as described in 
Sect.~\ref{subsec:radiation}, which, in 
turn, is a direct consequence of the finite disk plus bulge description of our 
geometrical distributions of stars and dust. We will discuss here both the 
change in the peak of the SEDs
as well as the change in the overall amplitude of the infrared brightness.
As expected, for a given grain size and composition the peak of the SEDs 
shifts towards longer wavelengths and its amplitude decreases in weaker 
radiation fields, especially at large galactic radii. However the wavelength 
shift of this peak and its amplitude as a function
of grain size have a more complex behaviour. \\

\noindent
{\it i) The wavelength variation of the peak of the infrared brightnesses with
grain size for a fixed galaxian position (radiation field).}

\noindent
For the stochastically heated grains there is a strong shift in the peak of 
the SED towards longer wavelengths with increasing grain size. This is seen 
for the small grains (Si and PAH; the green curves) placed at large galaxian 
radii (second and 4th row), which all exhibit strong stochastic heating 
(as also seen from the broad probability distribution of temperature in 
Fig.~\ref{fig:temp_distrib}, corresponding panels). Since most of the energy 
is radiated at the 
highest temperature side of the probability distributions, and since the 
increase in the grain size will decrease the width of the probability 
distributions, this means that grains with bigger sizes will reach 
systematically lower maximum temperatures in the probability distributions, 
thus radiating at longer wavelengths.  

By contrast, grains in
equilibrium temperature with the radiation fields will show a small shift in
the SED peak with
increasing grain size, and in the opposite direction, namely towards shorter
wavelengths (see the blue curves from the first and third row from the top, 
left column of Fig.~\ref{fig:infrared_brightness}). According to 
Eq.~\ref{eq:infraredbrightness}, the peak of the infrared brightness will
be determined by the wavelength dependence of $Q_{abs}$ in the far-infrared 
and by the equilibrium temperature at which they radiate. For a given 
radiation field $u_{rad}$, the equilibrium temperature depends only on the 
$Q_{abs}$ (see Eq.~\ref{eq:energyconservation}). Since bigger grains are more 
opaque at shorter wavelengths,
tending to a black-body case, they absorb more efficiently, and therefore their
equilibrium temperature is higher, providing the $Q_{abs}$ has the same
wavelength dependence in the far-infrared, independent of grain size. This is
indeed the case for silicate grains. The graphite grains however have some 
variation in the wavelength dependence in the far-infrared with grain size,
which produce the non-monotonic shift in the peak of their infrared brightness
(see first and third row from the top, middle column from 
Fig.~\ref{fig:infrared_brightness}, where the blue curves 
intersect).

A special case is represented by the smallest PAH molecules (the red curves on
the right column of Fig.~\ref{fig:infrared_brightness}). Their infrared 
brightnesses are dominated by the
emission bands due to vibrational transitions, and these features occur at the
same wavelengths, independent of the molecule size. This is because the
vibrations seen in the mid-infrared correspond to the fundamental stretching
or bending modes of the C-H and C-C bonds, and do not involve the molecule as a
whole.\\

\noindent
{\it ii) The wavelength variation of the peak of the infrared brightnesses with
grain size for a variable galaxian position (radiation fields).}

\noindent
The shift in the peak of the infrared brightness towards longer wavelengths 
with increasing grain size for stochastically heated grains (described at 
{\it i}) 
is a strong function of galaxian position. By comparing the first and 3rd rows
from the top with the second and the 4th rows for the small Si and PAH grains 
(green curves, Fig.~\ref{fig:infrared_brightness})
we see that the shift becomes less pronounced for grains at small galaxian
radii, where the radiation fields are stronger and redder. For the largest PAH 
molecules
(or corresponding silicate grains) there is almost no shift, their peak 
remaining
constant with wavelength. This shows that these grains, despite being small,
are in a transition phase towards equilibrium temperature, as also proven by
their narrower temperature distribution from the corresponding panels in
Fig.~\ref{fig:infrared_brightness}. So even PAH molecules can start to emit 
closer to equilibrium temperature if placed in the centre of the galaxy.

Conversely, the described shift of the peak of the SEDs towards shorter
wavelength with increasing grain size for grains heated at equilibrium
temperature is also a strong function of galaxian position. By comparing the
first and 3rd rows from the top with the second and the 4th rows for the big silicate and
graphite grains (blue curves) we see that the shift can be reversed in the
weaker and bluer radiation fields at large galactic radii. This means that 
even big grains can start to emit stochastically in the outer regions of
galaxies. Indeed, by looking at the corresponding panels from 
Fig.~\ref{fig:temp_distrib} we see that the big grains
have a non-negligible width in the probability distribution of temperature for
the outer disks.

To conclude, the shift of the peak of the infrared brightness as a function of
grain size strongly depends on the temperature of the grains, and therefore on
the stochastic or non-stochastic nature of the heating mechanism. Since the
heating mechanism depends both on grain size and on the intensity and colour of
the radiation fields, it is clear that the shift cannot be described in
terms of grain size only. This also shows that
models that have a fixed grain size for the transition between the main
heating mechanisms of dust, irrespective of the radiation fields, will
lead to systematic spurious shifts in the mid-infrared to FIR colours with 
increasing galactocentric radius.\\ 

\noindent
{\it iii) The variation of the amplitude of the infrared brightnesses with grain size
for a fixed galaxian position (radiation fields).}

\noindent
As apparent from Fig.~\ref{fig:infrared_brightness}, the amplitude of the 
infrared brightness increases with increasing grain size, with a small 
increase for the small grains and a 
bigger increase for the bigger grains. Since the width of the SEDs is 
approximately constant (for a constant set of parameters), the amplitude of 
the infrared brightness will scale 
with the area under their SEDs, namely with the energy absorbed 
(per unit $a^2$). For a given radiation field $u_{rad}$, 
Eq.~\ref{eq:energyconservation} tells us that 
this is determined only by the optical properties of the grains in the UV and 
optical regime. Indeed, overall the $Q_{abs}$ increases with increasing grain
size, with a smaller trend for smaller sizes, and a bigger trend for bigger
grains. Especially in the case of silicate, the big grains show
a tendency for higher efficiency in absorbing optical photons, which will boost
the amplitude of their SEDs due to a higher proportion of red photons being
absorbed. Thus, bigger grains will absorb more red photons than the smaller 
grains, which, for a fixed colour of the radiation fields, will allow big grains
to have a higher increase in their infrared brightnesses with increasing grain
size. \\

\begin{figure*}
\centering
\includegraphics[width=12.0cm]{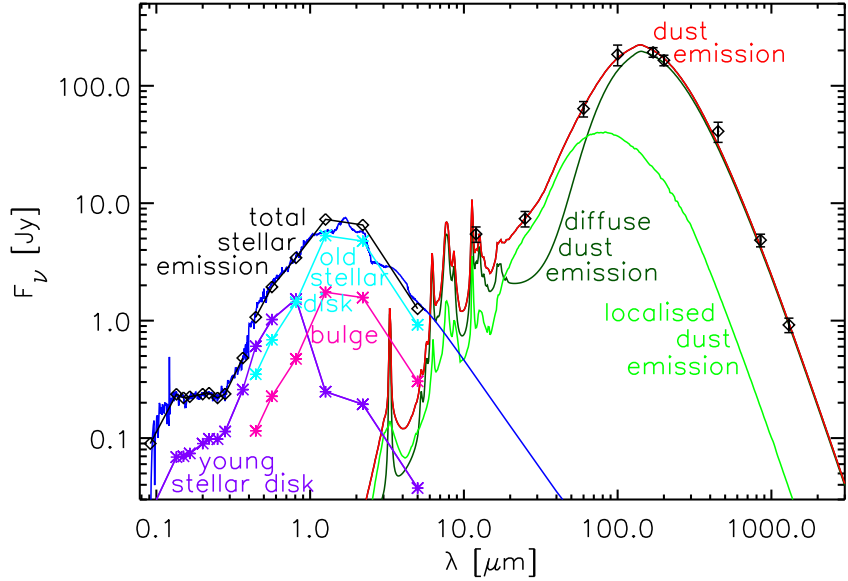}
\caption{The best fit model SED of NGC891, corresponding to $\tau^{f}_{B}=3.5$,
  $SFR=2.88$\,M$_{\odot}$/yr, $F=0.41$, together with the observed data in the
  MIR/FIR/submm (plotted as rhombus symbols with error bars). The observed flux
  densities and the corresponding references are given in Table~\ref{tab:ref}. 
  The
  calculated total intrinsic stellar SED is plotted as black line and rhombus 
 symbols. The
  predicted stellar SED given by population synthesis models is plotted 
  as dark blue line. The intrinsic stellar
  SED of the young stellar population in the diffuse component (the fraction 
$1-F$ escaping the star-forming complexes) is plotted with mauve line. The
intrinsic SED (only diffuse) of the old stellar population, as derived from the
optimisation of the optical/NIR images of NGC891, is plotted with light blue
line. The corresponding SED for the bulge is plotted with pink line. 
The different symbols overplotted on the UV/optical SEDs indicate the
wavelength at which the radiative transfer calculations were performed. The
predicted PAH and dust emission SED is given with red line. Also plotted are
the two main components of infrared emission: the predicted diffuse SED 
(dark-green line) and the predicted clumpy component associated with the 
star-forming regions (light-green line).}
\label{fig:sed_ngc891}
\end{figure*}

\noindent
{\it iv) The variation of the amplitude of the infrared brightnesses with grain size
for a variable galaxian position (radiation fields).}

\noindent
As mentioned in {\it iii)}, the bigger grains have a faster increase
in the amplitude
of their infrared SEDs with increasing size than the small grains, due to
the increase in the efficiency of absorbing red photons. If the radiation
fields will also change in colour, due to their spatial variation, this will 
induce an additional difference in the amplitude of the SEDs of big and small 
grains. Thus, at small galactocentric radii, where 
the radiation fields are redder (see Fig.~\ref{fig:radiation_radial}), more 
optical photons are available to boost the amplitude of the SEDs of the
big grains. 
The trend of increasing the relative contribution of big grain emission to the 
small grain emission with decreasing galactocentric radius is apparent from 
Fig.~\ref{fig:infrared_brightness},
especially for the case of silicates, which have a broad range in grain
sizes. Obviously the increase in the intensity of the radiation fields with
decreasing galactocentric radius will also increase the amplitude of the SEDs,
but this will produce an overall  boost for both small and big grains.

To conclude, the increase in the relative contribution of big grain emission 
to the small grain emission is a strong function of the colour of the radiation
fields, and, unlike the wavelength dependence, does not depend on the heating
mechanism of the grains. This also shows that models that assume a fixed colour
of the radiation fields (e.g. that of the local interstellar radiation fields)
will incur systematic errors in the mid-infrared to FIR colours with increasing
galactocentric radius.

Finally, the increase in the amplitude of the $8\,{\mu}$m PAH features with
respect to the $3.3\,{\mu}$m feature at small galactocentric radius is also due to 
the redder radiation fields, which will provide additional optical photons
capable of
exciting the vibrational transition at around $8\,{\mu}$m, but 
not energetic enough to excite the
transitions around $3.3\,{\mu}$m.

\subsection{The integrated SED}
\label{subsec:sed}

\begin{table*}
\caption{The total MIR/FIR/submm flux densities of NGC~891. The observed flux
  densities available only within a fixed aperture were extrapolated to the
  whole galaxy using the best fit model of NGC~891.} 
\label{tab:fluxes}
\centering
\begin{tabular}{rrrlll}
\hline \hline
wavelength & $F_{\nu}$ & error($F_{\nu})$ & telescope & reference &\\
${\mu}$m & Jy & Jy & & &\\
\hline
 12 & 5.46 & 0.82 & IRAS & Rice et al. (1988); Sanders et al. (2003) & average\\
 25 & 7.39 & 1.11 & IRAS & Rice et al. (1988); Sanders et al. (2003) & average\\
 60 & 63.8 & 9.6 & IRAS & Rice et al. (1988); Sanders et al. (2003) & average\\
 100 & 185 & 28 & IRAS & Rice et al. (1988); Sanders et al. (2003) & average\\
 170 & 193 & 18 & ISO & Popescu et al. (2004) & - \\
 200 & 165 & 17 & ISO & Popescu et al. (2004) & - \\
 450 & 40.95 & 8.0 & SCUBA & Israel et al. (1999) & extrapolated \\
 850 & 4.85 & 0.6 & SCUBA & Israel et al. (1999) & extrapolated\\
1200 & 0.92 & 0.13 & IRAM & Guelin et al. (1993) & extrapolated\\
\hline
\end{tabular}
\end{table*}

The best fit SED, obtained by spatially integrating
the solution described above for the diffuse dust emission component
and adding it to the solution for the dust emission from the
localised component from the star-formation regions,
is shown in Fig.~\ref{fig:sed_ngc891}, together with the 
observed FIR/submm data
that were used to constrain the model solution. Details on the observed flux
densities used in the plot are given in Table~\ref{tab:fluxes}. The best fit 
solution
corresponds to $\tau^{f}_{B}=3.5$, $SFR=2.88$\,M$_{\odot}$/yr and  $F=0.41$. 
Compared with the solution obtained in Paper~I ($\tau^{f}_{B}=4.1$,
  $SFR=3.8$\,M$_{\odot}$/yr and  $F=0.22$) there are changes
  in all three free parameters. The decrease in 
  $\tau^{f}_{B}$ is solely attributed to the fact that we used improved
  observational data to fit the submm points. If we had used the original
  data the solution for opacity would be unchanged. The decrease in $SFR$
  and the increase in $F$ is essentially due to the combination of adding the 
  contribution of the young stellar population in the optical,
  and of including the wavelength
  dependence of the fraction of photons escaping the star-forming regions. 

With the red line we show the model fit for the total dust emission of
NGC~891. 
We also show the main components of the dust emission, the diffuse
component (dark-green line) and the localised emission
from the clumpy component 
(light-green line). One 
can see that overall most of the dust
emission is powered by the diffuse component. Our solution gives for the
total dust luminosity $L_{dust}^{total}=9.94 \times 10^{36}$\,W, of which
$L_{dust}^{diff}=6.89 \times 10^{36}$\,W is emitted in the diffuse medium
($69\%$). From the figure it is apparent that the diffuse component dominates
the emission longwards of 60\,${\mu}$m and shortwards of $20\,{\mu}$m. Thus,
most of the emission in the FIR and in the NIR/MIR (PAH region) is
diffuse. It is only at intermediate wavelengths ($20-60\,\mu$m) where the
localised dust emission within the star-forming complexes dominates. In the
IRAS $25\,{\mu}$m and Spitzer $24\,{\mu}$m bands almost all the emission 
is predicted to come from the star-forming complexes, suggesting 
 the efficacy of this band as a direct tracer of SFR. Indeed empirically
 studies have shown that of the Spitzer bands it is the 24\,${\mu}$m band which
 is more closely related to star-formation (e.g. Calzetti et al. 2010).

In Fig.~\ref{fig:sed_ngc891} we also show the corresponding intrinsic stellar 
SED (as would be seen in the absence of dust) of NGC~891, together with the 
stellar emissivity components. In all cases the symbols indicate the 
wavelengths at which the radiative transfer calculations were performed. The 
black line
represents the total stellar emission produce in the galaxy. In the UV range
this is given by the population synthesis models (plotted as the blue line), as
explained in Sect.~\ref{subsubsec:templateyoung}. A fraction of this emission 
is locally absorbed in the
star-forming complexes, while the remaining $1-F$ escapes in the diffuse young
stellar disk (the mauve line). Thus, the difference between the black line and 
the mauve line in the UV represents the localised (and wavelength dependent) 
absorption of stellar light (see Eq.~\ref{eq:localised}). 
In the optical range the total emission is given
as a sum of the emission from the young stellar disk (mauve line plots only
the diffuse emission from the young stellar disk, which is slightly different 
from the total emission of the young stellar disk, due to the local 
absorption), the old stellar disk (light blue line) and the bulge (pink line).

One important result of such calculations  is to determine the 
fractional contribution of the different stellar components to the dust 
heating. For the case of NGC~891 we derive the following fractions: 
$11\%$ for bulge, $20\%$ for the old stellar disk, $38\%$ for the young stellar
disk and $31\%$ for the star forming complexes.  This means that the young
stellar populations are responsible for $69\%$ of the dust heating while the 
old stellar populations account for the remaining $31\%$.

\section{Fidelity of the model}
\label{sec:assumptions}

In order to fit observed dust/PAH emission SEDs of real galaxies like NGC~891
it was necessary to make some basic assumptions and approximations, which,
however, have implications not only for the SED of the integrated re-radiated 
light, but also for the geometrical characteristics of this light and of the 
direct UV/optical stellar light. The
main assumptions are the existence of a diffuse dust component associated with 
the young stellar population, the approximation of this dust in the form of an
exponential disk and the utilisation of a fixed spectral emissivity law for
the young and old stellar populations. Here we check whether our model is
consistent with the available observational constraints beyond those of
integrated dust emission SEDs and evaluate the limitations imposed by the
approximations. 

\subsection{The existence of a diffuse dust component associated with the young
stellar population}
\label{subsec:second-dustdisk}

A fundamental aspect of our model was the inclusion of a second disk of dust
associated with the young stellar population, which was taken to mimic the
diffuse dust that is known to exist in spiral arms
through direct observations in the FIR,  both in the Milky Way and in nearby 
well resolved galaxies. The
second dust disk was originally introduced in Paper~I to provide the observed 
level of submm emission in the spatially integrated SED. As discussed in
Paper~I and in
Popescu \& Tuffs (2005), the additional quantity of dust needed to fit the
observed submm data cannot be provided by clumpy optically thick dust, whether
such dust is in
star formation regions or in passive quiescent clumps. This conclusion has
been reinforced by the utilisation of the improved
dust emission templates for star formation regions introduced in
Sect.~\ref{subsec:infraredclumpy}, since these are now empirically constrained 
by submm and near-mm data.

More fundamentally, although one is in principle free to
add dust to the model in any form one likes to reproduce the
observed level of emission deep in the submm, if this dust is
self-shielded, it will in practice struggle to supply the necessary luminosity
to fit the FIR flux density peak of spiral galaxies at
around $160\,{\mu}$m. To peak at around
160\,micron the dust grains must be heated by radiation fields at around 1
Habing, which is indeed the illumination of the diffuse dust. Heating
at around 100 Habing (as at the PDR surface in embedded star-formation
regions) would provide the luminosity but with too blue dust FIR colours,
while the dust emission from the self-shielded dust
would be too red as well as not providing the luminosity\footnote{A possible
way out of this conundrum might be to invoke heating of the self-shielded
grains by the absorption of
secondary NIR/MIR photons emitted at the PDR surface for the star-formation
regions.
However, even if this process could
provide enough luminosity to account for the  $160\,{\mu}$m peak
in the integrated SED of galaxies, it could not simultaneously boost
the MIR PAH emission, which require UV or blue optical photons for
excitation. By contrast the integrated MIR PAH emission
is known to be statistically related to the
$160\,{\mu}$m emission from spiral galaxies (eg Bendo et al. 2008),
typically accounting for $\sim\,15\%$ of the total re-radiated
starlight.}. Of course, one solution would be to invoke
star-forming complexes extending to
radii of around 10 times their actual sizes into the diffuse component,
so that the dust illumination is reduced to the required levels. However
this is akin to spreading the dust around the spiral arms, which is
exactly what our second dust disk solution tries to mimic, so the
difference to a diffuse dust component then becomes semantic. 

In general, the fit to data provided by our two-dust disk 
model is one of relatively
large opacity for the central regions of local universe spiral galaxies
(${\tau}_B^f=3.5-4$; Paper~I; Driver et al. 2007; see also
Sect.~\ref{subsec:sed}), with most of the opacity provided by the second 
dust
disk. This is in contrast to the solution of low opacity
(${\tau}_B^f=1$) and one disk of dust (obtained by Xilouris et al. 1997
and Bianchi 2007 from modelling the optical data only) which
fails by a factor of $\sim 3$ to reproduce the submm emission (Popescu et
al. 2000; see also Baes et al. 2010).


Apart from the need to fit the submm data, and the physical considerations 
already outlined in Sect.~\ref{subsec:distribution} linking the second dust 
disk with the corresponding
star-forming disk of gas, there is further empirical evidence for
the existence of a second dust disk, as derived from comparison of the model
predictions with other data, as follows:\\

\noindent
{\it i) Comparison of model predictions for the stellar emissivity in the
  optical with population synthesis models}

\noindent
We checked whether the predicted intrinsic SED of
the old stellar populations in NGC~891, as derived from optimisation of the 
optical data
only (Xilouris et al. 1999), together with the corresponding SFR needed to
produce the total luminosity emitted by the dust, could be fitted by the
population synthesis models of Kotulla et al. (2009). In 
Fig.~\ref{fig:sed_ngc891} one can see that the 
predictions for 
the optical emission from the old stellar population (light blue line) in the 
B, V, and I bands fall severely below the predictions from the population 
synthesis model (dark blue line). In this plot we already included the second
dust disk. In its absence the SFR needed to reproduce the energy emitted in the
infrared would be even greater, so the discrepancy would increase. It is clear 
from here that there is a need for extra 
stellar luminosity presumably hidden by extra dust
(the second dust-disk in our formulation). In Popescu et al. (2000) we 
assumed that the young
stellar population was only emitting in the UV, while in the optical we only
had the contribution from the old stellar populations, as derived by Xilouris
et al. (1999). This assumption is not supported by the predictions of the
population synthesis models, and 
indicates that the optimisation of the optical images can only reveal 
information about the old stellar populations and associated dust, but 
completely hides the information about the
young stellar populations and associated dust.\\

\noindent
{\it ii) Comparison of model prediction for the spatial distribution of
  PAH emission with observations}

\begin{figure}
\centering
\includegraphics[width=9.0cm]{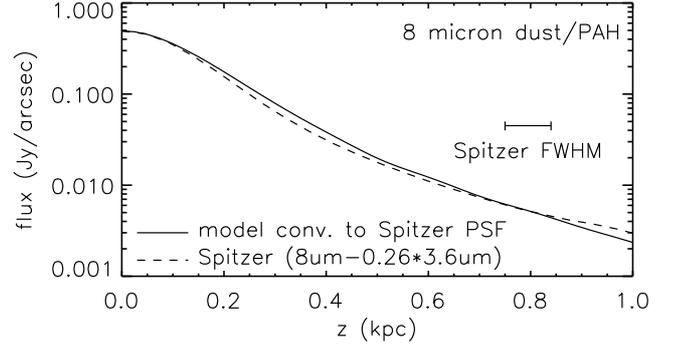}
\caption{Comparison between the $8\,{\mu}$m Spitzer vertical profile of NGC~891
integrated over longitude (dashed line) and our corresponding model 
predictions (solid line). Before comparison with the model, the
$8\,{\mu}$m image had the stellar component subtracted off.}
\label{fig:zprofilespitzer76}
\end{figure}

\begin{figure*}
\centering
\includegraphics{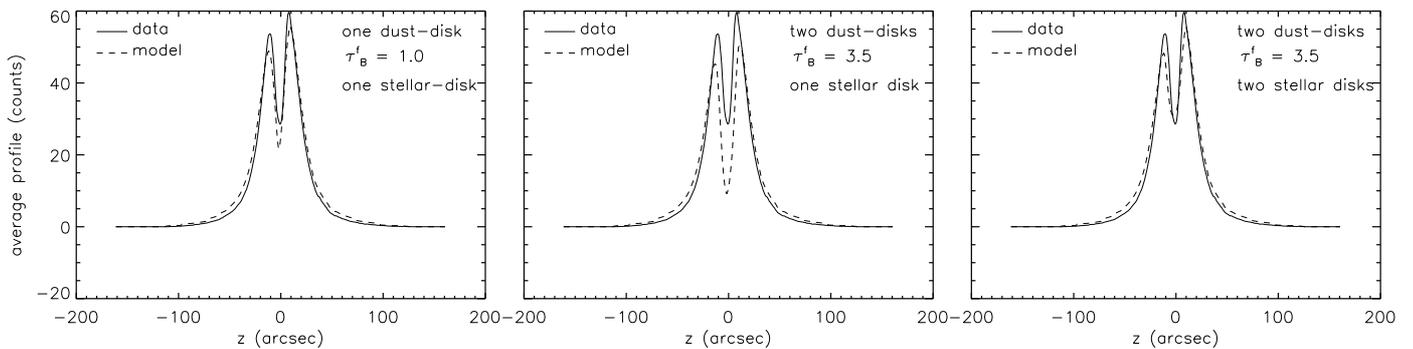}
\caption{A comparison between the observed averaged vertical profile of NGC~891
in the I band (solid line) and the corresponding model predictions 
 (dashed line) for three cases: the one dust disk model (left panel) with
${\tau}_B^f = 1.0$ and one stellar disk (the old stellar disk); the two 
dust-disk model (middle panel) with ${\tau}_B^f = 3.5$ 
and one stellar disk (the old stellar disk); the 
two dust-disk model (right panel) with ${\tau}_B^f = 3.5$ and two stellar disks
(the old and the young stellar disk).}
\label{fig:profilei}
\end{figure*}

\noindent
Since our model with the second disk of dust and stellar emissivity was
only fitted to the spatially integrated dust emission, it is possible to use
the spatial information provided by the observations to check the predictions
of the model. We have already used the predictions from Popescu et
al. (2000) to compare the predictions of the model for the spatial
distribution of infrared emission with the ISOPHOT data of NGC~891 at 170 and 
$200\,{\mu}$m 
(Popescu et al. 2004). The comparison was done for the radial profiles, as the
observed ISOPHOT images were only resolved in radial direction. The model
predictions were found to be in excellent agreement with the observations. 
More recently Spitzer images of NGC~891 became available, which have a linear
resolution of 100\,pc at $8\,{\mu}$m, comparable to the thickness of the 
second dust disk. Since
we have now included the PAH features in the model, which dominate the 
$8\,{\mu}$m emission, these data can be used to test the
predictions for the vertical distribution of PAH emission in NGC~891. 
For this comparison we first subtracted the component of direct stellar light
at $8\,{\mu}$m (estimated by multiplying a Spitzer $3.6\,{\mu}$m image by a
factor of 0.26; Helou et al. 2004). Fig.~\ref{fig:zprofilespitzer76} shows
again an excellent agreement.
Thus the model containing a second
dust disk can predict both the overall level of emission and the shape of the 
predicted vertical profile.\\




\noindent
{\it iii) Comparison of model predictions for the attenuation-inclination
  relation with observations}

\noindent
A strong independent evidence for the existence of a second dust disk came from
the attenuation-inclination relation. It was shown in Popescu \& Tuffs (2009) 
that while a two dust-disk model with higher central face-on opacity can
reproduce the observed attenuation-inclination relation, a single dust
disk model with ${\tau}_B^f = 1.0$ completely fails to reproduce the observed
data. The attenuation-inclination relation is an especially sensitive test, as
the rise in attenuation with inclination will very strongly depend on the 
relative scaleheights of the assumed dust layers and stellar populations. In 
particular it is an independent test for the existence of the second component 
of dust represented by the second dust disk. We should also mention here that 
one of the strength of this test is that it is completely independent of the 
assumed dust emission properties. Thus, this would also seem to rule out a 
one-dust disk 
model with low opacity but with dust grains having modified optical properties 
in the submm (e.g. enhanced submm emissivity), as proposed by Alton et al. 
(2004) and Dasyra et al. (2005).\\

\noindent
{\it iv) Comparison of model predictions for the vertical profiles in the 
optical and NIR in NGC~891 with observations}  

\noindent
The solution with a second dust-disk and extra luminosity coming from the
young stellar population emitting in the optical bands was further used to
predict the appearance of the galaxy in the optical bands. In particular the
average vertical profiles have been used to test whether the dip produced in the
plane of the galaxy by the dust layer has the right deepness. 
Fig.~\ref{fig:profilei} shows a
comparison between the I band averaged vertical profile obtained from the
observed images of NGC~891 (Xilouris et al. 1998) and the corresponding model
predictions for three cases. The first case (left panel) is the original 
solution obtained by Xilouris et al. (1999), which 
only includes the old stellar disk and associated dust. The second case 
(middle panel) is our two dust-disk model, but without the inclusion of the
young stellar disk. Finally, the right panel shows again the predictions for 
our two-dust disk model, but with the inclusion of extra luminosity 
coming from the young stellar disk. One can immediately see that the inclusion
of a second dust disk which is not accompanied by a stellar luminosity
component would produce a stronger dust lane (a larger depth in the vertical
profile) than shown in the observed images. This is a clear indication that 
there
is a need for extra luminosity, as also indicated by population synthesis
models (see i) above). In fact the two dust disk model with two 
(old and young) stellar
components is able to reproduce the observed data better than the one disk
model, especially in the B band. 

In the K band the model with the two dust
disks and two stellar components predicts a somewhat more prominent
dust lane than observed. This would indicate either a
shorter scalelength for the second dust disk (though this would be difficult to
reconcile with the excellent fit we found in Paper~I for the radial profile at 
850\,${\mu}$m), or more luminosity in
the young stellar disk than predicted by the population synthesis model. It 
was already noted by Dasyra et al. (2005), Bianchi (2007) and
Bianchi (2008) that the second dust disk shows its maximum effect in the K 
band, since it is only there that the first disk of dust becomes transparent 
and is therefore not shielding the second disk of dust. A possible alternative
reason for the difficulty in fitting the vertical profile in K-band
is that, due to its very small scale height, 
the appearance of the second dust disk at these long wavelengths would be 
easily blurred in real galaxies if perturbations from co-planarity occur, even
if any such perturbations had relatively small amplitudes. Although the 
scaleheight of the molecular layer in Milky
Way is around 90\,pc, as adopted in our model for the second dust layer, it is
well known that CO and other tracers of star-formation exhibit systematic
vertical displacements from the mean place known as ``corrugations'' (Spicker
\& Feitzinger 1986 and references therein). The discovery of Matthews \& Uson 
(2008) of a non-planar disk in star-formation tracers in an isolated galaxy 
other than the Milky Way, is a strong evidence that non-planarity is a 
rather frequent phenomena. Matthews \& Uson found that undulated
patterns  with amplitude of $\sim 250$ pc are visible in particular in the
distribution of the young stellar population and the dust, suggesting that
the process leading to the vertical displacements may be linked with the
regulation of star formation in galaxies. The effect of corrugations will
mean that the edge-on thin disk will not be seen as
such in the optical, but only the distribution corresponding to the amplitude 
of the corrugations. This would be more than enough to blur and hide 
any dust layer in the NIR. Nevertheless, the young stars will be still perfectly
correlated with the thin layer of dust and gas with a (local) scaleheight of
90\,pc, since the effect of corrugations is purely gravitational. So the basic
vertical geometrical coupling between stars and dust which gives rise to the
dust emission from the thin disk, as calculated in this model, still applies. 
To conclude, if corrugations occur, the second dust-disk will tend to be
blurred, making dust lanes in edge-on galaxies in the NIR less prominent. 
 
\subsection{Approximating the spiral arm component with an exponential
  disk}
\label{subsec:spiral}

Even if it is accepted that there is a diffuse dust layer
associated with the young stellar population, we must still evaluate the
effect of artificially distributing this stellar emissivity and dust opacity
in an exponential disk instead of a spiral arm pattern. It is clear that this
will completely prohibit a comparison of surface brightness distributions in
face-on systems, but here we are only concerned with the spatially integrated
dust emission SEDs and the effects of this approximation on these.

To do this we ran a simulation
with the parameters corresponding to the best fit solution of NGC~891 (see
Sect.~\ref{subsec:sed}), i.e. we
kept the same luminosity for the young stellar population and the same amount
of dust, but we redistributed the corresponding stellar emissivities and dust
opacities in a spiral pattern. We modeled the spiral pattern using 3 circular
arms, with a radial distribution as given in Fig.~\ref{fig:arm}. The vertical 
distribution is the same as for the case of the thin exponential disk. This 
means that for
any given line of sight through the spiral arm the opacity is higher than for
the case of a second dust disk, and for any given line of sight through the
interarm regions the opacity is lower than for the case of the second dust disk
- being just the opacity of the first dust disk. So we will have a solution 
with a
high contrast between arm and interarm regions, that can be still characterised
by the same central face-on opacity - which is the effective central face-on
opacity if all the dust were distributed in an exponential disk instead of in a
spiral arm. 

\begin{figure}
\centering
\includegraphics[width=9.3cm]{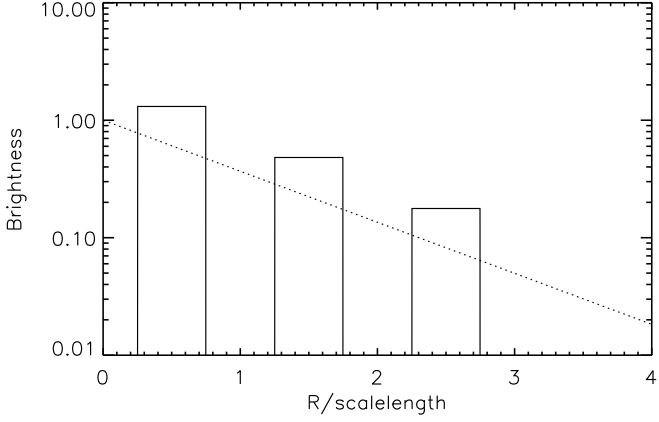}
\caption{The radial distribution used to test the effect of a spiral pattern 
for the young stars and associated diffuse dust (solid line), overplotted on 
  the distribution for the standard model with an exponential 
  disk (dotted line). The radial coordinate is given in units of exponential 
  scalelengths, where the scalelength of the exponential disk is 
  $h^{\rm tdisk}=5670\,$pc.}
\label{fig:arm}
\end{figure}

\begin{figure}
\centering
\includegraphics[width=9.3cm]{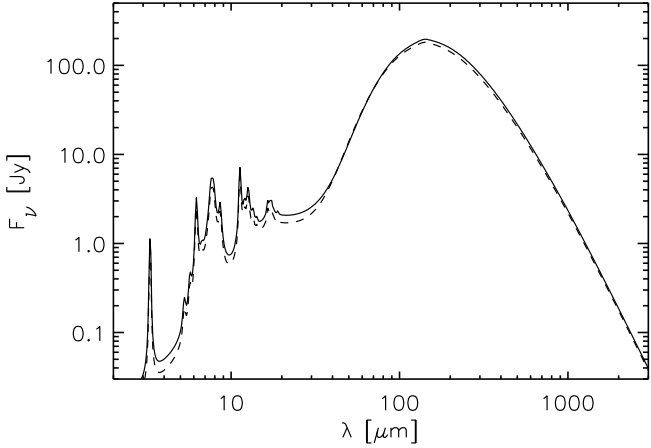}
\caption{Comparison between model solutions for the dust emission SEDs 
  calculated under the assumption that the emissivity of the young
  stellar population and the opacity of the associated dust are distributed in
  a thin disk (solid line) and in a spiral pattern
  (dashed line), respectively. The solution is for the best fit of NGC~891.}
\label{fig:compare_sed}
\end{figure}

After calculating the radiation fields and the infrared emissivities using 
the same procedure outlined in Sect~\ref{sec:calculation} and 
\ref{sec:illustration} we obtain a total 
integrated infrared SED 
that looks very similar to that obtained for the exponential disk case (see
Fig.~\ref{fig:compare_sed}). The integrated dust luminosity is only $5.5\%$
lower than for the standard model. Here we should mention that the 
attenuation-inclination curve (not plotted
in this paper) obtained for the solution with a spiral pattern is also almost 
identical to that obtained for the case of an exponential disk. The 
insensitivity of the shape of the attenuation-inclination curve to the 
inclusion of a spiral pattern was already demonstrated by Semionov et
al. (2006). In passing, we also note that Misiriotis et al. (2000) showed that 
the appearance of simulated dusty spiral galaxies seen edge-on, calculated 
using a spiral structure, does not differ from that calculated using a pure 
exponential disk.

The similarity in the solutions obtained for both the integrated infrared SEDs
and for the attenuation-inclination relation reassures us that the 
approximation of a second exponential
thin disk of stellar emissivity and dust opacity is an excellent one when
making predictions for the spatially integrated SEDs. This is in contrast to
the large change in the shape and amplitude of the infrared SEDs 
(as we will show in Sect.~\ref{sec:library}) 
and attenuations of stellar light (see Fig.~\ref{fig:compare_attenuation}) 
when changing the main free parameters of our model.

\subsection{The approximation of a fixed spectral shape for the SEDs of the 
old and young stellar populations}
\label{subsec:approximation}

\begin{figure}
\centering
\includegraphics[width=9.3cm]{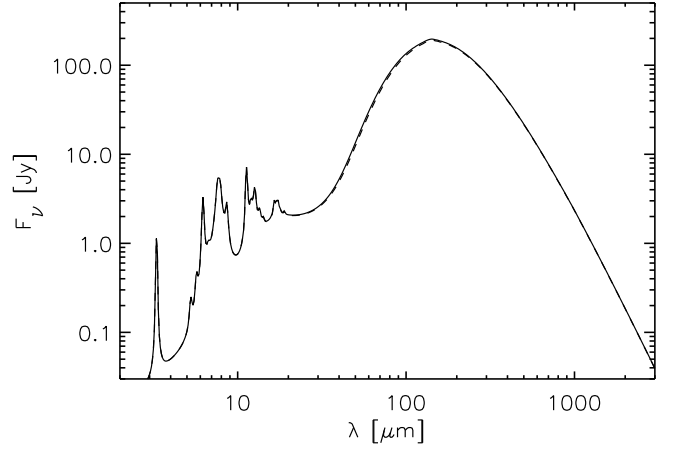}
\caption{Comparison between model solutions for dust emission SEDs calculated 
 using intrinsic SEDs of the old stellar population peaking in the J (solid
 line) and K band (dashed line),
 respectively. The solution is for the best fit of NGC~891.}
\label{fig:compare_sed_J_K}
\end{figure}

As discussed in Sect.~\ref{subsec:intrinsic}, the dust emission SEDs are
calculated by dividing the stellar emission SED into two fixed
spectral templates with differing spatial distributions -
one UV-dominated corresponding to the young stellar
population, weighted by the model parameter $SFR$,
and one optically dominated, corresponding to the
old population and weighted by the model parameter $old$.
This is a radically different approach to the
handling of stellar emissivity SEDs compared to previous
models for the panchromatic UV-submm emission
of galaxies. It enables us to obtain the same solution for the dust emission,
and therefore the reddening of any given galaxy, without the need to
input trial population synthesis solutions for the full UV/optical/NIR stellar
emission SED to the calculation of the dust emission.
This new approach requires however that the solution to the dust
emission is invarient to the assumed shape of the stellar emissivity
SED within each of the two templates. Here we test this assumption in turn
for the old and the young stellar emission templates.

For the old
stellar population the spectral shape was empirically derived from fitting the
optical images of NGC~891. The spectral shape was consistent with the SED
having a flux density per unit frequency peaking in the J band. Reasonable
variations from this shape are to consider spectral templates peaking towards
longer wavelengths, e.g. in the K band. We therefore consider a calculation in
which  the spectral shape was altered to allow for a brightening of the K band
luminosity by $58\%$ and a corresponding dimming of the J band 
luminosity by $48\%$, change that preserves the spectral integrated 
luminosity of the old stellar population. The result of this calculation from
Fig.~\ref{fig:compare_sed_J_K} shows that the predicted dust emission SEDs are
completely insensitive to (possible) changes in the spectral shape of the 
old stellar populations, as indicated from the overlap of the infrared SEDs. 

\begin{figure}
\centering
\includegraphics[width=9.3cm]{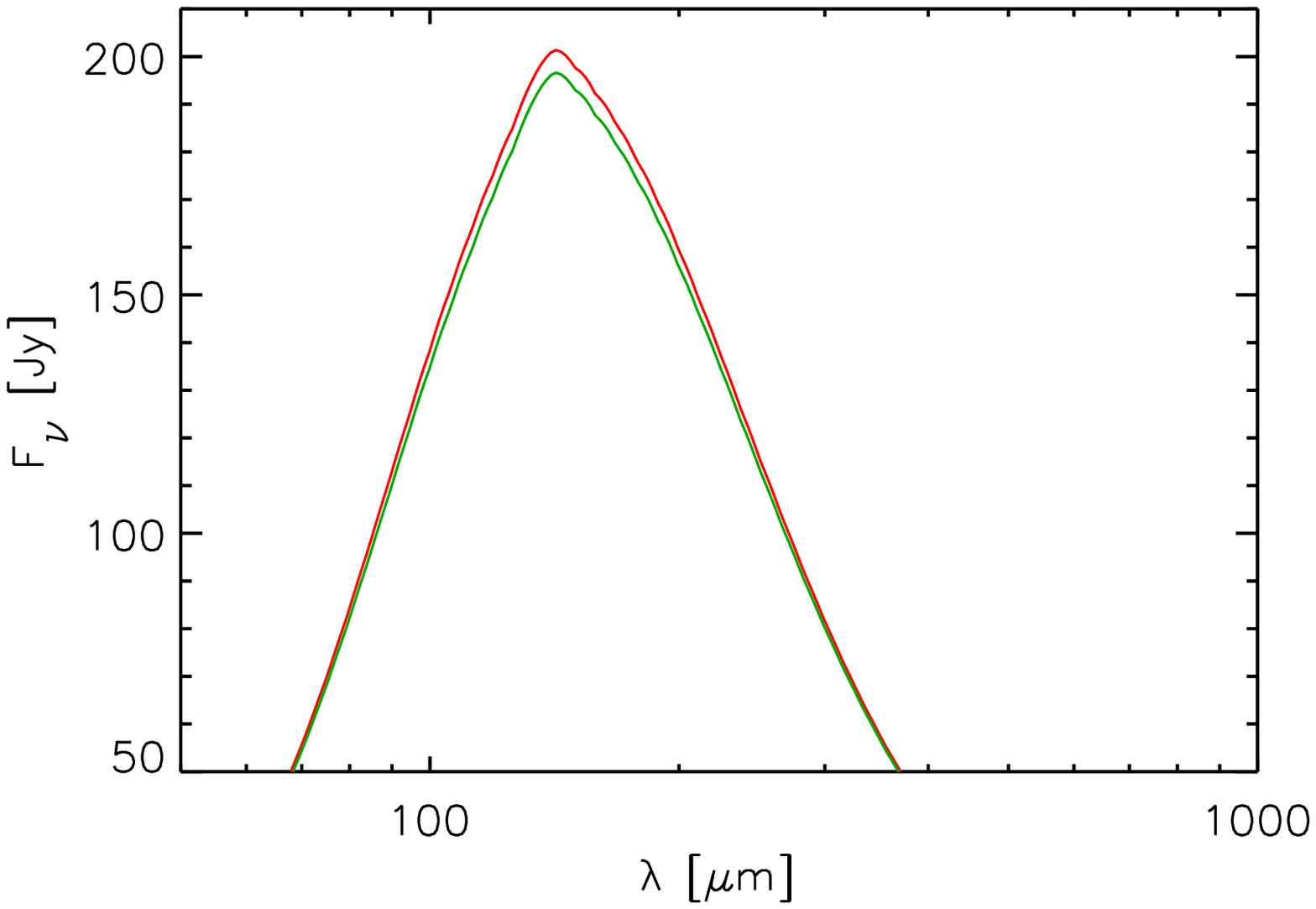}
\includegraphics[width=9.3cm]{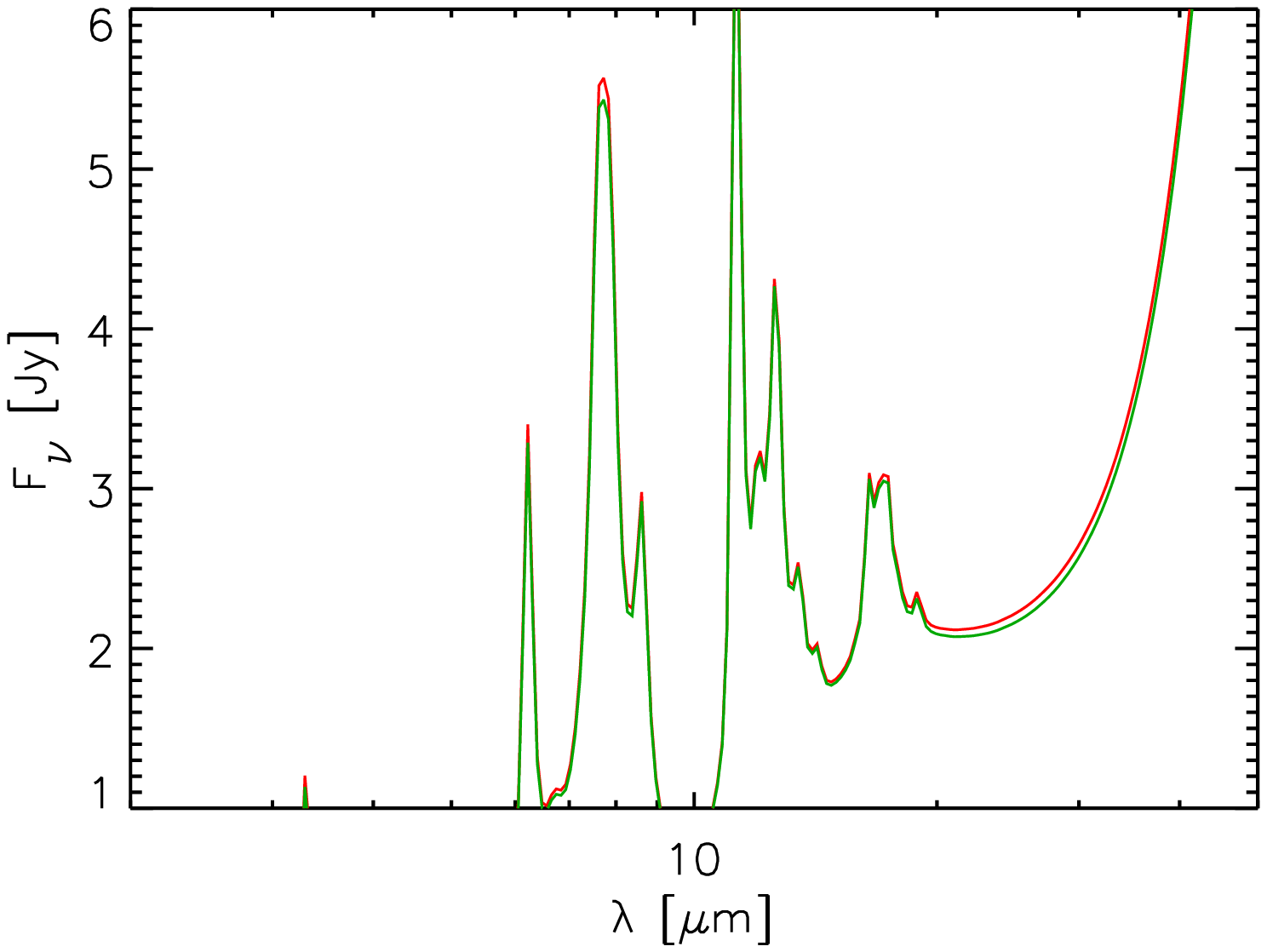}
\caption{Comparison between model solutions for dust emission SEDs calculated 
 using different spectral templates for the intrinsic SEDs of the young 
stellar population in the UV: our standard model 
(green line) and a model with bluer colours (red line),
 respectively. The solution is for the best fit of NGC~891. In both cases the
 the total luminosity of the young stellar population was kept fixed.}
\label{fig:compare_sed_young}
\end{figure}

\begin{figure}
\centering
\includegraphics[width=9.3cm]{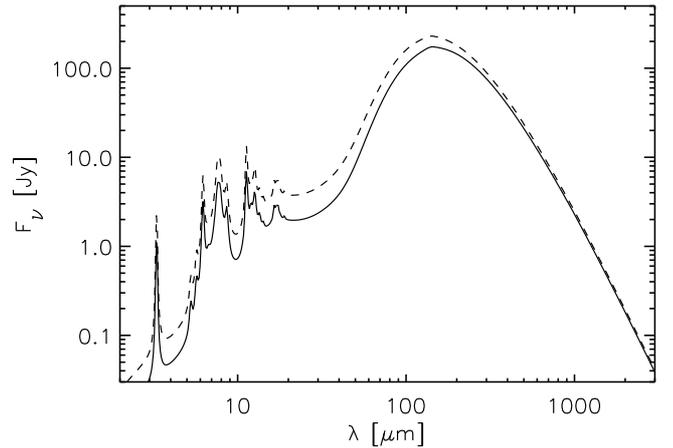}
\caption{Comparison between model solutions for diffuse dust emission SEDs 
  calculated
  using different contributions of the young and old stellar populations, but
  preserving the total stellar luminosity. The solid line is the solution
  corresponding to the best fit of NGC891, but without the bulge
  contribution. The dashed line corresponds to an increase in the luminosity of
the young stellar population in the diffuse stellar disk by a factor of 2 and a
decrease in the luminosity of the old disk stellar population by a factor of
1.68.}
\label{fig:compare_sed_young_old}
\end{figure}

For the young stellar population the spectral shape was fixed from a
combination of population synthesis models and empirical fitting of the optical
images. This resulted in a spectral template having a flux density per unit
frequency peaking in the I band. Reasonable variations from this shape are to
consider spectral templates peaking towards shorter wavelengths, e.g. in the V
band. We therefore consider a calculation in which the spectral shape was
altered to allow for a brightening of the V band luminosity by $34\%$ and a
dimming of the I band luminosity by $25\%$. As before this change was chosen
to preserve the spectral integrated luminosity of the young stellar
population. The resulting dust emission SED (not shown in this paper) 
is essentially indistinguishable from the one calculated using our standard
model. There is only an increase in the predicted dust luminosity by $0.2\%$,
which is to be expected due to the shift in the spectral peak of the young 
stellar population to shorter wavelengths, which resulted in a larger fraction 
of stellar photons being absorbed by dust. So changing the colours of the young
stellar population in the optical by as much as $59\%$ produces changes of 
less of a percent for the predicted dust emission luminosity. 

Finally, it is
important to check what the effect of changing the colours of the young
stellar population in the UV is, since there it is these photons that make a
significant contribution to the dust heating. We therefore made a more 
drastic change, by changing the slope of the UV SED from nearly flat (in 
$F_{\nu}$) to a
monotonically decreasing slope between the $1350\AA$ and $2800\AA$ spectral
sampling points. We therefore brightened the $1350\AA$ data point by $24\%$
and dimmed the $2800\AA$ flux by $38\%$, thus producing a very blue 
spectrum.
As before we kept the overall luminosity of the young stellar population
constant. The resulting dust emission SED is shown in the two panels of 
Fig.~\ref{fig:compare_sed_young}, plotted in a linear scale and for different 
cuts in luminosity to allow a better visualisation of the small differences 
in the SEDs. Again, since a bluer
stellar SED was considered, this resulted in more stellar photons being
absorbed by dust and a larger predicted dust luminosity, as seen
from the plots. The overall increase in the dust luminosity was $1.7\%$, which
is still a minor variation taking into account the dramatic change by $62\%$ 
in the
colours of the UV stellar SED. In fact this is almost surprising, but we should
keep in mind that the bluer UV photons have higher probability of absorption in
the star forming complexes, and therefore the escaping radiation illuminating
the diffuse component would still be strongly reddened due to the local
absorption.  

By contrast we also did a calculation in which we changed the relative 
contribution of the young to old stellar populations and kept fixed the total 
stellar luminosity. We have used the solution for NGC~891 for the diffuse 
component, set the bulge-to-disk ratio to 0 and used this as a reference SED. 
We then increased the luminosity of the young stellar population by a factor 
of 2 and decreased the luminosity of the old stellar populations in the disk 
by a factor of 1.68, which preserves the total stellar luminosity. The
resulting SEDs are shown in Fig.~\ref{fig:compare_sed_young_old}. This time
there is a significant change in the SEDs, followed by an increase in the
total dust luminosity by $33\%$ and an increase in the MIR emission by a
factor of $\sim 2$.
This shows
again that the main factors influencing the dust emission SEDs are the overall
luminosities of the old and young stellar populations, and not (possible) 
variations in the spectral shape of the template stellar SEDs.
 
\subsection{The relative opacity of the first and second dust disk}

A further approximation of the model is the fixed ratio between the
  opacity of the first and second dust disks. While  most of the fixed
  parameters have been calibrated to empirical relations, we acknowledge that
  this ratio has been only calibrated to the value of NGC~891 (close to 
NGC~5907). The validity of this assumption has been succesfully tested in a 
statistical sense on the attenuation-inclination relation in Driver et al. 
(2007). However
 this assumption still needs to be proven on an object-by-object basis by 
fitting the panchromatic SEDs of galaxies. As we will show in 
Appendix~\ref{sec:howto} we have
 identified a parametric test to potentially flag out galaxies that may not
 follow the colour-luminosity relation predicted by our model due to other
 geometrical characteristics.

\section{The library of SEDs for the diffuse component}
\label{sec:library}

We have created a library of diffuse SEDs that spans the parameter
space of 4 parameters: $\tau^f_B$, $SFR'$, $old$, and $B/D$\footnote{We
  note that the fixed (and calibrated) parameters of the model are given in the
  Appendix~\ref{sec:tables}.} With the exception
of $SFR'$, these are the main parameters of the model. $SFR'$ is defined in
terms of the primary parameters $SFR$ and $F$:
\begin{eqnarray}\label{eq:sffectivesfr}
SFR' = SFR\times (1-F)
\end{eqnarray}
which is equivalent with the $SFR$ for the case that $F=0$. This definition is
possible because in our model we made the assumption that 
the wavelength dependence of the fraction of escape photons from the clumpy 
component into the diffuse one is fixed, allowing us to separately calculate 
the SEDs for the diffuse component and for the clumpy component. In this
formulation $SFR'$ is an effective star formation rate powering the diffuse
dust emission.

In Table~\ref{tab:param} we give the parameter values at which the model for the diffuse component has been
sampled. In total we have 7 (for $\tau^f_B$) x 9 (for $SFR'$) x 9 (for $old$) x
5 (for  $B/D$) = 2835 combinations, corresponding to 2835 simulated SEDs. 
All
the model SEDs are available in electronic form. 
The choice of the parameter values was done such that it covers
  the parameter space of local universe galaxies but also the asymptotic
  values of these parameters. Thus for $\tau^f_B$ we considered the range 
[0.1,8.], which means galaxies with almost no dust at all to galaxies having
over twice as much opacity as the average value found from the statistical 
analysis
of the Millennium Galaxy Catalogue (Driver et al. 2007). For $SFR'$ we
considered the range [0.,20.]\,$M_{\odot}/yr$, which means galaxies with no
recent star-formation activity  (no heating from the young stellar population)
to galaxies having very high star-formation rate (typical of starburst
galaxies which should be outside the range of spiral galaxies addressed by this
paper). For the old stellar population the parameter $old$ was scanned
analogous to $SFR$. Finally, we considered galaxies spanning the whole
Hubble sequence, from bulgeless galaxies ($B/D=0.0$) to bulge dominated 
galaxies ($B/D=2.0$). This
choice of parameters allows a smooth overlap with the parameter space of
starburst galaxies, very quiescent galaxies, early type galaxies, and higher
redshift galaxies. It also allows rare (unexpected) cases to be considered. As 
we
will show in Sect.~\ref{subsec:sedcolour}, the locus of the FIR colours
corresponding to the parameter space defined by the values in 
Table~\ref{tab:param} overlaps quite well with the locus of observed FIR
colours of real life spiral galaxies.

\begin{table}
\caption{The parameter space sampled by our library of dust and PAH emission
  SEDs (for the diffuse component only).} 
\label{tab:param}
\begin{tabular}{rrrrr}
\hline
& $\tau^f_B$ & $SFR'$ & $old$ & $B/D$ \\
&            & $M_{\odot}/yr$ & & \\
\hline\hline
1 & 0.1 &  0.0 &  0.0 & 0.00\\
2 & 0.3 &  0.1 &  0.1 & 0.25 \\
3 & 0.5 &  0.2 &  0.2 & 0.50 \\
4 & 1.0 &  0.5 &  0.5 & 1.00 \\
5 & 2.0 &  1.0 &  1.0 & 2.00 \\
6 & 4.0 &  2.0 &  2.0 & \\
7 & 8.0 &  5.0 &  5.0 & \\
8 &     & 10.0 & 10.0 & \\
9 &     & 20.0 & 20.0 & \\
\hline
\end{tabular}
\end{table}

\begin{figure*}
\centering
\includegraphics{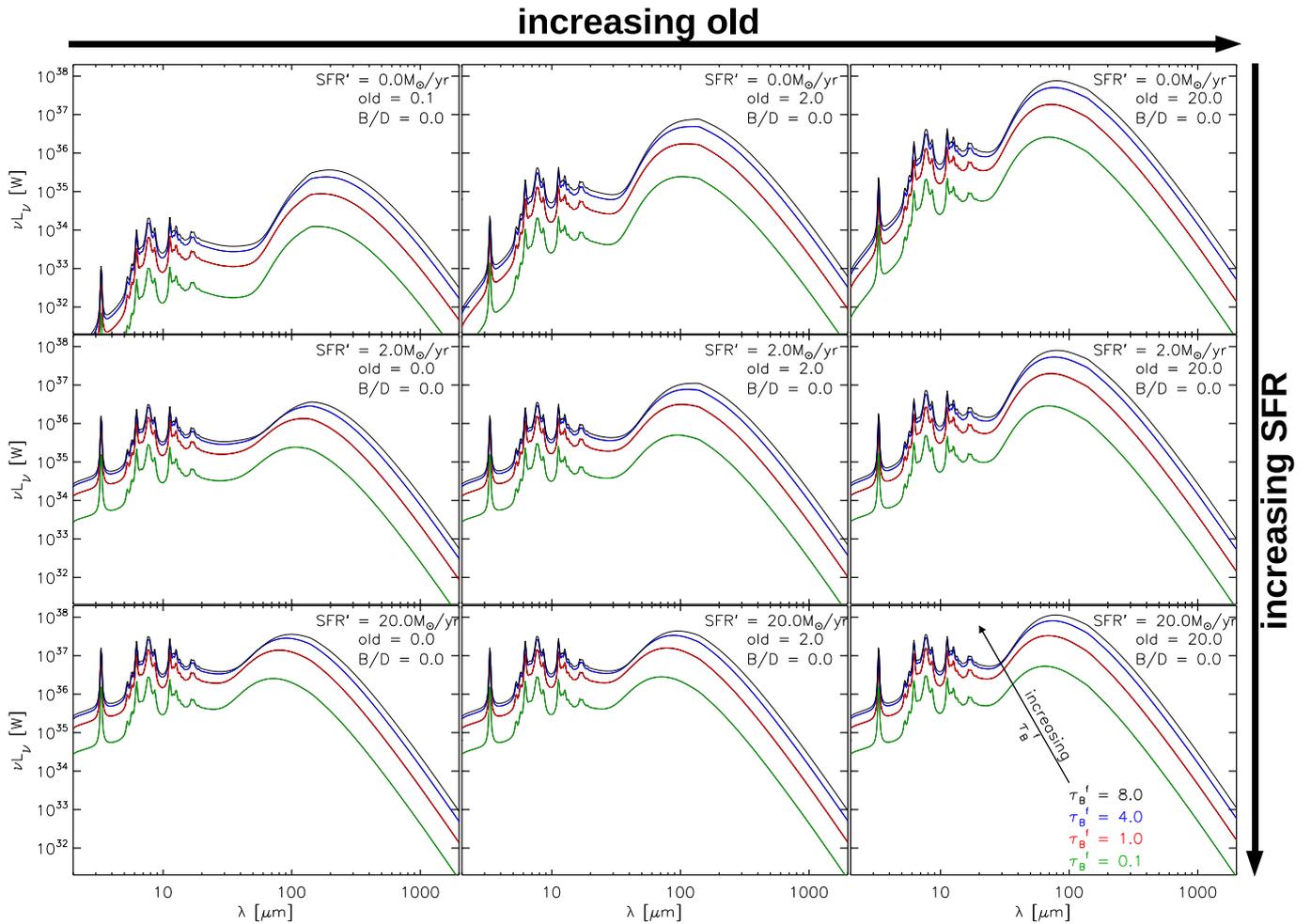}
\caption{Integrated dust and PAH emission SEDs for the diffuse component, for 
  model galaxies with
  different face-on opacities (plotted as different curves in each panel). All
  models shown in this figure are for pure disk galaxies ($B/D=0.0)$. From
  left to right the panels show model galaxies with various levels of
  contribution from the old stellar populations (${\rm old}=0.1$ (0.0), 
${\rm old}=2.0$ and  ${\rm old}=20.0$). From top to bottom the panels show 
models with various levels of SFR (${\rm SFR'}=0.0, 2.0,
20.0\,M_{\odot}$/yr). The colour coding is as follows: green is for 
  $\tau^{f}_{B}=0.1$, red is for $\tau^{f}_{B}=1.0$, blue is for 
  $\tau^{f}_{B}=4.0$ and black is for $\tau^{f}_{B}=8.0$.}
\label{fig:sed_tau}
\end{figure*}

To compute the library of diffuse SEDs we first calculated a library
of radiation fields, computed for the main stellar components of our model:
young stellar disk, old stellar disk and bulge, for the 7 values of opacity
used in Table~\ref{tab:param} and for the 16 UV-optical wavelengths detailed in
Sect.~\ref{subsec:intrinsic}. For the bulge and the old stellar disk only 6 
optical wavelengths were
used, as in our model we assume that these stellar components have negligible
emission in the UV range. In total we created a library of 196 data cubes
of radiation fields, sampled at 22 radial positions and 12 vertical positions
(264 spatial points within the model galaxy).  We then created the library of
temperature distributions, for the 4 grain compositions used in our model
(silicate, graphite, PAH$^0$, PAH$^{+}$), the 7 values of opacity, 9 values of
$SFR'$, 9 values of $old$ and 5 values of $B/D$. In total we computed a library
of 11340 data cubes of temperature distributions, each sampled at 264 spatial
points within the model galaxy and for all grain sizes contained in the dust
model. For each of the temperature distribution data cubes we created
corresponding cubes of infrared emissivity. In total a library of 11340
files of infrared emissivities were calculated.

When used to fit observed panchromatic SEDs, the library of simulated dust and 
PAH emission SEDs should be used in conjunction with the library of simulated
attenuations of stellar light recalculated in this paper, taking into account 
the $B/D$ ratio of the galaxy and its inclination, as described in 
Appendix~\ref{sec:howto}.

\section{Predicted variation of the dust emission
SEDs with the main parameters of the model.}
\label{sec:predictions}
 
\subsection{Variation of the SEDs with ${\tau}^f_B$}
\label{subsec:sedtau}

\noindent
{\it i) Amplitude of the SED}

As expected, the amplitude of the diffuse
dust and PAH emission SEDs increases with
increasing optical depth for the optical thin cases and tends to a saturation
value for the optically thick cases. This is seen in Fig.~\ref{fig:sed_tau} 
from the bunching 
of the blue and black curves on one hand (${\tau}^f_B=4,8$) and from the big gap
  between the green and red curves on the other hand (${\tau}^f_B=0.1,1$). An 
interesting
    feature of the SEDs is the fact that the ratio between the FIR and MIR
    (PAH) amplitudes increases with increasing opacity for the models where the
    stellar luminosity has a higher contribution from the old stellar
    population with respect to the young stellar population (but where the
    young stellar population has still a non-negligible contribution). This can
    be seen for example on the right bottom panel of Fig.~\ref{fig:sed_tau}, 
    where the green
    curve (${\tau}^f_B=0.1$) shows almost identical levels in the FIR peak 
    and in the PAH features, while the black curve  (${\tau}^f_B=8$) shows two
      order of magnitude difference in the FIR to MIR levels. This change in
      the FIR to MIR colour is due to the fact that with increasing optical
      depth the disk becomes first optically thick to the UV radiation
      (provided by the young stellar population),
      while still being relatively transparent to the optical photons (mainly
      provided by the old stellar population). This
      will have the consequence that the PAH emission (and small grain
      emission), which is mainly heated by the UV photons will tend to a
      saturation level, as expected for the optically thick case, while the FIR
      emission will be still boosted by the optical photons, which have not
      reached the optically thick limit yet. Apart from this effect, which is 
      simply
      related to the optical properties of the grains, there is an additional
      geometrical effect which will boost the FIR-to-MIR colour with increasing
      opacity. This is due to the fact that the old stellar populations have 
      larger
      scale-heights than the young stellar populations, which will mean that a
      larger proportion of the optical photons will be less confined to the
      regions of higher optical depth, and will therefore provide extra 
      heating in the optically thin regions of the galaxy.\\

\noindent
{\it ii) Peak of the SED}

Another interesting feature of the plots is the fact that the peak of the
SEDs from diffuse dust
shifts towards longer wavelengths with increasing opacity for the optically 
thick regime, for the models where
the stellar luminosity has a higher contribution from the young stellar
populations with respect to the old stellar population. This can be seen for
example in the bottom row of Fig.~\ref{fig:sed_tau}, where the trend of 
shifting the peak of
the SEDs becomes increasingly less strong in moving from the left to the right
panel, where the solution changes from a young to an old stellar population
dominance of the stellar SED. As before, when moving to a solution where more
optical photons are available, these will provide extra heating to the dust,
due to their more optically thin regime, both because of their
optical properties and because of their spatial distribution (as discussed
before), which in turn will cancel the shift towards cooler SEDs.

\begin{figure*}
\centering
\includegraphics{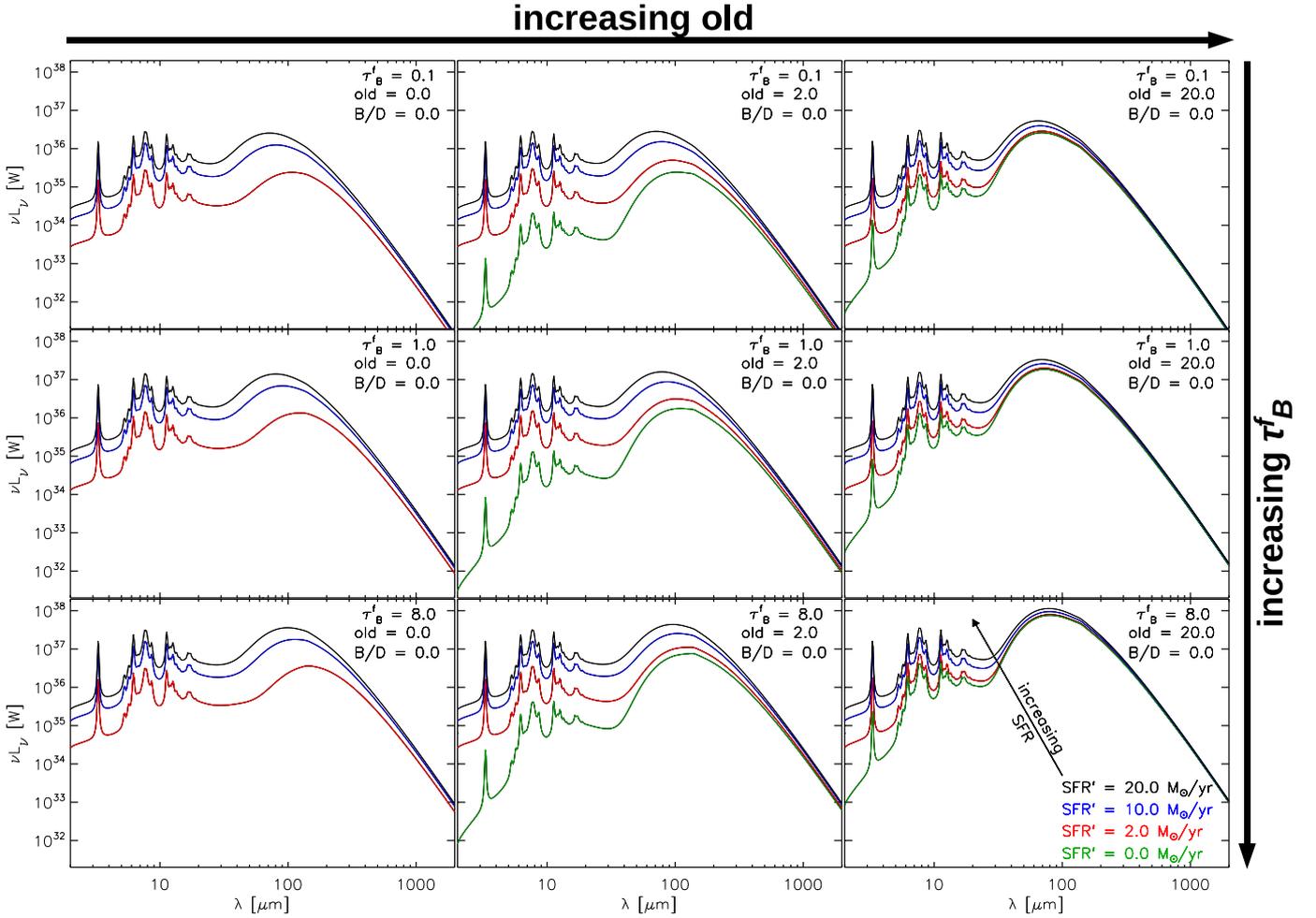}
\caption{Integrated dust and PAH emission SEDs for the diffuse component, for model galaxies with
  different levels of SFR (plotted as different curves in each panel). All
  models shown in this figure are for pure disk galaxies ($B/D=0.0)$. From
  left to right the panels show model galaxies with various levels of
  contribution from the old stellar populations (${\rm old}=0.1$, 
${\rm old}=2.0$ and  ${\rm old}=20.0$). From top to bottom the panels show 
models with various face-on B band opacities ($\tau^{f}_{B}=0.1$,
$\tau^{f}_{B}=1.0$, and $\tau^{f}_{B}=8.0$). The colour coding is as follows: 
  green is for 
  ${\rm SFR'}=0.0\,M_{\odot}$/yr, red is for ${\rm SFR'}=2.0\,M_{\odot}$/yr, blue is for 
  ${\rm SFR'}=10.0\,M_{\odot}$/yr and black is for ${\rm SFR'}=20.0\,M_{\odot}$/yr.}
\label{sed_sfr}
\end{figure*}

\subsection{Variation of the SEDs with $SFR$ (young stellar population)}
\label{subsec:sedsfr}

The increase in the contribution of the young stellar population to the stellar
SEDs will have the effect of increasing the MIR to FIR level of the
diffuse dust emission
SEDs, with essentially no change in the submm level (see Fig.~\ref{sed_sfr}). 
At the same
time the peak of the dust emission SEDs will shift towards shorter
wavelengths. Overall this will result in warmer SEDs, with both FIR SED peaks
and MIR-to-FIR colours becoming systematically bluer. It is interesting to
 note that the
effect of increasing $SFR$ is completely different from the effect of
increasing ${\tau}^f_B$, showing that the two parameters are completely
orthogonal. One should also notice that the trend in bluer SEDs with
increasing $SFR$ becomes less pronounced for models with higher contribution
from the old stellar populations to the stellar SEDs (see trends in moving from
the left column to the right column of Fig.~\ref{sed_sfr}).

\subsection{Variation of the SEDs with $old$ (the old stellar population)}
\label{subsec:sedold}

The increase in the luminosity of the old stellar population produces a 
shift of the peak of the infrared SED of the diffuse dust towards shorter wavelengths 
(see Fig.~\ref{fig:sed_old}), similar to the case of increasing the luminosity of 
the young 
stellar population. However, the shift is accompanied by a decrease in 
the MIR-to-FIR ratio, opposite to the effect obtained in the case of increasing
the luminosity of the young stellar population. Indeed, the MIR-to-FIR colours
become cooler, due to the fact that the additional optical photons will mainly
boost the FIR emission and not the PAH emission. Obviously this effect is less
pronounced for models with higher luminosities of the young stellar
populations (see trends from going from the left to the right hand column in
Fig.~\ref{fig:sed_old}) and is enhanced for models with higher dust opacity (see trends for
going from the top to the bottom row in Fig.~\ref{fig:sed_old}).

\begin{figure*}
\centering
\includegraphics{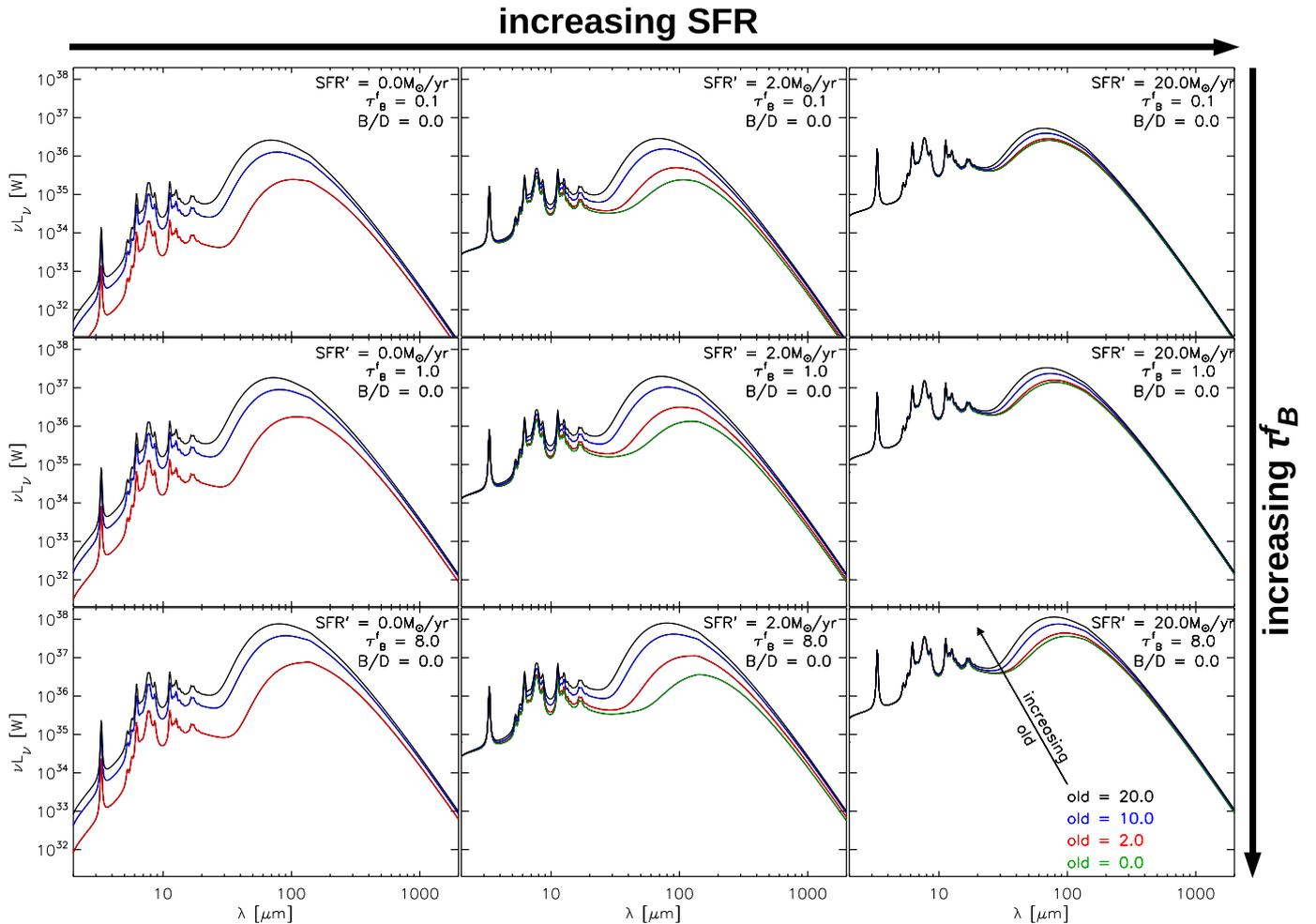}
\caption{Integrated dust and PAH emission SEDs for the diffuse component, for model galaxies with
  different contribution from the old stellar population (plotted as different 
  curves in each panel). All
  models shown in this figure are for pure disk galaxies ($B/D=0.0)$. From
  left to right the panels show model galaxies with various levels of
  SFR (${\rm SFR'}=0.0$, 
${\rm SFR'}=2.0$ and  ${\rm SFR'}=20.0$). From top to bottom the panels show 
models with various face-on B band opacities ($\tau^{f}_{B}=0.1$,
$\tau^{f}_{B}=1.0$, and $\tau^{f}_{B}=8.0$). The colour coding is as follows: 
  green is for 
  ${\rm old}=0.0$, red is for ${\rm old}=2.0$, blue is for 
  ${\rm old}=10.0$ and black is for ${\rm old}=20.0$.}
\label{fig:sed_old}
\end{figure*}

\subsection{Variation of the SEDs with $B/D$ (bulge-to-disk ratio)}
\label{subsec:sedbd}

The variation in the bulge-to-disk ratio produces variations in the infrared
SEDs with less dynamical range (see Fig.~\ref{fig:sed_bd}), simply due to the 
smaller dynamical range in
the values of the parameter $B/D$. Nevertheless,
the peak of the SEDs is shifted to shorter infrared wavelengths with increasing
$B/D$ and the MIR-to-FIR colours get cooler, following the trends expected for
a variation in the luminosity of an old stellar population. Thus the variation
induced by the $B/D$ ratio follows similar trends with the variation induced by
the $old$ parameter. It is however expected that in most cases the value of 
the $B/D$ to be  inputted from observations, and thus not to be a free 
parameter of the model.

Overall it is interesting to observe that the 3 parameters of the model:
$\tau^f_B$, $SFR$ and $old$ are orthogonal parameters, producing quite
different effects in shaping the dust emission SEDs. For example it is clear
that the only way to increase the level of the submm emission is through a
variation of $\tau^f_B$, as neither an increase in the luminosity of the old or
of the young stellar populations could account for this. In other words in the
submm we are tracing dust column densities, while in the MIR and FIR we are
tracing both dust column densities and heating sources. It is also obvious that
if we want to have warmer MIR-to-FIR colours we have to increase the luminosity
of the young stellar populations, while if we want to have cooler MIR-to-FIR
colours we need to increase the luminosity of the old stellar populations, with
different degrees of modulations due to variations in $\tau^f_B$.  Here we
need to remember that for fitting total SEDs the template for the star-forming
complexes must be added to the diffuse SEDs, boosting the MIR
emission (see Sect.~\ref{subsec:sedcolour}, below).

\begin{figure*}
\centering
\includegraphics{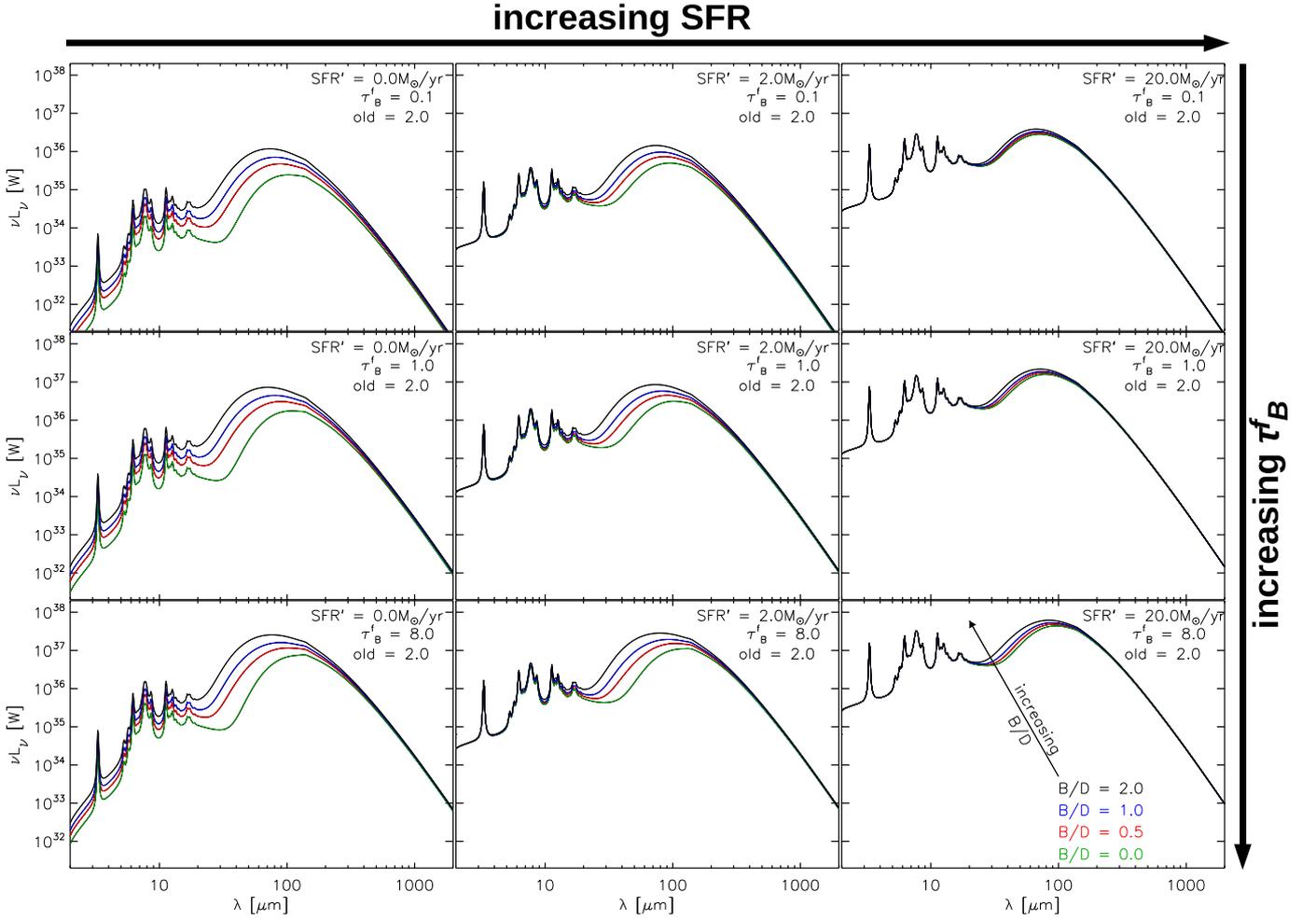}
\caption{Integrated dust and PAH emission SEDs for the diffuse component, for model galaxies with
  different bulge-to-disk ratios (plotted as different 
  curves in each panel). All
  models shown in this figure are for galaxies with a fixed contribution from
  the old stellar population (${\rm old}=2.0$). From
  left to right the panels show model galaxies with various levels of
  SFR (${\rm SFR'}=0.0$, 
${\rm SFR'}=2.0$ and  ${\rm SFR'}=20.0$). From top to bottom the panels show 
models with various face-on B band opacities ($\tau^{f}_{B}=0.1$,
$\tau^{f}_{B}=1.0$, and $\tau^{f}_{B}=8.0$). The colour coding is as follows: 
  green is for 
  ${\rm B/D}=0.0$, red is for ${\rm B/D}=0.5$, blue is for 
  ${\rm B/D}=1.0$ and black is for ${\rm B/D}=2.0$.}
\label{fig:sed_bd}
\end{figure*}

\subsection{Colour variation of the combined diffuse and
localised dust emission SEDs}
\label{subsec:sedcolour}

Adding the emission from the localised dust emission template of
star-formation complexes to the solution for the diffuse emission
to obtain the total emission adds extra variations to the dust
emission SEDs, especially boosting the warm dust emission
from larger grains. This effect, controlled by the $F$
parameter, is in practice rather orthogonal to the variations in the SED
of the diffuse component discussed above, due to the different behaviour
of the UV/optical attenuation as a function of $F$
(see Appendix~\ref{sec:composite}) and due to the fact that,
over the parameter range of the model, the localised dust emission template
is almost always much warmer in the FIR than the predicted FIR colours
of the diffuse dust emission. In fact the FIR colours of the diffuse
dust emission only approach those of the PDR template
for the extreme case $SFR=20$ and $old=20$ in an optically
thin galaxy in which the UV illuminates the diffuse dust,
as can be seen by comparing the top RH panel of Fig. 18
with the template SED in Fig. 5. A further reason for the
orthogonality of the template SED to the SED of the
diffuse emission is that dust in the diffuse ISM is
more efficient in chanelling absorbed UV energy into the 
PAH features than dust in the star-formation regions, due to
the photo-destruction of PAH in star-formation regions
(see Sect.~\ref{subsec:infraredclumpy}). 
Overall,
solutions for the total infrared emission
require most (though not all) of the emission in the
$20\,-\,60\,{\mu}$m range to be localised emission from
star-formation regions, with the MIR and FIR/submm emission
flanking this range being predominantly diffuse in origin.

The combined effect of the 5 main parameters of the model $\tau^{f}_{B}$, 
$SFR$, $old$, $B/D$ and $F$  will produce a large variation in the colours of
the simulated SEDs. In Fig.~\ref{fig:color-color} we plot the $170/100\,{\mu}$m
colour versus the $100/60\,{\mu}$m
colour for our model SEDs. The SEDs were derived by
combining all the simulated SEDs of the diffuse component from our library with
the template SED for star-forming regions, whereas the combination was done for
different values of the $F$ factor (including the asymptotic values). This
defines the locus in the colour-colour diagram occupied by our model SEDs. To
check that the parameter space covered by the models
overlaps with the parameter space  of real life spiral galaxies, we 
overplotted the corresponding
colours of the spiral galaxies with Hubble type earlier than Sd from the 
ISOPHOT Virgo Cluster Deep Survey
(IVCDS; Tuffs et al. 2002a, 2002b; Popescu et al. 2002) and from the catalog of
compact sources of the ISOPHOT Serendipity Survey (Stickel et al. 2000). One
can see that except from a few outliers, the observed galaxies lie on the
locus defined by our models. One can also observe that the position of NGC~891
in this diagram does not have any preferential place, but lies somewhere towards
the very quiescent region of the models (as expected). More detailed
comparisons with data and applications of the model will be given in a series
of future papers. 

From Fig.~\ref{fig:color-color} it is also obvious that our models embrace a 
somewhat larger area in parameter space than
that defined by real life spiral galaxies. This was done on purpose, from the choice of
the minimum and maximum values of the main parameters of the model. We tried 
to push the limited values towards the asymptotic values, in order to cover 
even rare (unexpected) cases that could occur in real life (see 
Sect.~\ref{sec:library} and Table~\ref{tab:param} for the range of parameter
space).

Most fundamentally, however, we emphasise that in the context of a 
RT solution it is the combination of the observed amplitude and
colour of the dust emission SED with the self-consistently applied
constraint on the attenuation in the UV/optical that
leads to an inherently non-degenerate solution.

\begin{figure}
\centering
\includegraphics[width=9.cm]{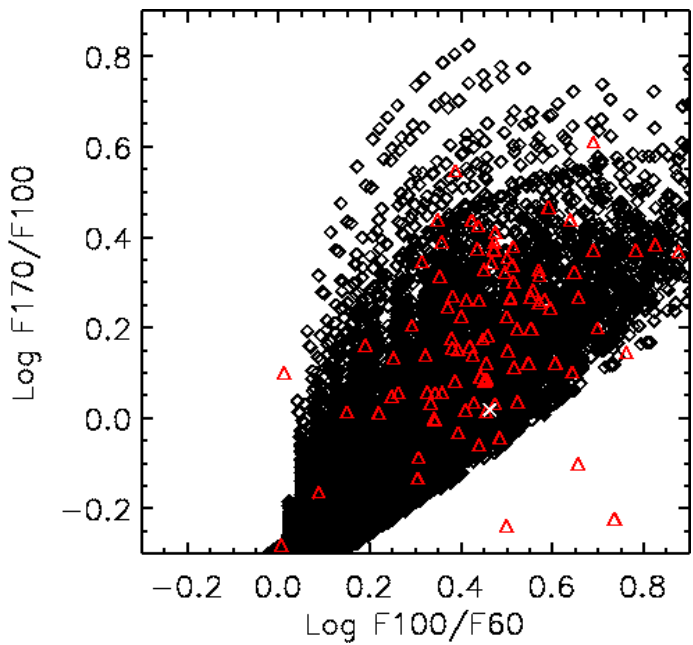}
\caption{The 170/100 versus 100/60 micron colour-colour plot for our simulated
  SEDs (black diamonds). Overplotted with red triangles are the observed
  colours of the spiral galaxies from the ISOPHOT Virgo Cluster Deep Survey 
(IVCDS; Tuffs et al. 2002a, 2002b; Popescu et al. 2002) and
the ISOPHOT Serendipity Survey (Stickel et al. 2000). Only spiral galaxies with
Hubble type earlier than Sd are plotted. The white star represents the position
of NGC~891 in the colour-colour diagram.}
\label{fig:color-color}
\end{figure}

\section{Discussion}
\label{sec:discussion}
In this paper we have assembled the components from which we can 
combine the predictions
for the attenuation of starlight (as originally
formulated in Paper~III and re-expressed here in Appendix~\ref{sec:attenuation} 
and Appendix~\ref{sec:composite})
with the predictions for the SED of the dust/PAH re-radiated starlight
(self-consistently calculated as described in Sect.~\ref{sec:calculation})
to extract the physical parameters of a star-forming galaxy
from an observed UV/optical/NIR - mid-IR/far-IR/submm SED. 

The detailed mathematical prescription of how to combine the library of
dust emission SEDs with the library of attenuation to fit UV/optical to
infrared/submm SEDs is given in Appendix~\ref{sec:howto}. Here we discuss some
of the issues regarding the applicability and physical relevance of our model.

Firstly, our model is designed for data sets that have good enough optical
images to derive $B/D$ and $i$ parameters. For the $i$ parameter this should be
the case for most optical surveys of local universe galaxies. For example, the
SDSS, with an angular resolution of 1.6$^{\prime\prime}$, was able to derive
inclination angles for galaxies with recession velocities ranging to
$\approx 10000$\,km/s. With modern surveys routinely having
$\approx 0.5^{\prime\prime}$ resolution, inclinations should be available 
up to $z\approx 0.1$, which encompasses the local Universe galaxies. In 
cases where inclination angles cannot be derived, an expectation value may be 
assumed instead, and the quartile range of uncertainty in
derived physical parameters resulting from a
lack of knowledge of $i$ can be derived by
running separate optimisations for the
quartile values of $i$. It should be noted that the inclination angle should 
not be
considered as a free parameter, as its effect on the SED would not be orthogonal
to the  $\tau^f_B$ parameter. The other parameter $B/D$ should also be an input
parameter for the routine and not a free parameter. In cases where information
on $B/D$ is missing, users are advised to use the correlation between $B/D$ and
apparent optical-NIR colours to estimate the $B/D$.

The use in the model of the third parameter accessible through  optical
observations, ${\theta}_{gal}$, is qualitatively different to
that of $B/D$ and $i$. Whereas $B/D$ and $i$ can (for practical applications)
only be used as constraints, ${\theta}_{gal}$ may in practice
be used either as a constraint (as in the example given above)
or as a free parameter. When ${\theta}_{gal}$ is used as a
free parameter, the angular surface area
${\theta}_{gal}^2$ effectively serves as free amplitude scaling
factor for the amplitude of the diffuse dust emission
(see  Eq.~\ref{eq:conversion2}).

Secondly, our model is in its present form only designed to
analyse star-forming galaxies. As noted above, when performing an
optimisation with the optical structural parameters fixed by direct observation,
the optimisation no longer has any degree of freedom in terms of
scaling parameters. This makes the model
sensitive to a test of the fundamental assumption, that the
dust is exclusively heated by stellar photons. The fits will be
systematically biased if dust is heated by photons from non-stellar
sources like AGN or by some completely different channel.
For example, by incorporating the optical and UV constraints, the model best fit
parameters for $SFR$ and $\tau^f_B$ will be likely to be skewed upwards
if an AGN heating dust with a  harder  UV photon spectrum than expected 
from a pure stellar source is present. In this case the
dereddened UV/optical spectrum from step {\it vii} will however provide a direct
flag for the presence of a bluer spectral emissivity in the UV than could be
provided by stars. Potentially, therefore, our SED fitting technique
can provide a method for recognising the presence of
dust-obscured AGN activity.
 
Thirdly, we note that the spectral form of the diffuse component
of the dust emission SED given by
Eq.~\ref{eq:conversion2} predicts that galaxies will lie on
specific locii in a colour - surface brightness diagram for
the dust emission, dependent on the surface density of UV/optical
emissivity and $\tau^f_B$. In cases where ${\theta}_{gal}$
is known from optical measurements, any deviation of the observed 
positions of galaxies in colour -  surface brightness space
from the positions predicted by the model can be therefore 
used as a non-parametric
test of the fidelity of the geometry used in the model
without having very specific a priori knowledge of the
physical properties of the observed sources. This would also
serve as the most direct verification of the major role played by
emission from diffusely distributed dust in our model.

Finally, we caution the potential user of this model that,
although RT solutions applied in this way are in principle
a very powerful way to constrain physical properties of galaxies,
their predictive power is ultimately reliant on an priori knowledge of the
geometry of the system. Although we have specified this
geometry using empirical constraints derived from spiral
galaxies of intermediate morphological type, and have
endeavoured to extend the applicability to galaxies of earlier
or later types through the B/D parameter, constrained from
optical imaging, the practical range of morphological types
where the advantages of the RT treatment outweigh deviations
of real from assumed geometry still remain to be determined.
This question can only be answered by statistical analysis,
such as potentially realisable in a non-parametric way vis the
colour-surface-brightness relation for the diffuse dust emission component
(as proposed above) for different morphological classes
of galaxies, and will be the subject of future studies.

\section{Summary}
\label{sec:summary}

In this paper we presented a radiative transfer model for spiral galaxies
that self-consistently
accounts for the attenuation of stellar light in the UV/optical and for the dust
re-emission in the MIR/FIR/submm. We used this model to create a comprehensive 
library  of dust and PAH emission SEDs and corresponding attenuations of
stellar light (originally presented and described in Paper ~III and given in 
revised form in this paper) that can
be used to routinely fit the panchromatic SEDs of large statistical samples of 
spiral galaxies.

Our model has been calibrated for local Universe galaxies, and therefore its 
applicability should mainly lie within low redshift galaxies. The models are 
also targeted to spiral galaxies. We have not
attempted to model elliptical galaxies, starburst galaxies, dwarf galaxies or 
AGN nuclei.

The free parameters of our model for the calculation of the infrared SEDs are:

\begin{itemize}
\item
the central face-on opacity in the
B-band $\tau^f_B$,
\item
the clumpiness factor $F$ which defines the fraction of UV photons locally
absorbed in star-formation regions and the corresponding fraction available to
illuminate the diffuse dust, 
\item
the star-formation rate $SFR$,
\item
the normalised luminosity of the old stellar  population $old$, 
\item
the bulge-to-disk ratio $B/D$. 
\end{itemize}
For the parameter $\tau^f_B$ we have provided a relation for converting to 
total dust masses (see Eq.~\ref{eq:dustmass}).

The free parameters $\tau^f_B$, $F$, and $B/D$ are common to the set of
parameters needed to predict the attenuation of the UV/optical stellar light 
(the fourth free parameter that affects the attenuation
is the inclination $i$ of the disk). Therefore the simulated dust emission SEDs 
presented here can be used in conjunction with the predictions for attenuation 
given in this paper for a self-consistent modelling of the UV/optical-IR/submm 
SEDs. 
Our model has also a parameter the angular size ${\theta}_{gal}$, which 
can be used either 
as a constraint (fixed via optical images) or as a free parameter. When 
${\theta}_{gal}$ is used as a free parameter it determines the amplitude
scaling of the fit.

Since in our model we made the assumption that 
the wavelength dependence of the fraction of escape photons from the clumpy 
component into the diffuse one is fixed, we can separately calculate the
SEDs for the diffuse component and for the clumpy component. For the clumpy
component we adopted the model of Groves et al. (2008), which we tuned to fit 
the total 
MIR/FIR/submm SED of the ensemble of star-forming complexes in the Milky Way
(Sect.~\ref{subsec:infraredclumpy}).
For the diffuse component we defined an effective star-formation rate $SFR'$
(corresponding to the total illumination by the young stellar population of
the diffuse dust) and
created a library of diffuse SEDs that spans the parameter
space of $\tau^f_B$, $SFR'$, $old$, and $B/D$. In total we sampled 7 values in
$\tau^f_B$, 9 in $SFR'$, 9 in $old$ and 5 in $B/D$, making a total of 2835
simulated SEDs. This corresponds to a library of 196 data cubes
of radiation fields, sampled at 22 radial positions and 12 vertical positions
(264 spatial points within the model galaxy), a library of 11340 data cubes of 
temperature distributions and a library of 11340 files of infrared emissivities.
We described (Appendix~\ref{sec:howto}) how in practice one can combine the 
predictions for the attenuation of starlight
with the predictions for the SED of the dust/PAH re-radiated starlight
to extract the physical parameters of a star-forming galaxy
from an observed UV/optical/NIR - mid-IR/far-IR/submm SED.  

The analysis of the library of diffuse integrated SEDs
(Sect.~\ref{sec:predictions}) showed that the main
parameters of the model are quite orthogonal, producing different
effects in shaping the dust emission SEDs. Specifically we have shown that:
\begin{itemize}
\item
The amplitude of the dust and PAH emission SEDs increases with
increasing optical depth for the optical thin cases and tends to a saturation
value for the optically thick cases.
\item
The ratio between the FIR and MIR (PAH) amplitudes increases with increasing 
opacity for the models where the  stellar luminosity has a higher contribution 
from the old stellar population with respect to the young stellar population.
\item
The peak of the SEDs
shifts towards longer wavelengths with increasing opacity for the optical thick
regime, for the models where
the stellar luminosity has a higher contribution from the young stellar
populations with respect to the old stellar population.
\item
The increase in the contribution of the young stellar population to the stellar
SEDs will have the effect of producing warmer SEDs, with both FIR SED peaks
and MIR-to-FIR colours becoming systematically bluer.
\item
The increase in the luminosity of the old stellar population produces SEDs with
warmer FIR SED peaks but with redder MIR-to-FIR colours.
\item
The variation in the bulge-to-disk ratio produces variations in the infrared
SEDs with less dynamical range, simply due to the 
smaller dynamical range in the values of the parameter $B/D$. Overall the peak 
of the SEDs is shifted to shorter infrared wavelengths with increasing
$B/D$ and the MIR-to-FIR colours get cooler, following the trends expected for
a variation in the luminosity of an old stellar population.
\end{itemize}

Overall we found that the effect of increasing $SFR$ is completely different 
from the effect of increasing ${\tau}^f_B$ and works in the opposite direction 
from that produced by an increase in $old$. We also verified that the 
only way to markedly increase the level of the submm emission is through a 
variation of $\tau^f_B$.

The spatially integrated SEDs were derived from simulated
images of dust emission, themselves calculated from 3D data cubes of infrared
emissivity representing the response of grains (integrated over size
distribution and composition) to the radiation fields derived from radiative
transfer calculations. Our calculations take full account of the
large variations in colour of the radiation fields as a function of position in
the galaxy, which we show to be important (Sect.~\ref{subsec:radiation}). 
Such variations lead
to large trends in the grain temperature distributions (Sect.~\ref{subsec:temperature}) used in the calculation
of stochastic emission, leading to corresponding trends in the
infrared brightnesses (Sect.~\ref{subsec:infrared}) with position in the 
galaxy. In particular we concluded that: 

i)
The shift of the peak of the infrared 
brightness as a function of grain size strongly depends on the temperature of 
the grains, and therefore on
the stochastic or non-stochastic nature of the heating mechanism. Since the
heating mechanism depends both on grain size and on the intensity and colour of
the radiation fields, it is clear that the shift cannot be described in
terms of grain size only. This also shows that
models that have a fixed grain size for the transition between the main
heating mechanisms of dust, irrespective of the radiation fields, will
lead to systematic spurious shifts in the mid-infrared to FIR colours with 
increasing galactocentric radius.

ii) The increase in the relative contribution of big grain emission 
to the small grain emission is a strong function of the colour of the radiation
fields, and, unlike the wavelength dependence, does not depend on the heating
mechanism of the grains. This also shows that models that assume a fixed colour
of the radiation fields (e.g. that of the local interstellar radiation fields)
will incur systematic errors in the mid-infrared to FIR colours with increasing
galactocentric radius.

Basis for the radiative transfer calculations were empirical constraints for
the geometry of the large scale distributions of stellar emissivity and dust 
opacity of the translucent and opaque components of galaxies 
(Sect.~\ref{subsec:distribution}). The translucent
components were constrained using the results from the radiation transfer
analysis of Xilouris et al., while the geometry of the optically thick 
components is constrained from physical considerations with a posteriori
checks of the model predictions with observational data. The main checks
concern the nature of the diffuse distribution of dust associated with the
young stellar population which is needed to predict the observed
FIR/submm emission of real life galaxies. These checks
(Sect.~\ref{sec:assumptions}) showed that:

i) Comparison of model
predictions for the stellar emissivity in the optical with population synthesis
models is not consistent with one dust disk model with one stellar disk
component, and indicates the need for extra luminosity hidden by extra dust -
the second dust disk and young stellar disk in our model formulation.

ii) Comparison of model predictions for the vertical profiles of PAH emission
with observations at 100\,pc linear resolution reveals a good agreement,
consistent with the existence of a thin dust layer, as represented by the
second dust disk.

iii) Comparison of model predictions for the attenuation-inclination relation
with observations is also consistent with the existence of a thin dust layer,
as represented by the second dust disk. 

iv) Comparison of model predictions for the optical surface brightness
distributions of the edge-on galaxy NGC~891 with observations indicates that a 
two dust disk model with two
(old and young) stellar components is able to reproduce the observed data
better than the one disk model with one stellar disk. We also showed that a two
dust disk model which is not accompanied by an extra stellar luminosity
component produces a stronger dust lane than observed in real images. In the K
band the model with the two dust disks and two stellar components predicts a 
somewhat more prominent dust lane than observed, indicating either a shorter
scalelength for the dust disk (though this would be difficult to
reconcile with the excellent fit we found in Paper~I for the radial profile at 
850\,${\mu}$m) or more luminosity in the young stellar disk than
predicted by the population synthesis models. A possible alternative
reason for the difficulty in fitting the vertical profile in K-band
is that, due to its very small scaleheight, the 
appearance of the second dust disk in the K band, where the first dust disk
becomes transparent, would be easily blurred 
in real galaxies if perturbations from co-planarity occur, as has been 
observed in the form of ``corrugations'' in the
Milky Way and external galaxies. 

We also investigated the approximation of the spiral arm component with 
an exponential disk (Sect.~\ref{subsec:spiral}), and showed that this 
approximation has negligible effect
on the predicted dust emission SEDs when the amount of dust is preserved. 

A particular feature of our model is the representation of the stellar
emissivity SEDs in terms of just two parameters, $SFR$ and $old$, since this
enables UV/optical SEDs to be dereddened without recourse to population
synthesis models, thereby avoiding bias due to the
age/metallicity-opacity degeneracy. This feature entailed the approximation of  
a fixed spectral shape for the SEDs of the old and young stellar populations.
We quantified this approximation showing it has a negligible effect on the dust
emission SEDs (Sect.~\ref{subsec:approximation}).

We have derived updated and improved solutions for the edge-on galaxy NGC~891
(Sect.~\ref{sec:illustration}),
one of the galaxies from the set used to constrain the geometry of the model.
Since the luminosity of the old stellar populations has been fixed to
the values derived by Xilouris et al. (1999) from the optimisation of the
optical/NIR images, $old=0.792$ and $B/D=0.33$, only 3 free parameters were 
needed to fit the 
dust emission SED of NGC~891, namely $\tau^f_B$, $SFR$, $F$.  
For the integrated SED of NGC~891 the best fit solution was found for
$\tau^{f}_{B}=3.5$, $SFR=2.88$\,M$_{\odot}$/yr and  $F=0.41$, which corresponds
to a total dust luminosity of $L_{dust}^{total}=9.94 \times 10^{36}$\,W, of
which $L_{dust}^{diff}=6.89 \times 10^{36}$\,W is emitted in the diffuse medium
($69\%$). The results indicate that the diffuse component dominates
the emission longwards of 60\,${\mu}$m and shortwards of $20\,{\mu}$m, while
the localised dust emission within the star-forming complexes dominates at
intermediate wavelengths ($20-60\,\mu$m). It is only at intermediate 
wavelengths ($20-60\,\mu$m) where the localised dust emission within the 
star-forming complexes dominates. The total dust emission of NGC~891 
is predominantly powered by the young stellar populations, which contribute 
$69\%$ to the dust heating. The detailed input from the different stellar 
components is as follows: $11\%$ from the bulge, $20\%$ from the old stellar 
disk, $38\%$ from the young stellar disk and $31\%$ from the star forming 
complexes.

\begin{acknowledgements}
We would like to thank an anonymous referee for constructive and useful
comments that helped improve the quality of the paper.
We would also like to acknowledge Aigen Li for providing us with the data for 
the optical properties of carbonaceous grains used in Draine \& Li (2007).

One of us (NDK) acknowledges support from EU Marie Courie grant 39965 and EU
REGPOT grant 206469.  
\end{acknowledgements}

\newpage

\appendix

\section{Calculation of the extinction cross-section and scattering phase
  function}
\label{sec:extinction}

A prerequisite for the calculation of radiation fields and infrared emission 
is knowledge of the extinction cross-section of the population of grains, and, 
since we consider anisotropic scattering in the radiation transfer 
calculations, the scattering phase-function needs to be known as well. 

The absorption cross-section of grains of composition 
$i=\{$Si,Gra,PAH$^0$,PAH$^{+}\}$, $C_{abs,\, i}$, is obtained by integrating the
absorption efficiencies $Q_{abs,\,i}$ over the grain size distribution $n(a)$:

\begin{eqnarray}\label{eq:abscross-section}
C_{abs,\, i}(\lambda) & = & \int_{a_{min}}^{a_{max}} \pi\, a^2\,n(a)\,Q_{abs,\,i}(a,\lambda)\,da
\end{eqnarray}
where $C_{abs,\, i}$ is given in units of $[{\rm cm}^2\,{\rm H}^{-1}]$, $a_{min}$ is the minimum grain size 
and $a_{max}$ is the maximum grain size. 

\begin{figure}
\centering
\includegraphics[width=9cm]{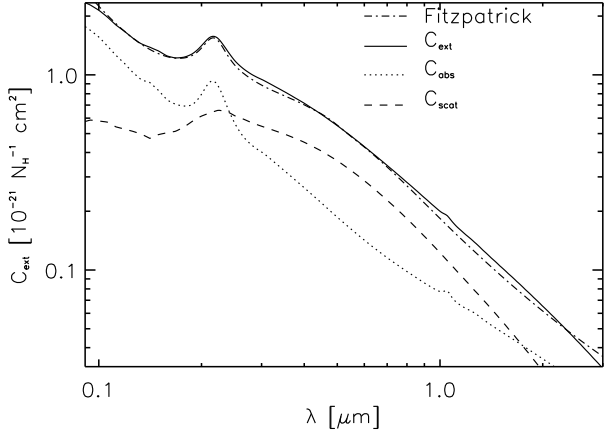}
\caption{The extinction curve for the dust model used in this paper (solid
  line), which is a Milky Way dust model. With dashed-dotted line we plotted the
  observed mean extinction curve of our galaxy (Fitzpatrick 1999). Also 
   plotted are
  the two components of extinction, the model absorption curve (dotted line) 
and the model scattering curve (dashed line).}
\label{fig:extinction_abs_sca_curve}
\end{figure}

Similarly, the scattering cross-section of grains of composition 
$i$, $C_{sca,\, i}$, is obtained by integrating the
scattering efficiencies $Q_{sca,\, i}$ over the grain size distribution $n(a)$:
\begin{eqnarray}\label{scacross-section}
C_{sca,\, i}(\lambda) & = & \int_{a_{min}}^{a_{max}} \pi\, a^2\,n(a)\,
Q_{sca,\, i}(a,\lambda)\,da
\end{eqnarray}
Then, by summing over the grain composition $i$ we obtain the total absorption 
and scattering cross-sections, $C_{abs}$ and $C_{sca}$:
\begin{eqnarray}\label{eq:totalabscross-section}
C_{abs}(\lambda) & = & \sum_i C_{abs,\, i}(\lambda)\\
\label{eq:totalscacross-section}
C_{sca}(\lambda) & = & \sum_i C_{sca,\, i}(\lambda)
\end{eqnarray}
The extinction cross-section $C_{ext}$ is the sum of the absorption and 
scattering cross-sections:
\begin{eqnarray}\label{eq:extcross-section}
C_{ext}(\lambda) & = & C_{abs}(\lambda) + C_{sca}(\lambda)
\end{eqnarray}

We note that the extinction cross section $C_{ext}$ is defined as per 
unit H. In some applications it is useful to define a cross section per unit 
dust mass, which we denote here $C_{ext}^{m}$: 
\begin{eqnarray}\label{eq:extinction_mass_H}
C_{ext}^{m}(\lambda) = \frac{C_{ext}(\lambda)}{\int_{a_{min}}^{a_{max}} (4/3)\pi\,
  a^3\,n(a)\,\rho_g\,da}
\end{eqnarray}
where $\rho_g$ is the density of the grain material and $C_{ext}^{m}$ is in
units of ${\rm cm}^2/{\rm g}$.

In turn, $C_{ext}^{m}$ is related to the extinction coefficient 
$\kappa_{\rm ext}$, as used in the mathematical prescription of the dust 
distribution from 
Eqs.~\ref{eq:dustemissivity1} and \ref{eq:dustemissivity2} using:
\begin{eqnarray}\label{eq:extinction_mass_length}
\kappa_{ext}(\lambda,R,z) = \rho_{dust}(R,z) \times C_{ext}^{m}
\end{eqnarray}
where $\rho_{dust}(R,z)$ is the dust mass density at position (R,z) in the
galaxy in units of ${\rm g}\,{\rm cm}^{-3}$ and $\kappa_{ext}(\lambda,R,z)$ is in 
units of $cm^{-1}$.

 Figs.~\ref{fig:extinction_abs_sca_curve} and 
\ref{fig:extinction_composition_curve} show the resulting extinction 
curve of the dust model adopted here, together with the absorption and 
scattering components (Fig.~\ref{fig:extinction_abs_sca_curve}) and the 
components given by the different grain composition 
(Fig.~\ref{fig:extinction_composition_curve}). As expected, the
figures confirm that the model extinction curve fits well the observed mean 
extinction curve of our galaxy.

The averaged anisotropy of the scattering phase function g needed in the
radiative transfer calculation is obtained in a similar manner to
Eqs.~\ref{eq:abscross-section}, \ref{scacross-section}, 
\ref{eq:totalabscross-section}, and \ref{eq:totalscacross-section}.
\begin{eqnarray}\label{eq:anisotropy}
g_i(\lambda) & = \int_{a_{min}}^{a_{max}} \pi\,
a^2\,n(a)\,Q_{sca,\,i}(a,\lambda)\,Q_{phase,\, i}(a,\lambda)\,da
\end{eqnarray}
\begin{eqnarray}\label{eq:totalanisotropy}
g(\lambda) & = & \sum_i g_{i}(\lambda)
\end{eqnarray}
where $Q_{phase,\, i}$ is the anisotropy efficiency.

\begin{figure}
\centering
\includegraphics[width=9cm]{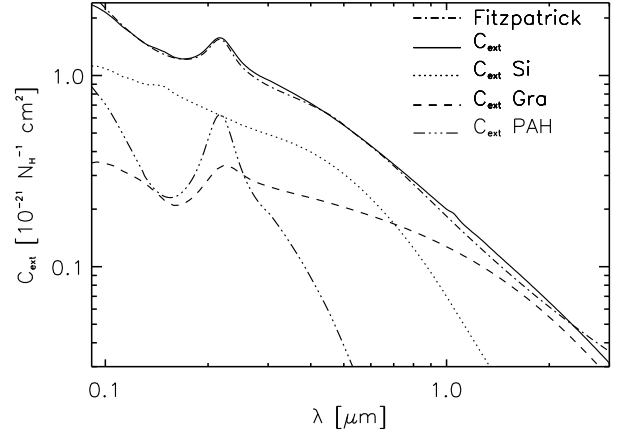}
\caption{The extinction curve for the dust model used in this paper 
(solid line) together with the contributions to the 
extinction from the different dust compositions used in the model: Si 
(dotted line), Gra (dashed line), PAH (dashed-three-dotted line). As in 
Fig.~\ref{fig:extinction_abs_sca_curve}, the observed mean extinction curve of
our galaxy is plotted with dashed-dotted line.}
\label{fig:extinction_composition_curve}
\end{figure}

\section{The library of attenuations of stellar light for the diffuse
  stellar components}
\label{sec:attenuation}

\begin{figure}
\centering
\includegraphics[width=9.5cm]{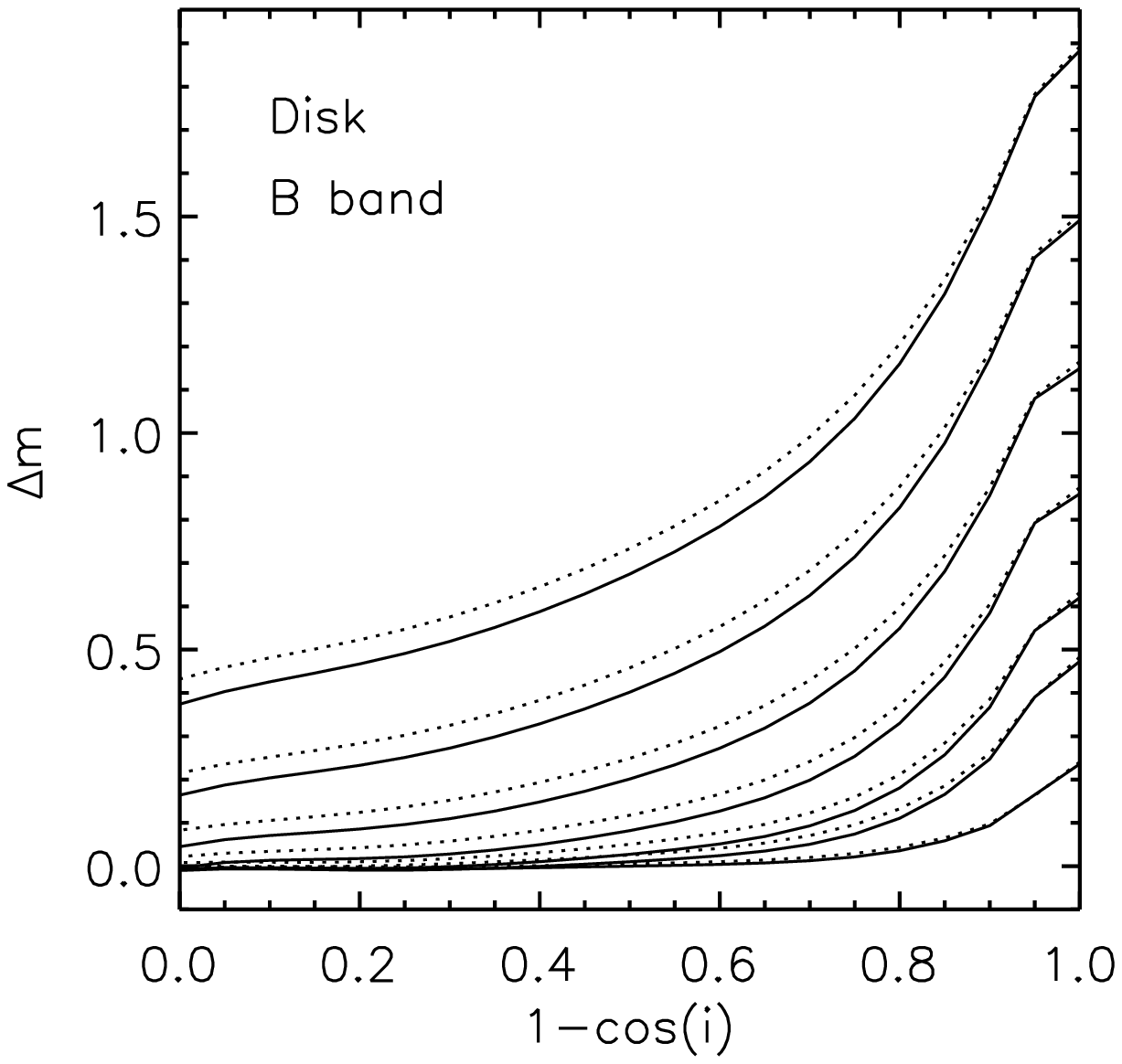}
\includegraphics[width=9.5cm]{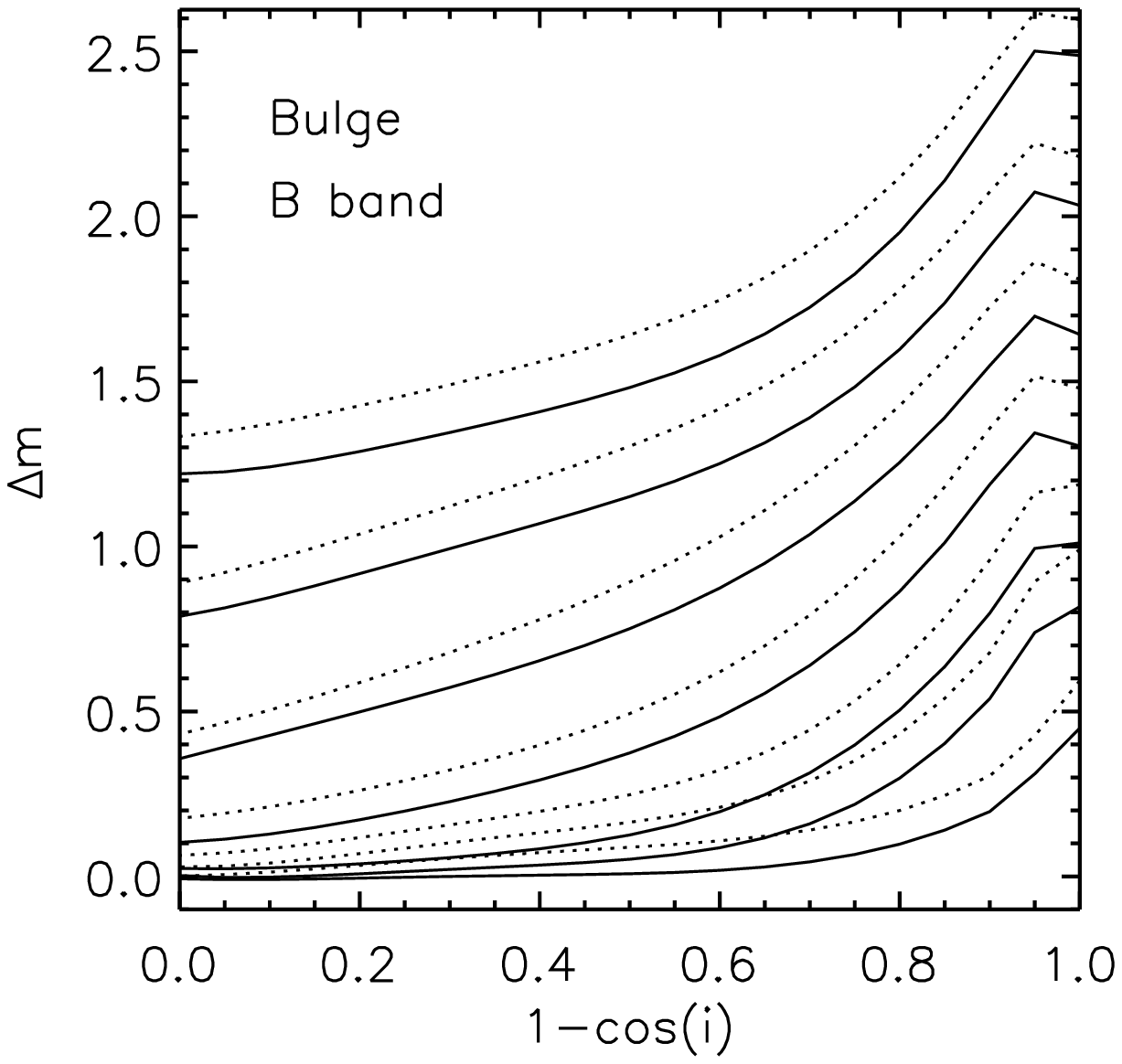}
\includegraphics[width=9.5cm]{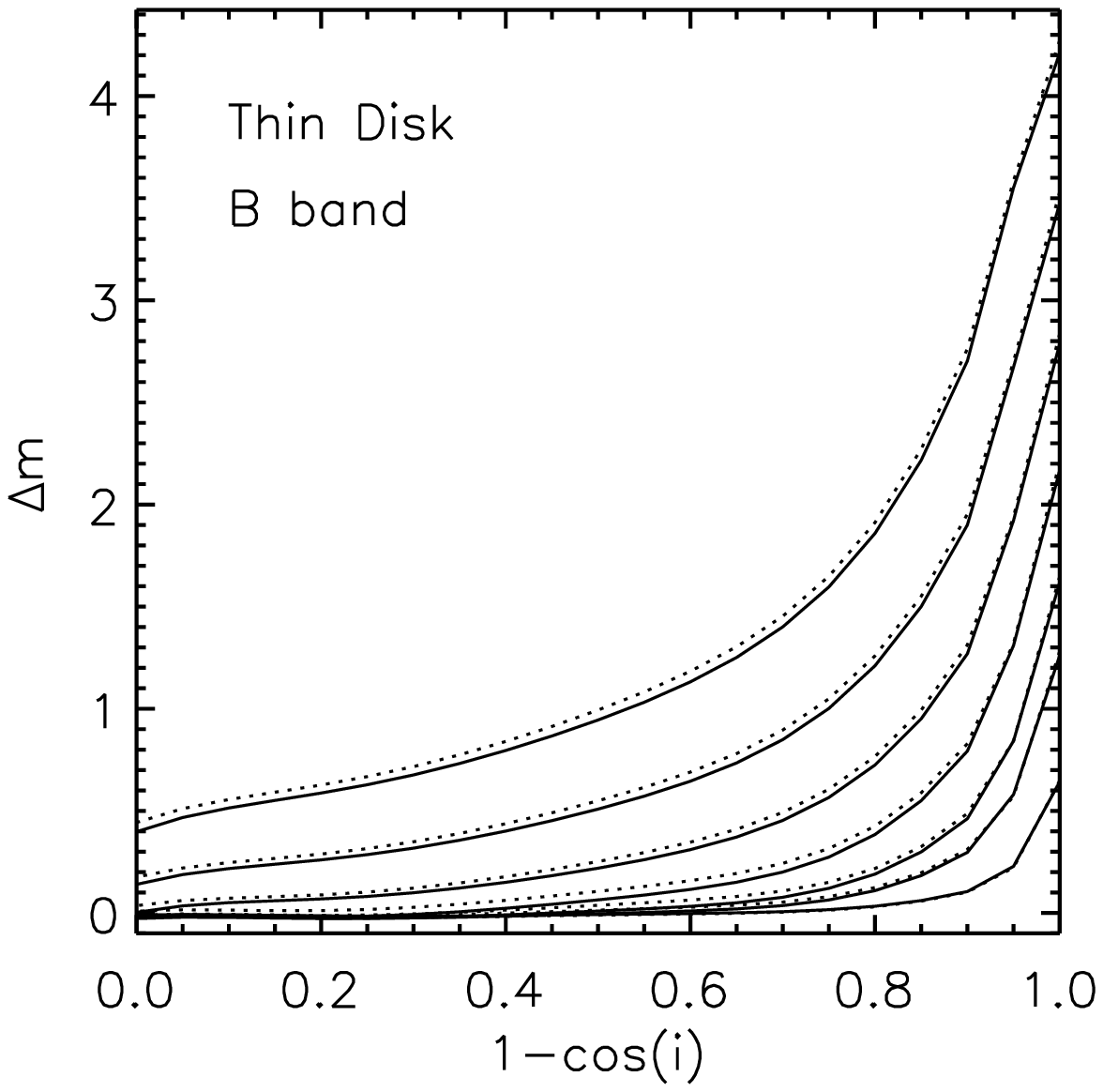}
\caption{The attenuation-inclination relation for the disk (top), bulge
  (middle) and thin disk (bottom) in the B band. In each panel the solid curves
  represent the attenuations calculated with the dust model used in this paper,
  incorporating a mixture of silicate, graphite and PAH molecules. The dotted
  curves represent the corresponding attenuations calculated with the dust
  model used in Paper~III, incorporating only a mixture of silicates and
  graphites. In each panel the 7 different curves represent the
  attenuation-inclination relations for different $\tau^{f}_{B}$: 0.1, 0.3,
  0.5, 1.0, 2.0, 4.0, 8.0, with the values increasing from bottom curves to top
curves.}
\label{fig:compare_attenuation}
\end{figure}

The second set of simulated data needed to fit the panchromatic SEDs is the
library of attenuations in the UV/optical/NIR as a function of $\tau^{f}_{B}$
and $i$. 
As mentioned in Sect.~\ref{subsec:dustmodel}, the revision of the dust model
required a recalculation of the database for the attenuation of stellar light,
as presented in Paper~III. The overall concept and characteristics of the
calculations are the same as in Paper~III, but a small change, mainly in the 
zero
point of the calculations, was apparent due to the change in the relative
contribution of absorption and scattering to the total extinction. Here we only
give an example of a comparison between the attenuation-inclination curves
obtained using the new and the old dust model
(Fig.~\ref{fig:compare_attenuation}). 

The attenuation-inclination relations for disks show a systematic change with
inclination when changing the dust model. Thus the attenuation for the low
inclinations is decreased more than for the high inclinations, with a tendency 
for the curves to converge at the
edge-on inclinations. This means that the shape of the attenuation-inclination
is steepened for the present dust model. The curves for bulges show the biggest
offset when changing the dust model, but in most of the cases there is no change
in the shape of the curves. As one can see the curves run almost parallel,
except perhaps for the lowest values of opacity. The smaller change is seen for
the thin disk component, where neither the shape nor the zero point are
changed significantly.

We also did some tests to quantify the effect of the change in the dust model
to the overall energy balance. By integrating the attenuation over all angles
we obtained an estimate of the total energy absorbed by dust in a
galaxy. This absorbed energy was found to be on average $10\%$ smaller for the
attenuations calculated using the new dust model than for those from Paper~III. 

\section{Formulation of composite attenuation of stellar light}
\label{sec:composite}

In Sect.~5.1 of Paper~III a generalised formula was given, showing 
how the composite attenuation (that is the overall attenuation of an arbitrary
combination of luminosity components from stellar populations in the young
stellar disk, the old stellar disk and the bulge) can be derived from the
library of attenuation of stellar light from the diffuse component and the
attenuation of the clumpy component. At any wavelength, the composite
attenuation depends on the relative luminosity of the three stellar components,
which we have described in this paper in terms of the parameters $SFR$, $F$,
$old$ and $B/D$, which we used to describe the dust emission. Here we re-write 
the generalised expression for the composite attenuation 
(Eq.~16 from Paper~III) for the specific parameterisation adopted in this paper.

At a given wavelength $\lambda$, the composite attenuation 
${\Delta}m_{\lambda}$ in a galaxy is given by:

\begin{eqnarray}\label{eq:composite_attenuation}
{\Delta}m_{\lambda} = -2.5\,\log\frac{L_\lambda}{L^{0}_\lambda}
\end{eqnarray}
where ${L^{0}_\lambda}$ and ${L_\lambda}$ are the intrinsic and the apparent
luminosity densities. The quantities ${L^{0}_\lambda}$ and ${L_\lambda}$ can be
further expressed as a summation of the corresponding quantities for the disk,
thin disk and bulge:

\begin{eqnarray}\label{eq:intrinsic_total}
L^{0}_\lambda & = & L^{0,\, disk}_\lambda + L^{0,\, tdisk}_\lambda + L^{0,\,
  bulge}_\lambda\\
L_\lambda & = & L^{disk}_\lambda + L^{tdisk}_\lambda + L^{bulge}_\lambda
\label{eq:apparent_total}
\end{eqnarray}

The apparent and intrinsic luminosity densities for the disk, thin disk and
bulge are related as follows:

\begin{eqnarray}\label{eq:intrinsic_apparent_disk}
L^{ disk}_\lambda & = & L^{0,\,disk}_\lambda 10^{- \frac{\displaystyle 
\Delta m^{disk}_{\lambda}}{\displaystyle 2.5}} = L^{0,\,disk}_\lambda\,A^{ disk}_\lambda\\
\label{eq:intrinsic_apparent_tdisk}
L^{ tdisk}_\lambda & = & L^{0,\,tdisk}_\lambda\,(1-F\,f_{\lambda})\,10^{- \frac{\displaystyle 
\Delta m^{tdisk}_{\lambda}}{\displaystyle 2.5}} = L^{0,\,tdisk}_\lambda\,A^{ tdisk}_\lambda\\
\label{eq:intrinsic_apparent_bulge}
L^{ bulge}_\lambda & = & L^{0,\,bulge}_\lambda 10^{- \frac{\displaystyle 
\Delta m^{bulge}_{\lambda}}{\displaystyle 2.5}} =  L^{0,\,bulge}_\lambda\,A^{ bulge}_\lambda
\end{eqnarray}
where $\Delta m^{disk}_{\lambda}$,  $\Delta m^{tdisk}_{\lambda}$ and  $\Delta
m^{bulge}_{\lambda}$ are the attenuation values expressed in magnitudes for the 
diffuse component in the disk, thin disk and bulge and  $A^{disk}_{\lambda}$,  
$A^{tdisk}_{\lambda}$ and  $A^{bulge}_{\lambda}$ are the corresponding
attenuations expressed in linear form. Using Eqs.~\ref{eq:oldsed}, 
\ref{eq:bulgesed},
\ref{eq:youngsed} and Eqs.~\ref{eq:intrinsic_apparent_disk},
\ref{eq:intrinsic_apparent_tdisk}, \ref{eq:intrinsic_apparent_bulge} 
we can rewrite Eqs.~\ref{eq:intrinsic_total} and \ref{eq:apparent_total} as:

\begin{eqnarray}\label{eq:intrinsic_total_param}
L^{0}_\lambda & = & old\,L^{old}_{\lambda,\, unit} +
\frac{SFR}{1M_{\odot}\,yr^{-1}}\,
L^{young}_{\lambda,\, unit} + (B/D)\,old\,L^{old}_{\lambda,\,unit}\\
\label{eq:apparent_total_param}
L_\lambda & = & old\,L^{old}_{\lambda,\, unit}\,A^{disk}_{\lambda} + 
\frac{SFR}{1M_{\odot}\,yr^{-1}}\,
L^{young}_{\lambda,\, unit}\,(1-F\,f_{\lambda})\, A^{tdisk}_{\lambda}+ \nonumber\\
& & (B/D)\,old\,L^{old}_{\lambda,\,unit}\,A^{bulge}_{\lambda}
\end{eqnarray}
By making the notation:
\begin{eqnarray}\label{eq:notation}
{\epsilon}_{\lambda}(SFR,old) =
\frac{SFR}{1M_{\odot}\,yr^{-1}}\,\frac{1}{old}\,
\frac{L^{young}_{\lambda,\,unit}}{L^{old}_{\lambda,\,unit}}
\end{eqnarray}
the composite attenuation from Eq.~\ref{eq:composite_attenuation} becomes:
\begin{eqnarray}\label{eq:composite_attenuation_final}
{\Delta}m_{\lambda} = -2.5\,\log\,A_{\lambda}
\end{eqnarray}
where
\begin{eqnarray}\label{eq:xi1}
A_{\lambda} & = & \frac{A^{disk}_{\lambda} +
(1-F\,f_{\lambda})\,{\epsilon}_{\lambda}(SFR,old)\,A^{tdisk}_{\lambda} + 
(B/D)\,A^{bulge}_{\lambda}}
{1 + {\epsilon}_{\lambda}(SFR,old) + B/D}
\end{eqnarray}
For the case that $old=0$ in the optical and in the UV 
Eq.~\ref{eq:composite_attenuation_final} becomes:
\begin{eqnarray}\label{eq:composite_attenuation_UV}
{\Delta}m_{\lambda}  = -2.5\,\log (1-F\,f_{\lambda}) + \Delta m^{tdisk}_{\lambda}
\end{eqnarray}
The use of the calibration factor $F_{cal}$ in our procedure means that in
practice the equations describing the attenuation due to the diffuse dust
illuminated by the young stellar disk need to be rescaled to accommodate 
different values of $F$ than those used in the calibration. For this we need to
use the correction factor for the diffuse component $corr^{d}(F)$ as defined in
Eq.~\ref{eq:correction_diffuse_F}, and rescale  
Eq.~\ref{eq:intrinsic_apparent_tdisk} in a similar way to the formulation of 
the radiation fields in Sect.~\ref{subsubsec:combining_radiation}:
\begin{eqnarray}\label{eq:intrinsic_apparent_tdisk_corr}
L^{ tdisk}_\lambda & = & L^{0,\,tdisk}_\lambda\,(1-F_{cal}\,f_{\lambda})\,
corr^{d}(F)\,A^{tdisk}_{\lambda}
\end{eqnarray}
In this case Eq.~\ref{eq:xi1} becomes:
\begin{eqnarray}\label{eq:xi1_corr}
A_{\lambda} = [A^{disk}_{\lambda} +
(1-F_{cal}\,f_{\lambda})\,corr^{d}(F)\,{\epsilon}_{\lambda}(SFR,old)\,
A^{tdisk}_{\lambda} + \nonumber \\
(B/D)\,A^{bulge}_{\lambda}]/
[1 + {\epsilon}_{\lambda}(SFR,old) + B/D]
\end{eqnarray}
and Eq.~\ref{eq:composite_attenuation_UV} becomes:
\begin{eqnarray}\label{eq:composite_attenuation_UV_corr}
{\Delta}m_{\lambda}  = -2.5\,\log (1-F_{cal}\,f_{\lambda})\,corr^{d}(F) + \Delta m^{tdisk}_{\lambda}
\end{eqnarray}
Eq.~\ref{eq:xi1_corr}, together with Eqs.~\ref{eq:composite_attenuation_final}
and \ref{eq:xi1} are the analog of the original expression for the composite
attenuation from Eq.~16 in Paper~III. 
Eq.~\ref{eq:composite_attenuation_UV_corr} together with 
Eq.~\ref{eq:composite_attenuation_UV} are
the analog of Eq.~17 from Paper~III.

\section{How to use the model to fit UV/optical to infrared/submm SEDs}
\label{sec:howto}

The six physical parameters determining our model prediction
of the SED are:
\begin{itemize}
\item
$\tau^{f}_{B}$, $SFR$ , $F$, $old$, $B/D$ and $i$\footnote{$i$ is the
  inclination of the disk, as used in Paper~III} 
\end{itemize}
whereby 
\begin{itemize}
\item
$\tau^{f}_{B}$, $SFR$, $F$, $old$, and $B/D$
\end{itemize}
 determine 
$L_{\lambda,\, dust}^{model}$,
the model prediction for the dust luminosity density in the IR/submm
as given by Eq.~\ref{eq:infraredsed_total} of this paper,
whereas 
\begin{itemize}
\item
$\tau^{f}_{B}$, $SFR$, $F$, $old$, $B/D$ and $i$
\end{itemize}
determine the attenuation in the measured UV/optical/NIR emission (whereby the
dependence on $SFR$ and $old$ is a weaker dependence due to the
dependence of the composite attenuation on the relative amplitudes
of the young and old stellar populations, as described in 
Appendix~\ref{sec:composite}). This 
attenuation is given in magnitude form,
${\Delta}m_{\lambda}$, by
Eqs.~\ref{eq:composite_attenuation_final},\ref{eq:xi1} and
\ref{eq:xi1_corr}. In the following it is convenient
to express it in the linear form:
\begin{eqnarray}\label{eq:atten}
A_{\lambda}(\tau^{f}_{B},SFR,F,old,B/D,i) =  10^{(-0.4\times{\Delta}m_{\lambda})}
\end{eqnarray}

We note that two of the six physical parameters, $SFR$ and $old$, are extrinsic
(that is, the quantity scales with the amount of material in an object).
This is a consequence of our model galaxy having a fixed size, 
expressed in terms of the fixed reference scale length of
the old stellar population in B-band of $h^{\rm disk}_{s, ref} = h^{disk}_s(B)
= 5.67$\,kpc.

The dust emission SED of the diffuse component of a galaxy with a 
value\footnote{Throughout this paper 
$h^{disk}_{s}$ is taken 
as the true B-band scalelength of the stellar population.
We note however that in any dusty galaxy $h^{disk}_{s}$ will not actually be
a directly observable quantity, even if the distance to the galaxy is known.
As described and quantified in Paper~IV the apparent size 
of a galaxian disk obtained from photometric fits will differ from the 
true size due to the stronger attenuation of light at the centre
of galaxies compared to the outer regions.
Due to the appearance of size (expressed by $\zeta$) 
as a quadratic term in Eq.~\ref{eq:conversion} determining the amplitude
of the SEDs for a given value of $SFR$ and $old$,
this effect can appreciably
influence the solutions obtained from fitting SEDs. Therefore,
if using the  measured angular sizes of galaxies as a constraint
in fitting the dust emission SEDs the apparent sizes should be converted
into true sizes using the correction factors tabulated in
Tables~1-5 of Paper~IV. Similarly, the $B/D$ ratio should be
converted from an observed ratio, as determined in photometric
fits to data, into the intrinsic ratio, using correction factors tabulated in 
Paper~III.
In principle, $i$, as derived from measurements of axial ratios, should also 
be corrected.} for
$h^{disk}_{s}$differing from $h^{disk}_{s, ref}$ will be:

\begin{eqnarray}\label{eq:conversion}
L^{diff}_{\lambda,\,dust}(SFR,old,\tau^f_{B},F,B/D) 
  =  {\zeta}^2 \times \nonumber\\
\times L_{\lambda,\,dust}^{diff, \, model}(SFR^{model},old^{model},\tau^f_B,F,B/D) 
\end{eqnarray}
where
\begin{eqnarray}\label{eq:conversion1}
SFR & = & {SFR}^{model}\times {\zeta}^2 \nonumber\\
old & = & {old}^{model}\times {\zeta}^2
\end{eqnarray}
where $SFR^{model}$ is the star-formation rate of the model galaxy having the 
reference size $h^{disk}_{s,\,ref}$, $old^{model}$ is the normalised luminosity 
of the old stellar disk population of the model galaxy having the reference
size $h^{disk}_{s,\,ref}$, $SFR$ is the real star formation rate of the 
galaxy that we want to model, $old$ is the real normalised 
luminosity of the old stellar disk population of the galaxy that we want to 
model, and

\begin{eqnarray}\label{eq:zeta}
\zeta = \frac{h^{disk}_{s}}{h^{disk}_{s,\,ref}}  
\end{eqnarray}
Eq.~\ref{eq:conversion} expresses the fact that radiation field energy
density, and hence the colours of the dust emission, varies according
to surface density of luminosity. In cases where a galaxy is
unresolved $\zeta$ is unconstrained
by the data, and becomes a further free parameter of the model.
Since we may also not know the distance $D$ to the galaxy, it
is convenient to express the dust emission SED of the diffuse component 
from Eq.~\ref{eq:conversion}
as a flux density by dividing throughout by $4{\pi}D^2$, to obtain:

\begin{eqnarray}\label{eq:conversion2}
S^{diff}_{\lambda,\,dust}(SFR,old,\tau^f_B,F,B/D) = 
\frac{1}{4{\pi}}\times
\left(\frac{\theta_{gal}}{h^{disk}_{s,\,ref}}\right)^2\times   \nonumber\\
     \times L_{\lambda,\,dust}^{diff, \,model}(SFR^{model},old^{model},\tau^f_B,F,B/D)
\end{eqnarray}
where 
$\theta_{gal}$ is the half angle subtended at the Earth by the actual
B-band scalelength of the galaxy: 

\begin{eqnarray}\label{eq:theta}
{\theta}_{gal} = \frac{h^{disk}_s}{D}  
\end{eqnarray}
We note that, provided the galaxy is sufficiently
resolved for ${\theta}_{gal}$ to be measured, $S^{diff}_{\lambda,\,dust}$ as 
expressed
by Eq.~\ref{eq:conversion2} depends on the value of $\zeta$ 
(via $SFR$ and $old$ -cf. Eq.~\ref{eq:conversion1}), but is 
independent of the distance $D$.

Our model is constructed such that the total emitted luminosity 
$L_{\lambda,\,star}$
of UV and optical light powering the dust emission can be directly constrained
from available measured apparent UV/optical spatially integrated
fluxes, $S_{\lambda,\,star}^{obs}$, corrected for attenuation (expressed here 
in linear form by Eq.~\ref{eq:atten}) as a function
of $\tau^f_B$, $SFR$, $F$, $old$, $B/D$, $i$ and distance $D$:

\begin{eqnarray}\label{eq:obs_intrinsic}
L_{\lambda,\,star}(\tau^f_B,SFR,F,old,B/D,i,D) 
=  \nonumber\\
4{\pi}D^2\,\frac{S_{\lambda,\,star}^{obs}} 
{A_{\lambda}(\tau^f_B,SFR,F,old,B/D,i)} 
\end{eqnarray}
where, as outlined in Sect.~5 of Paper~III and in Appendix~\ref{sec:composite} of
this paper the dependence on $SFR$, $old$ and $B/D$ is for the
optical range only. 
Depending on the exact range of wavelengths for which 
$S_{\lambda,\,star}^{obs}$ is available, $L_{\lambda,\,star}$ can be integrated 
over wavelength to obtain constraints on the parameters $SFR = SFR^{con}$ 
and/or $old = old^{con}$ as a function of $\tau^f_B$ and $i$.
Specifically, if UV data is available we can 
require that

\begin{eqnarray}\label{eq:conversion3}
\int_{UV} L_{\lambda,\,star}(\tau^f_B,F,i,D)\,d\lambda
= \int_{UV} L^{tdisk}_{\lambda} d\lambda
\end{eqnarray}
Combining Eq.~\ref{eq:sfr_UV} from Sect.~\ref{subsubsec:templateyoung} and 
Eqs.~\ref{eq:obs_intrinsic} and \ref{eq:conversion3}, this leads to:
\begin{eqnarray}\label{eq:sfr_con}
\frac{SFR}{1{\rm M}_{\odot}\,{\rm yr}^{-1}} =
\frac{SFR^{con}(\tau^f_B,F,i,D)}{1{\rm M}_{\odot}
\,{\rm yr}^{-1}} = 4{\pi}D^2 \times \frac{1}{L^{young}_{unit,\,UV}}\nonumber\\ 
 \times \int_{UV} \frac{S_{\lambda,\,star}^{obs}}
   {A_{\lambda}(\tau^f_B,F,i)}\,d\lambda
\end{eqnarray}

Correspondingly, if optical data is available, we can require that
\begin{eqnarray}
\int_{optical}L_{\lambda,\,star}(\tau^f_B,SFR,F,old,B/D,i,D)\,d\lambda = 
\nonumber\\
= \int_{optical}( L^{disk}_{\lambda} +  L^{tdisk}_{\lambda} +
L^{bulge}_{\lambda})\,d\lambda
\end{eqnarray}

After manipulation of Eqs.~\ref{eq:old}, \ref{eq:oldsed}, \ref{eq:bulgesed} and 
\ref{eq:youngsed}
from Sects.~\ref{subsubsec:templatebulge},  \ref{subsubsec:templatebulge} and 
\ref{subsubsec:templateyoung} and Eq.~\ref{eq:obs_intrinsic}, this leads to:
\begin{eqnarray}\label{eq:old_con}
old = old^{con}(\tau^f_B,SFR,F,B/D,i) = 
\frac{(FUNC1 + FUNC2)}{FUNC3}
\end{eqnarray}
where 
\begin{eqnarray}
FUNC1 = 
4\pi\,D^2 \times \int_{optical}\frac{S_{\lambda,\,star}^{obs}}
{A_{\lambda}(\tau^f_B,SFR,F,old,B/D,i)}d\lambda
\end{eqnarray}
\begin{eqnarray}
FUNC2 = - \frac{SFR}{1{\rm M}_{\odot}\,{\rm yr}^{-1}} \times 
\int_{optical}L_{\lambda,\,unit}^{young}d\lambda
\end{eqnarray}
\begin{eqnarray}
FUNC3 = (1 + B/D) \times L_{unit}^{old}
\end{eqnarray}

Note that $old^{con}$ is a weak function of $SFR$ as well as $\tau^f_B$, $B/D$ 
and $i$ due to the need to take into account the contribution of
optical photons by the young stellar population, as described 
in Sect.~\ref{subsubsec:templateold}.

We are now in a position to determine the physical model
parameters from a combined set of measured 
UV/optical flux densities $S_{\lambda,\,star}^{obs}$, measured at wavelengths
${\lambda}^{iobs,\,star}$, and IR/submm flux densities of the pure dust component
(corrected for contamination by direct stellar light at short infrared 
wavelengths) $S_{\lambda,\,dust}^{obs}$, measured at wavelengths 
${\lambda}^{iobs,\,dust}$.
To do so, we minimise the function

\begin{eqnarray}\label{eq:chi2}
\chi2(SFR,old,\tau^f_B,F,B/D)
= \nonumber\\
\sum_{iobs,\,dust}
\left(\frac{S_{\lambda,\,dust}^{obs} -S_{\lambda,\,dust}^{}(SFR,old,\tau^f_B,F,B/D)
 }
{\sigma_{iobs,\,dust}}\right)^2
\end{eqnarray}
subject, if UV data is available, to the constraint
$SFR = SFR^{con}(\tau^f_B,F,i,D)$ from Eq.~\ref{eq:sfr_con} and, if 
optical data is available,
$old = old^{con}(\tau^f_B,SFR,F,B/D,i,D)$ from Eq.~\ref{eq:old_con}. 
$\sigma_{iobs,\,dust}$ are the 1 $\sigma$ uncertainties in the measurements and
$S_{\lambda,\,dust}$ is given by 

\begin{eqnarray}
S_{\lambda,\,dust}(SFR,old,\tau^f_B,F,B/D) = \nonumber \\ 
 S^{diff}_{\lambda,\,dust}(SFR,old,\tau^f_B,F,B/D) +
\frac{L^{local}_{\lambda,\,dust}(SFR,F)}{4\pi D^2}
\end{eqnarray}
where  $S^{diff}_{\lambda,\,dust}$ and $L^{local}_{\lambda,\,dust}$ are given 
by Eq.~\ref{eq:conversion2} and Eq.~\ref{eq:HIItemplate}, respectively.
In the case that the distance, the optical structure and
orientation parameters $D$, $\theta_{gal}$, ${\zeta}$, $B/D$ and $i$ are known
the optimization problem posed by Eq.~\ref{eq:chi2} is reduced from the
parameter set ($SFR$,$old$,$\tau^f_B$,$F$,$B/D$,$\zeta$) to the 
parameter set
($SFR$,$old$,$\tau^f_B$,$F$). With just 4 parameters, this might 
potentially be solvable
purely considering the dust emission data, bearing in mind the orthogonal effect
of these parameters on the amplitude and colour of the dust emission SEDs, 
(as described in Sect.~\ref{sec:predictions}) if at least 4 data points are 
available well sampling the whole MIR/FIR/submm range. However,
this will often be the exception rather than the rule, and in any case
the fit is primarily and more robustly constrained through the 
optical and UV measurements. This is firstly because,
if both UV and optical measurements are available the number of 
the primary search parameters would be reduced from four to just 
two - $\tau^f_B$ and $F$. Secondly, and perhaps more significantly, the model 
would
no longer have any degree of freedom in terms of scaling parameters,
due to the fact that, as noted above, $SFR$ and $old$ are the only
extrinsic parameters in the full parameter set.

Below we illustrate how to use the optical and UV constraints
by giving a possible processing path for a galaxy with a known
(spectroscopically determined) redshift,  for which integrated UV and
optical photometry were available, and for which
$B/D$, $i$ and $\theta_{gal}$ are known from optical imaging.
The parameter set to be determined is thus $\tau^f_B$, $SFR$, $old$, $F$:

\begin{itemize}
\item
step {\it i):}   choose a trial value for each of $\tau^f_B$ and $F$\\
\item
step {\it ii):}  set $SFR$ to $SFR^{con}$ and $old$ to $old^{con}$ from 
                Eqs.~\ref{eq:sfr_con} and \ref{eq:old_con} for the trial 
                values of $\tau^f_B$ and $F$.\\
\item
step {\it iii):} find $SFR^{model}$ and  $old^{model}$ from 
                Eq.~\ref{eq:conversion1}, substituting for $\zeta$ as defined 
                in Eq.~\ref{eq:zeta}. The value of $h^{disk}_s$ used in 
                Eq.~\ref{eq:zeta} should be derived from the optically measured 
                value using the correction factors tabulated as a function of 
                $i$ and $\tau^f_B$ in Tables~1-5 of Paper~IV. \\  
\item
step {\it iv):} using the trial values of $\tau^f_B$ and $F$, together with
                $SFR^{model}$ and $old^{model}$ and $\zeta$ from step {\it iii)} 
                find 
                $L_{\lambda,\,dust}^{diff, \,model}(SFR^{model},old^{model},\tau^f_B,F,B/D)$.\\
\item
step {\it v):}   substitute $L_{\lambda,\,dust}^{diff, \,model}$ from step 
                {\it iv)} into 
                Eq.~\ref{eq:conversion2} to compute
                $S^{diff}_{\lambda,\,dust}$. 
                Compute $L^{local}_{\lambda,\,dust}(SFR^{con},F)$ (note the use
                of the extrinsic value of $SFR=SFR^{con}$ here). 
                Substitute the values for $S^{diff}_{\lambda,\,dust}$ and 
                $L^{local}_{\lambda,\,dust}$ determined at the 
                wavelengths of the IR/submm observations in Eq.~\ref{eq:chi2} 
                to determine $\chi2$ for the trial combination of  $\tau^f_B$ 
                and $F$.\\
\item
step {\it vi):}  repeat steps {\it i)} to {\it v)}, until the combination of
                $\tau^f_B$ and $F$ that minimises $\chi2$ is found. The values
                of $SFR$ and $old$ from steps {\it ii)} to {\it vi)} found for 
                the pair of $\tau^f_B$ and $F$ that minimises $\chi2$ are then 
                best fit parameters in $SFR$ and $old$.\\
\end{itemize}

We note that in the case of edge-on galaxies UV data should not be used to
constrain $SFR$, since typically only a few percent of the total UV disk
luminosity will be seen, and the solution can be subjected to stochastic
variations, because the received photons may only be emitted by a small numbers
of star-forming regions. In this case the $SFR$ is better constrained from the
FIR emission, as modelled for NGC~891 (Paper~I and 
Sect.~\ref{sec:illustration}).

Having found the best fit parameter set, the following two further
steps can be made:

\begin{itemize}
\item
step {\it vii}:  deredden the UV/optical/NIR spectrum using the fitted values 
                 of $\tau^f_B$, $F$, and the measured $B/D$ and $i$\\
\item
step {\it viii}: apply a population synthesis modelling fit to the dereddened 
                 UV/optical spectrum from {\it vii} to find the star formation 
                 history.\\ 
\end{itemize}

\section{Tables describing the fixed parameters of our model}
\label{sec:tables}
\begin{table}[hbt] 
\caption{The geometrical parameters of the model, as defined in Paper III
and Sect.~\ref{subsec:distribution}. All length parameters are 
normalised to the B-band scalelength of the disk of the standard
modal galaxy, as derived for NGC~891:
$h^{disk}_s(B)=h^{\rm disk}_{s, ref}=5670$\,pc .}
\label{tab:geometry}
\begin{tabular}{ll}
\hline 
$h^{\rm disk}_{\rm s}$(B) & 1.000 \\
$h^{\rm disk}_{\rm s}$(V) & 0.966 \\
$h^{\rm disk}_{\rm s}$(I) & 0.869 \\
$h^{\rm disk}_{\rm s}$(J) & 0.776 \\
$h^{\rm disk}_{\rm s}$(K) & 0.683 \\
$z^{\rm disk}_{\rm s}$ & 0.074 \\  
$h^{\rm disk}_{\rm d}$ & 1.406 \\  
$z^{\rm disk}_{\rm d}$ & 0.048 \\   
$h^{\rm tdisk}_{\rm s}$& 1.000 \\
$z^{\rm tdisk}_{\rm s}$& 0.016 \\
$h^{\rm tdisk}_{\rm d}$& 1.000 \\
$z^{\rm tdisk}_{\rm d}$& 0.016 \\
$R_e$                  & 0.229 \\
$b/a$                  & 0.6 \\
\hline
$\frac{\displaystyle \tau^{\rm f, disk}_{\rm B}}
{\displaystyle \tau^{\rm f, tdisk}_{\rm B}}$&  0.387\\
\hline 
\end{tabular}
\end{table}

\begin{table}[hbt]
\caption{The intrinsic spectral luminosity densities
of the old and young stellar populations, as
defined in Sect.~\ref{subsec:intrinsic}, for $old=1$
and $SFR=1\,{\rm M}_{\odot}\,{\rm yr}^{-1}$,  respectively, and
tabulated at the sampling wavelengths of the radiation fields.
}
\label{tab:stellarsed}
\begin{tabular}{rcl}
\hline\hline
            & $old=1$         & $SFR=1\,{\rm M}_{\odot}\,{\rm yr}^{-1}$      \\   
\hline
${\lambda}$ & $L^{old}_{\nu,\,unit}$ & $L^{young}_{\nu,\,unit}$ \\
$\AA$      & W/Hz            & W/Hz             \\   
\hline\hline
            &                      & \\
  912.      &  ...                 & $0.344 \times 10^{21}$\\
 1350.      &  ...                 & $0.905 \times 10^{21}$\\
 1500.      &  ...                 & $0.844 \times 10^{21}$\\
 1650.      &  ...                 & $0.863 \times 10^{21}$\\
 2000.      &  ...                 & $0.908 \times 10^{21}$\\
 2200.      &  ...                 & $0.926 \times 10^{21}$\\
 2500.      &  ...                 & $0.843 \times 10^{21}$\\
 2800.      &  ...                 & $0.910 \times 10^{21}$\\
 3650.      &  ...                 & $1.842 \times 10^{21}$\\
 4430.      & $4.771 \times 10^{21}$ & $2.271 \times 10^{21}$\\
 5640.      & $9.382 \times 10^{21}$ & $3.837 \times 10^{21}$\\
 8090.      & $19.54 \times 10^{21}$ & $5.734 \times 10^{21}$\\
12590.      & $72.20 \times 10^{21}$ & $0.931 \times 10^{21}$\\
22000.      & $64.97 \times 10^{21}$ & $0.728 \times 10^{21}$\\
50000.      & $12.58 \times 10^{21}$ & $0.141 \times 10^{21}$\\
\hline
\end{tabular}
\end{table}

\begin{table}
\caption{The parameters of the dust model considered in this paper
(see Sect.~\ref{subsec:dustmodel}). For 
parameter definition see Weingartner \& Draine (2001).}
\label{tab:dustparam}
\centering
\begin{tabular}{ll}
\hline\hline
\multicolumn{2}{l}{C}\\
parameter &  value \\
\hline
$b_c$ (atoms/H) & $6.0\times 10^{-5}\times 0.93$\\
$b_{c,1}$ & 0.75\\
$b_{c,1}$ & 0.25\\
$a_{01}$ ($\AA$) & 4.0\\
$a_{02}$ ($\AA$) & 20.0\\
${\sigma}_1$ & 0.4 \\
${\sigma}_2$ & 0.55 \\
$C_g$ & $9.99 \times 10^{-12} \times 0.93$ \\
$a_{t,g}$ (${\mu}$m) & 0.0107 \\  
$a_{c,g}$ (${\mu}$m) & 0.428 \\  
${\alpha}_g$ & -1.54 \\ 
${\beta}_g$ & -0.165 \\ 
${\rho}_g$(g/cm$^3$) & 2.24 \\ 
\hline\hline
\multicolumn{2}{l}{Si}\\
parameter &  value \\
\hline
$C_s$ & $1.00 \times 10^{-13} \times 0.93$ \\
$a_{t,s}$ (${\mu}$m) & 0.164\\
$a_{c,s}$ (${\mu}$m) & 0.1\\
${\alpha}_s$ & -2.21\\
${\rho}_s$(g/cm$^3$) & 3.2\\
${\beta}_s$ & 0.3\\
\hline
\end{tabular}
\end{table}

\begin{table}
\caption{The wavelength dependence of the fraction of photons escaping from the
  clumpy component into the diffuse medium as described in
Sect.~\ref{subsec:clumpy}, for $F=F_{cal}=0.35$.}
\label{tab:flambda}
\begin{tabular}{rl}
\hline\hline
${\lambda}$ & $(1-F_{cal}f_{\lambda})$ \\
$\AA$      & \\
\hline\hline
912. & 0.427\\
1350. & 0.484\\
1500. & 0.527\\
1650. & 0.545\\
2000. & 0.628\\
2200. & 0.676\\
2500. & 0.739\\
2800. & 0.794\\
3650. & 0.892\\
4430. & 0.932\\
5640. & 0.962\\
8090. & 0.984\\
12590. & 0.991\\
22000. & 0.994\\
50000. & 0.999\\
\hline
\end{tabular}
\end{table}

\end{document}